%% file: SK-paper2.tex
\documentclass[a4paper,aps,prb,reprint,showpacs,superscriptaddress]{revtex4-1}
\usepackage{amsmath,amssymb,mathrsfs}
\usepackage[pdftex]{graphicx}
\usepackage{epstopdf}

\newcommand{\ii}{\text{i}}
\newcommand{\dd}{\text{d}}
\newcommand{\abs}[1]{\left|#1\right|}
\newcommand{\comm}[2]{\left[#1,#2\right]}

\newcommand{\average}[1]{\langle#1\rangle}

\DeclareMathOperator{\tr}{Tr}
\DeclareMathOperator{\sign}{sign}
\DeclareMathOperator{\re}{Re}
\DeclareMathOperator{\im}{Im}

\allowdisplaybreaks[4]
\usepackage{color}
\input{diagrams/tikz-settings.tikz}

\begin{document}
\title{Transport properties of fully screened Kondo models} 
\author{Christoph B. M. H\"orig}
\affiliation{Institute for Theory of Statistical Physics, RWTH Aachen University and JARA-Fundamentals of Future Information Technology, 52056 Aachen, Germany}
\affiliation{Institute for Theoretical Physics, Center for Extreme Matter and Emergent Phenomena, Utrecht University, Leuvenlaan 4, 3584 CE Utrecht, The Netherlands}
\author{Christophe Mora}
\affiliation{Laboratoire Pierre Aigrain, \'{E}cole Normale Sup\'{e}rieure, Universit\'{e} Paris 7 Diderot, CNRS; 24 rue Lhomond, 75005 Paris, France}
\author{Dirk Schuricht}
\affiliation{Institute for Theoretical Physics, Center for Extreme Matter and Emergent Phenomena, Utrecht University, Leuvenlaan 4, 3584 CE Utrecht, The Netherlands}
\date{\today}
\pagestyle{plain}

\begin{abstract}
We study the non-equilibrium transport properties of fully (exactly) screened Kondo quantum dots subject to a finite bias voltage or a finite temperature. Firstly, we calculate the Fermi-liquid coefficients of the conductance for models with arbitrary spin, i.e. its leading behaviour for small bias voltages or temperatures. Secondly, we determine the low-temperature behaviour of the static susceptibility from the exactly known Bethe Ansatz results for the magnetisation. Thirdly, we study the crossover from strong to weak coupling in the spin-1/2 and the spin-1 models coupled to one or two screening channels respectively. Using a real-time renormalisation group method we calculate the static and dynamical spin-spin correlation functions for the spin-1/2 model as well as the linear and differential conductance and the static susceptibility for the spin-1 model. We define various Kondo scales and discuss their relations. We assess the validity of the renormalisation-group treatment by comparing with known results for the temperature dependence of the linear conductance and static susceptibility as well as the Fermi-liquid behaviour at low energies.
\end{abstract}
\maketitle

\section{Introduction}\label{sec:intro}
The Kondo effect\cite{Hewson93} can be regarded as a paradigm for correlated many-body phenomena in quantum impurities. In the basic setup a localised spin is screened collectively by the spins of itinerant electrons in the surrounding bulk material. The thus formed non-trivial many-body state requires the application of sophisticated many-body methods for its theoretical description. By the mid 1980s the developments of such methods, including perturbative\cite{Anderson70} and numerical\cite{Wilson75,Bulla-08} renormalisation-group (RG) techniques, Fermi-liquid (FL) theory\cite{Nozieres74,Yamada75a,YosidaYamada75,Yamada75b} and the Bethe Ansatz,\cite{Andrei-83,TsvelickWiegmann83,Schlottmann89} had uncovered the essential physics behind the formation of the Kondo singlet, in particular the dynamical generation of a new, non-perturbative energy scale termed the Kondo temperature $T_\text{K}$. Yet the Kondo problem took a revival after it was realised\cite{GlazmanRaikh88,NgLee88,Kawabata91} that it can also be applied to describe transport experiments through quantum dots in the presence of strong Coulomb repulsions, a regime which became experimentally accessible\cite{Goldhaber-98,Cronenwett-98,Schmid-98,Goldhaber-98prl,Simmel-99,Wiel-00,Nygard-00} around the turn of the millennium.

This in turn triggered much interest in the theoretical investigation of the transport properties of quantum dots in the Kondo regime. There are essentially two different parameter regimes. In the first, at high energies compared to the Kondo temperature, e.g. when the applied bias voltage $V$ and/or temperature $T$ is much larger than $T_\text{K}$, the Kondo singlet has not yet formed. The non-equilibrium transport properties can be studied perturbatively in the coupling between the dot and the leads for example using perturbative\cite{Rosch-03prl,Rosch-05,DoyonAndrei06,Schoeller09,SchoellerReininghaus09,Moca-11,PS11} and functional\cite{Gezzi-07,SchmidtWoelfle10,Jakobs-10,Metzner-12}  RG methods or flow-equation techniques.\cite{Kehrein05,Kehreinbook} In the second regime, when all external energies are much lower than the Kondo temperature, $T,V,\ldots\ll T_\text{K}$, the Kondo singlet is fully developed. The spin degree of freedom is frozen out and the dot influences the transport properties merely as a potential scatterer. This allows the application of FL theory, which in particular yields an expansion of the conductance through the dot for small temperatures and voltages encoded in the FL coefficients $c_T$ and $c_V$ respectively.\cite{Oguri05,Rincon-09,Mora-09,SelaMalecki09,Roura-Bas10,Kretinin-11,Munoz-13}

The study of the crossover regime is naturally much more difficult. If the system is driven from the strong-coupling to the weak-coupling regime by increasing the temperature one can resort to standard non-perturbative techniques in equilibrium, in particular the Bethe Ansatz and the numerical RG method. Both methods are well developed and exact up to numerical approximations, while the latter can also be applied very flexibly to different physical setups governed by Kondo correlations. In contrast, if the driving is done by increasing the applied bias voltage the situation is much less understood. To overcome this Pletyukhov and Schoeller\cite{PletyukhovSchoeller12} developed a real-time RG (RTRG) scheme, called $E$-flow since the Laplace variable $E$ is used as flow parameter, which takes into account the generation of the spin relaxation rate and its feedback into the RG flow of the exchange coupling. This allowed the calculation of the differential conductance in the full crossover regime, which was found to be in excellent agreement with perturbative results at weak coupling, the exactly known FL relations, the temperature driven crossover behaviour obtained via the numerical RG method as well as recent experimental data.\cite{Kretinin-12,Klochan-13} 

Another approach based on a slave-boson representation of the Keldysh field integral was recently put forward\cite{SmirnovGrifoni13} by Smirnov and Grifoni. They obtained good agreement with both the numerical RG and RTRG results for the temperature and voltage dependence of the differential conductance respectively. In contrast to the $E$-flow scheme of the RTRG method it has already been possible to extend the analysis to finite magnetic fields.\cite{SmirnovGrifoni13NJP} Both approaches were applied so far to the spin-1/2 Kondo dot or the corresponding single-impurity Anderson model.

In this article we go beyond this and investigate the transport properties of Kondo quantum dots with higher spin $S$. We consider fully screened models where the dot spin is coupled to $N=2S$ screening channels. We first apply FL theory to derive the differential conductance at small temperatures and bias voltages and in particular the FL coefficient $c_V$. Secondly, we determine the low-temperature behaviour of the static susceptibility from the exactly known Bethe Ansatz results for the magnetisation.\cite{TsvelickWiegmann84,TsvelickWiegmann85} We then treat the out-of-equilibrium properties of the spin-1/2 and spin-1 models using the $E$-flow scheme of the RTRG approach. We first generalise it to the calculation of the static and dynamical correlation functions of the spin localised on the dot. We extend the previous analysis\cite{PletyukhovSchoeller12} of the differential conductance to the spin-1 model. The knowledge of the static spin susceptibility in particular allows us to probe the FL behaviour in the RTRG framework quantitatively, i.e. we extract absolute values for the coefficients $c_T$ and $c_V$ which we compare with our FL results.

This article is organised as follows: In Sec.~\ref{sec:model} we define the model and correlation functions. Following this we derive the FL coefficient $c_V $ for general fully screened Kondo dots in Sec.~\ref{sec:FL}. Similarly, in Sec.~\ref{sec:susceptibility} we determine the low-temperature behaviour of the static susceptibility from the exactly known Bethe Ansatz results for the magnetisation. In Sec.~\ref{sec:method} we review the $E$-flow scheme of the RTRG method and generalise it to the calculation of the dynamical correlation functions. Readers who are mainly interested in the results can skip Sec.~\ref{sec:method} and directly proceed to Sec.~\ref{sec:correlations} where we discuss the static susceptibility and dynamical correlation functions. We define the Kondo scale $T_0$ which is used in Sec.~\ref{sec:conductance} to extract the FL coefficients and check them against the results of Sec.~\ref{sec:FL}. We conclude and discuss our results in Sec.~\ref{sec:conclusion}. Some technical details are presented in the appendices. 

\section{Fully screened Kondo dots}\label{sec:model}
\begin{figure}[t]
\centering
\includegraphics[width=0.45\textwidth]{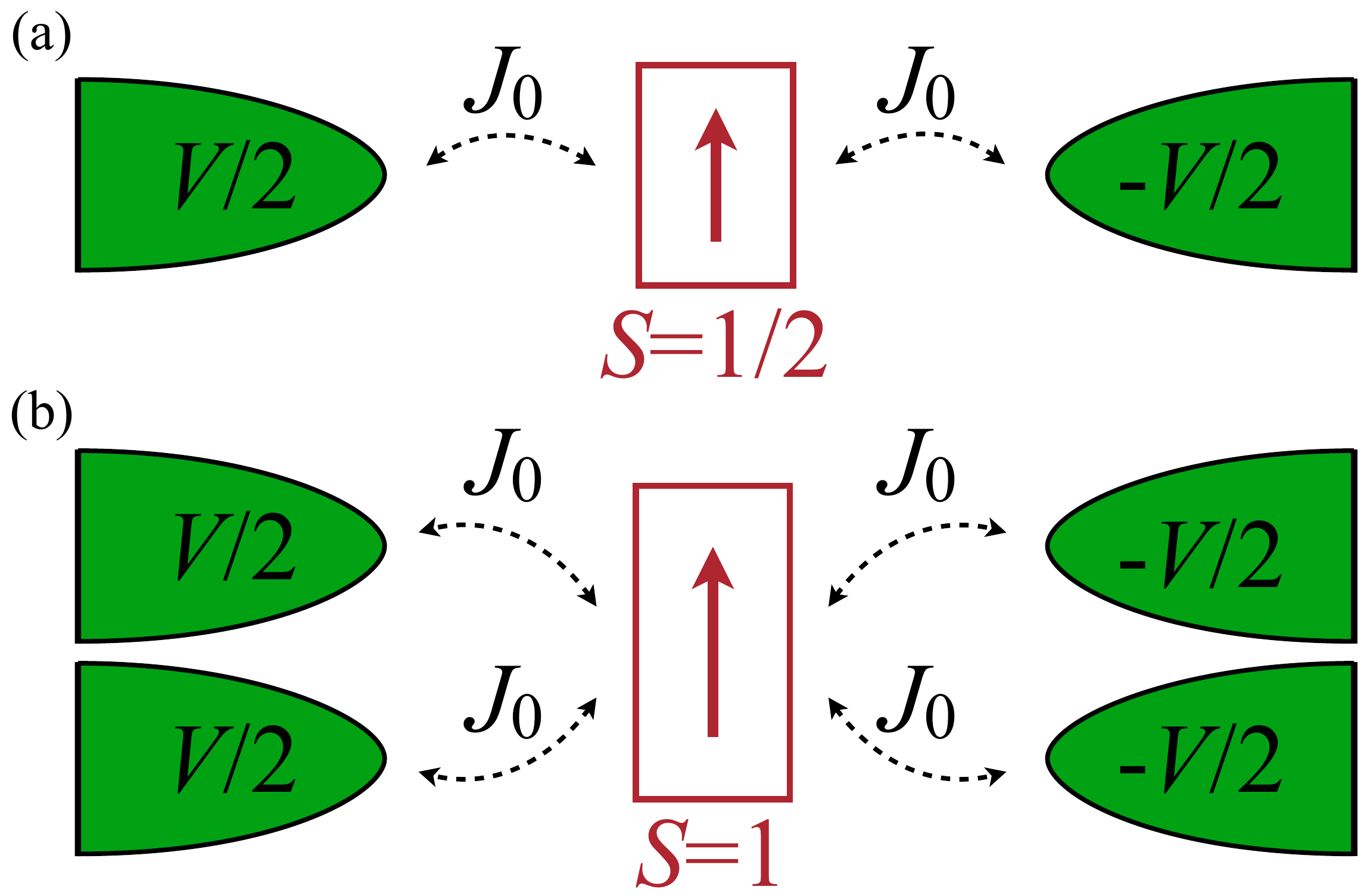}
\caption{(Colour online) Schematic picture of fully (exactly) screened Kondo models. (a) Spin-1/2 dot coupled to one screening channel ($N=2S=1$) which is divided in two leads held at chemical potentials $\mu_\text{L/R}=\pm V/2$. (b) Spin-1 dot coupled to two screening channels ($N=2$).  The left (right) leads of these channels are both held at $\mu_\text{L}=V/2$ ($\mu_\text{R}=-V/2$). We note that the exchange interaction does not mix the screening channels.}
\label{fig:model}
\end{figure}

In this paper we investigate the transport properties of fully screened Kondo quantum dots. Hereby the dot consists of a spin-$S$ which is coupled via exchange interactions to $4S$ electronic leads (see Fig.~\ref{fig:model} for a sketch for $S=1/2$ and $S=1$ respectively). Each pair of leads provides one screening channel, i.e. there are $N=2S$ screening channels in total. Thus the considered models are fully (or exactly) screened at sufficiently low energies. Specifically we consider the unified Hamiltonian
\begin{equation}
	H=\sum_{i\alpha k\sigma}\epsilon_k\,c_{i\alpha k\sigma}^\dagger 
	c_{i\alpha k\sigma}
	+\frac{J_0}{2\nu_0} \sum_{\substack{i\alpha\alpha' \\ k k'\sigma\sigma'}}\vec{S}
	\cdot\vec{\sigma}_{\sigma\sigma'}
	c_{i\alpha k\sigma}^\dagger c_{i\alpha' k'\sigma'}.
  \label{eq:ham}
\end{equation}
Here $\vec{S}$ denotes the spin operator on the dot which is in the spin-$S$ representation of SU(2). $c_{i\alpha k\sigma}^{\dagger}$ and $c_{i\alpha k\sigma}$ create and annihilate electrons with momentum $k$ and spin $\sigma=\uparrow,\downarrow$ in cannel $i=1,\ldots, N$ of lead $\alpha$=L,R, where $N=2S$, and $\vec{\sigma}$ denotes the vector of Pauli matrices. For the leads we assume flat bands of bandwidth $2D$ with the density of states ${N(\omega)=\nu_0 D^2/(D^2+\omega^2)}$. We note that the exchange interaction preserves the channel index $i$ and that the exchange coupling $J_0$ is dimensionless in our convention. The system is subject to a finite bias voltage $V$ with left leads held at $\mu_\text{L}=V/2$ and right leads at $\mu_\text{R}=-V/2$. Alternatively, the leads may be at finite temperature $T$. We use units such that $e=\hbar=k_\text{B}=1$, but reinstate them when appropriate.

In Sec.~\ref{sec:FL} we will consider the low-energy behaviour of the model \eqref{eq:ham} and in particular its conductance for arbitrary spin, while in Secs.~\ref{sec:method}--\ref{sec:conductance} we will analyse the full crossover from low to high energies for the spin-1/2 and spin-1 models sketched in Fig.~\ref{fig:model}.

The observables we consider in this work are the current and the static and dynamical spin-spin correlation functions. The current operator is defined as change in the number of particles in the left leads
\begin{equation}
	\hat{I}_\text{L}\equiv-\frac{\dd}{\dd t}\hat{N}_\text{L}
	\label{eq:current}
\end{equation}
with $\hat{N}_\text{L}=\sum_{ik\sigma}c_{i\text{L}k\sigma}^\dagger c_{i\text{L}k\sigma}$. Alternatively one can use the right leads with $\hat{I}_\text{R}=\dd\hat{N}_\text{R}/\dd t$. 

Furthermore, the longitudinal spin-spin correlation function and susceptibility are given by
\begin{eqnarray}
	\label{eq:S-definition}
	S(t) &=& \frac{1}{2}\Big\langle\comm{S^z(t)_\text{H}-\average{S^z}_\text{st}}{S^z(0)_\text{H}-\average{S^z}_\text{st}}_+\Big\rangle_\text{st},\\
	\label{eq:chi-definition}
	\chi(t) &=& \ii\Theta(t)\Big\langle\comm{S^z(t)_\text{H}}{S^z(0)_\text{H}}_-\Big\rangle_\text{st}.
\end{eqnarray}
Here the spin operators act on the impurity only. The time-evolution of $S^z(t)_\text{H}$ is given in the Heisenberg picture, $S^z(t)_\text{H}=e^{\ii Ht}S^ze^{-\ii Ht}$, and the average $\average{\,\cdot\,}_\text{st}$ refers to the stationary state at finite bias or temperature. We investigate the correlation functions in frequency-space and thus apply the Fourier transformation
\begin{equation}
	S(\Omega)=\int_{-\infty}^\infty \dd t\,\exp(\ii\Omega t)S(t),
\end{equation}
where $\Omega\equiv\Omega\pm\ii\delta$ ($t\gtrless 0$). A similar definition holds for the susceptibility $\chi(\Omega)=\chi'(\Omega)+\ii\chi''(\Omega)$. The static susceptibility of the impurity spin can be obtained via
\begin{equation}
\chi=\lim_{\Omega\to 0}\chi'(\Omega),
\label{eq:staticchi}
\end{equation}
which can be used to define the Kondo scale $T_0$ via\cite{Hewson93,Bulla-08}
\begin{equation}
	\chi(T=0,V=0)\equiv\chi_0= \frac{S(S+1)}{3T_0}.
	\label{eq:defT0}
\end{equation}
Its relation to the Kondo temperature $T_\text{K}$ mentioned above and formally defined in Eq.~\eqref{eq:kondo-temperature} will be discussed in detail below.

In the following section we will first derive the FL coefficients of the conductance in the low-energy regime. Following this we determine the low-temperature behaviour of the susceptibility in Sec.~\ref{sec:susceptibility}. In Sec.~\ref{sec:method} we then present details of our calculations using the RTRG technique in the $E$-flow scheme. The results of the latter are discussed in Secs.~\ref{sec:correlations} and~\ref{sec:conductance}.

\section{Fermi-liquid approach}\label{sec:FL}
\subsection{Effective Hamiltonian}\label{eq:FLham}

Current algebras and the language of conformal field theory give a convenient platform to discuss low-energy physics in quantum impurity models. For the Kondo model, Affleck and Ludwig~\cite{Affleck90,AffleckLudwig91a,AffleckLudwig91b} have shown that the impurity spin is absorbed by the conduction electron spin current in the infrared, i.e. at strong coupling. This absorption causes a rearrangement of the spin sector (conformal towers) and the quasiparticles that emerge after mending the spin with the charge and possibly flavour sectors are not necessarily fermionic objects. Since then the conformal field theory approach was successfully applied to describe non-Fermi liquid features in overscreened versions of the Kondo model.\cite{Affleck90,AffleckLudwig91a,AffleckLudwig91b}

The situation is however much simpler for fully screened models, i.e. when the number of screening channels $N$ is twice the spin of the impurity $S$, $N=2S$. In this case, which we study in this article, the elementary quasiparticles at strong coupling are fermions with a phase shift of $\pi/2$ with respect to the original electrons. In our symmetric source-drain geometry (i.e. $J_0$ does not depend on the lead index $\alpha$=L,R), the unperturbed Hamiltonian at strong coupling reads\cite{Mora-09}
\begin{equation}\label{h0}
H_0 = \sum_{i k \sigma} \varepsilon_k \left( b_{i k \sigma}^\dagger b_{i k \sigma}
+  a_{i k \sigma}^\dagger a_{i k \sigma} \right).
\end{equation}
The operators $b_{i k \sigma}$ and $a_{i k \sigma}$ are respectively even and odd combinations of the original electrons, $c_{i \text{L} k \sigma}\pm c_{i \text{R} k \sigma}$. Only the even modes $b_{i k \sigma}$ carry the $\pi/2$ phase shift. The odd modes are decoupled from the dot variables and they are not involved in the Kondo screening. 

The low-energy behaviour is a FL. It is controlled by the leading irrelevant operator, irrespective of the spin size on the dot,\cite{AffleckLudwig93}
\begin{equation}\label{leading}
H_{\rm LIO} = -\lambda : {\vec J} (0) \cdot {\vec J} (0) :,
\end{equation}
involving only the spin current at the impurity site, $x=0$,
\begin{equation}
{\vec J} (0)  = \frac{1}{2}\sum_{i \sigma \sigma'} b_{i \sigma}^\dagger (0)\vec{\sigma}_{\sigma,\sigma'}b_{i \sigma'} (0)
\end{equation}
with $b_{i \sigma} (x) = \sum_k b_{i k \sigma}e^{\ii k x}$. The notation $:\ldots :$ corresponds to normal ordering where all divergencies stemming from bringing the two spin currents close to each other are subtracted. $\lambda$ is a coupling constant of order $\sim 1/T_\text{K}$. Following a standard point-splitting procedure,\cite{AffleckLudwig93,LeHur-07,Mora09} we obtain the Hamiltonian corrections 
$H_{\rm LIO} = H_{\rm el} + H_{\rm int}$ to the fixed point Eq.~\eqref{h0}, with
\begin{equation}\label{hel}
H_{\rm el} =  - \frac{\alpha_1}{2 \pi \nu_1} \sum_{i,\{ k_l \},\sigma} 
(\varepsilon_{k_1} + \varepsilon_{k_2}) : b_{i k_1 \sigma}^\dagger b_{i k_2 \sigma}:
\end{equation}
and
\begin{equation}\label{hint}
\begin{split}
 H_{\rm int} & = \frac{\phi_1}{3 \pi \nu_1^2} \sum_{i<j,\{ k_l \},\sigma} 
: b_{i k_1 \sigma}^\dagger b_{j k_2 \sigma}
b_{j k_3 \sigma}^\dagger b_{i k_4 \sigma}  : \\
& + \frac{\phi_1}{\pi \nu_1^2} \sum_{i,\{ k_l \}} 
: b_{i k_1 \uparrow}^\dagger b_{i k_2 \uparrow}
b_{i k_3 \downarrow}^\dagger b_{i k_4 \downarrow}  : \\
& + \frac{\phi_1}{3 \pi \nu_1^2} \sum_{i\ne j,\{ k_l \}}
\bigg(2: b_{i k_1 \uparrow}^\dagger b_{j k_2 \uparrow}
b_{j k_3 \downarrow}^\dagger b_{i k_4 \downarrow}  : \\
& \qquad \qquad \qquad+ : b_{i k_1 \uparrow}^\dagger b_{i k_2 \uparrow}
b_{j k_3 \downarrow}^\dagger b_{j k_4 \downarrow}  : \bigg)
\end{split}
\end{equation}
where $\alpha_1= \phi_1=(3 \lambda/2) \pi \nu_1^2$ and $\nu_1 = 1/(2 \pi \hbar v_\text{F})$ is the density of states for one-dimensional chiral fermions. The four terms in $H_{\rm int}$ describe two-electron scattering processes caused by the impurity. In the first term, the interacting electrons belong to the same spin species but to different channels while, in the second term, opposite spin electrons interact within the same channel. In the third and fourth term, both spins and channels are different, they are exchanged after scattering in the former but not in the latter.

The effect of the leading irrelevant operator $H_{\rm LIO}$ on observables can be separated into three types of corrections: (i) the elastic scattering due to $H_{\rm el}$, (ii) the Hartree contributions deduced from $H_{\rm int}$, they can be seen~\cite{Mora09} as elastic processes since the energy of the incoming electron is conserved, (iii) apart from Hartree diagrams, all other diagrams derived from  $H_{\rm int}$ describe inelastic processes in which the incoming electron changes its energy by exciting an electron-hole pair. The types (i) and (ii) can be gathered in the total phase-shift
\begin{equation}\label{pshift}
\begin{split}
\delta_{i \sigma} (\varepsilon,\delta n_{j,\sigma'})
=&\;\delta_0   + \alpha_1 \varepsilon 
  - \phi_1 \delta n_{i,\bar{\sigma} } \\[1mm] 
& + \frac{\phi_1}{3} \sum_{j\ne i}  \left(  \delta n_{j,\sigma} - \delta n_{j,\bar{\sigma}} \right)
\end{split}
\end{equation}
accumulated by a lead electron that is elastically scattered by the impurity. $\varepsilon$ is the energy of the electron measured with respect to some reference energy $\varepsilon_0=0$. $\delta n_{i,\sigma}$ is the total density of spin $\sigma$ electrons in channel $i$ with respect to the zero temperature Fermi sea with Fermi energy $\varepsilon_0$. The identity $\alpha_1= \phi_1$ is sufficient to ensure the invariance of the phase shift Eq.~\eqref{pshift} upon a shift of $\varepsilon_0$. The form of the Hamiltonian Eqs.~\eqref{hel},~\eqref{hint} and the phase shift Eq.~\eqref{pshift} were first anticipated by  Nozi\`eres and Blandin~\cite{NozieresBlandin80} on the basis of a phenomenological FL approach. 

Below, in Sec.~\ref{sec:elastic} and Sec.~\ref{sec:inelastic}, we compute the mean current by adapting the formalism developed in Refs.~\onlinecite{Mora-08,Mora-09}, see also Ref.~\onlinecite{Vitushinsky-08}.

\subsection{Current calculation, elastic part}\label{sec:elastic}
Instead of using the definition Eq.~\eqref{eq:current}, we start from an alternative expression for the current, discussed in App.~\ref{appen-current},
\begin{equation}\label{current}
\hat{I}  = \frac{1}{2 \nu_1 h} \sum_{i,\sigma}  
\Bigl[ a^\dagger_{i \sigma} (x) b_{i \sigma} (x) - a_{i \sigma}^\dagger(-x) {\cal S} b_{i \sigma}(-x) + {\rm h.c.} \Bigr] 
\end{equation}
where we use the symmetrised current $\hat{I} = (\hat{I}_\text{L}+\hat{I}_\text{R})/2$ and $h=2\pi\hbar=2\pi$. The choice of $x<0$ is arbitrary due to current conservation.  In the simplest approach, the ${\cal S}$ matrix contains solely the phase shift $\pi/2$. It is however possible to simplify the problem by including the types (i) and (ii) contributions directly in ${\cal S} = e^{2 \ii \delta}$ with $\delta$ given by Eq.~\eqref{pshift}.

In the absence of type (iii) contributions, the fields $b_{i k \sigma}$ and $a_{i k \sigma}$ are free (non-interacting) with occupations controlled by the left and right lead chemical potentials $\mu_\text{L/R} = \pm V/2$, see App.~\ref{appen-current}. The calculation of the mean current is straightforward, and takes a Landauer-B\"uttiker form\cite{BlanterBuettiker00}
\begin{equation}\label{landauer}
I_{\rm el} = \frac{2 N}{h}  \int_{-\infty}^{+\infty} d \varepsilon\, T(\varepsilon)
[ f_{\rm L} (\varepsilon) - f_{\rm R} (\varepsilon) ],
\end{equation}
with the  transmission $T(\varepsilon) = \sin^2 [\delta(\varepsilon)]$. $f_\text{L/R} (\varepsilon)$ are the Fermi functions for the left and right leads. In this configuration, the phase shift is obtained from  Eq.~\eqref{pshift}, 
\begin{equation}
\delta (\varepsilon) = \frac{\pi}{2} + \alpha_1 \varepsilon,
\end{equation}
since $\delta n_{j,\sigma} = 0$ when $\varepsilon_0=0$ is chosen in the middle of the lead chemical potentials. Expanding the elastic current Eq.~\eqref{landauer} up to second order in $\alpha_1$, we obtain reinstating the electrical charge $e$, 
\begin{equation}\label{elastic-current}
I_{\rm el} = \frac{4 S e^2 V}{h} \left[ 1 - \alpha_1^2 \left( \frac{(\pi T)^2}{3} + \frac{(e V)^2}{12} \right) \right],
\end{equation}
and the linear conductance
\begin{equation}
G_{\rm el} = \frac{4 S e^2}{h} \left[ 1 - \alpha_1^2 \left( \frac{(\pi T)^2}{3} + \frac{(e V)^2}{4} \right) \right].
\end{equation}

\subsection{Current calculation, inelastic part}\label{sec:inelastic}
\begin{figure}
\includegraphics[width=7.cm]{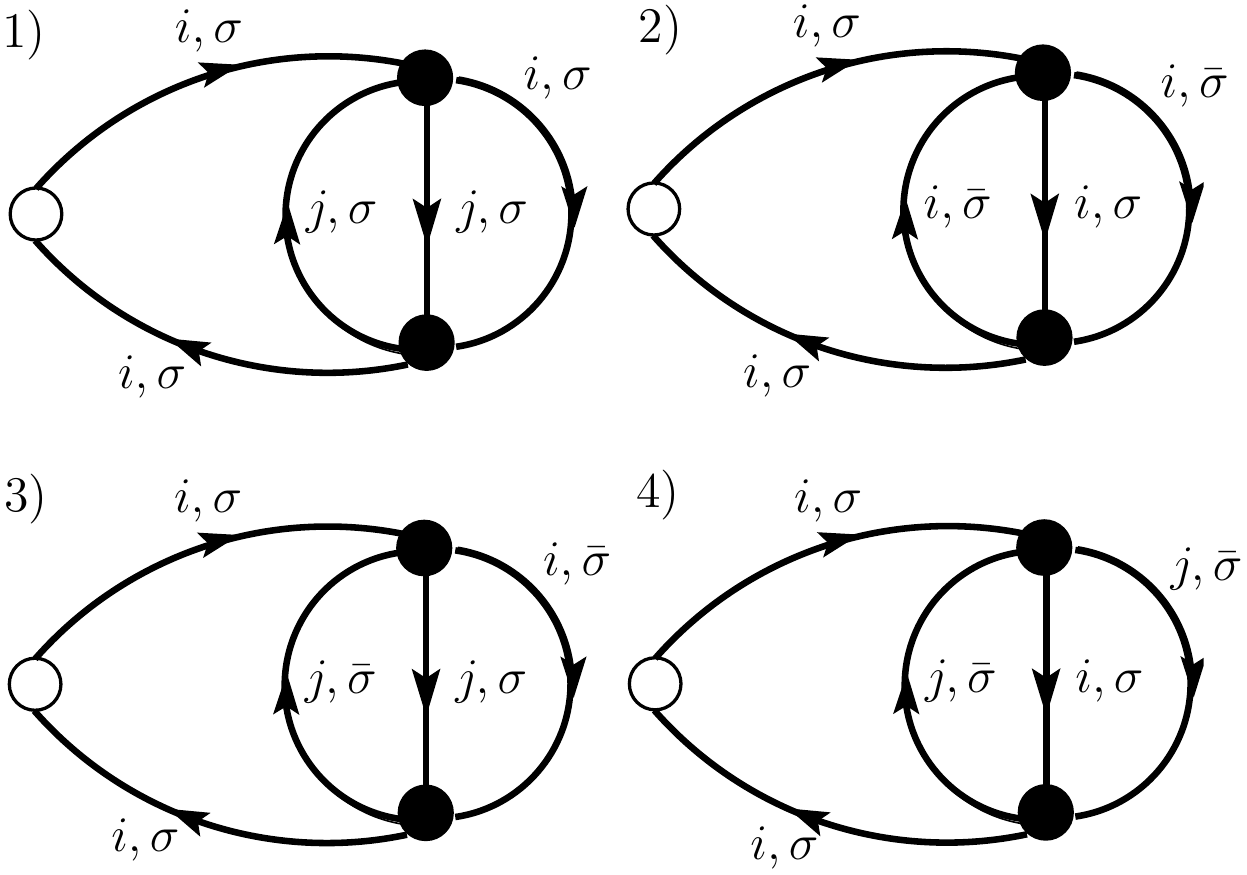}
\caption{\label{fig-diagrams} Second-order diagrams describing the inelastic processes, or type (iii), in the mean current calculation. The open circles represent a current vertex while filled black dots correspond to interaction vertices. The four diagrams correspond to the four terms in Eq.~\eqref{hint} with interaction vertices proportional to $\phi_1/3$, $\phi_1$, $2 \phi_1/3$ and $\phi_1/3$.}
\end{figure}
We use the Keldysh framework~\cite{KamenevLevchenko09} in order to compute the type (iii) contributions to the current. Quite generally, the mean current reads
\begin{equation}\label{current2}
I = \Big\langle T_\text{c} \hat{I} (t) e^{-\ii \int_{\cal C} d t' H_{\rm int} (t')}\Big\rangle,
\end{equation}
where the Keldysh contour ${\cal C}$ runs along the forward time direction on the branch $p=+$ followed by a backward evolution on the branch $p=-$, and $T_\text{c}$ is the corresponding time ordering operator. Evolution and mean values are determined by the free Hamiltonian $H_0$, Eq.~\eqref{h0}, in which all elastic processes have been incorporated. The calculation proceeds as follows: we expand the current Eq.~\eqref{current2} up to second order in $H_{\rm int}$ and compute the resulting mean values in Keldysh space. The zeroth order reproduces the elastic current derived in Eq.~\eqref{elastic-current}. The first order vanishes and the second-order diagrams are shown in Fig.~\ref{fig-diagrams} corresponding to the four terms in Eq.~\eqref{hint}. The calculation is the same for all diagrams, with a result proportional to
\begin{equation}
\frac{2 (\pi T)^2}{3} + \frac{5 (e V)^2}{12},
\end{equation}
but with different weight factors depending on spin and channel summations. The diagrams in Fig.~\ref{fig-diagrams} describe the sum of  uncorrelated processes~\cite{GogolinKomnik06} in which one or two electrons are incoherently transmitted from one lead to the other.~\cite{Sela-06}

Summing all terms, we obtain the current $I = I_{\rm el} + \delta I_{\rm inel}$ with
\begin{equation}
\delta I_{\rm inel} = - \frac{2 N e^2 V}{h} \phi_1^2 \, W_N \left[ \frac{2 (\pi T)^2}{3} + \frac{5 (e V)^2}{12} \right]
\end{equation}
and the numerical factor
\begin{eqnarray}
W_N&=&1 + (N-1) \left( \frac{1}{3} \right)^2 + (N-1) \left[ \left( \frac{1}{3} \right)^2  + \left( \frac{2}{3} \right)^2 \right]\nonumber\\
&=&\frac{1+2N}{3}
\end{eqnarray}
where $N=2S$ denotes the number of screening channels. With $\phi_1 = \alpha_1$, the final result for the linear conductance takes the form
\begin{equation}
G(T,V) = G_0 \left[ 1 - c_T' \left(\frac{T}{T_0}\right)^2 - c_V' \left(\frac{e V}{T_0}\right)^2 \right],
\label{eq:FLconductance}
\end{equation}
with the unitary conductance 
\begin{equation}
G(0,0)\equiv G_0=\frac{2e^2N}{h}=\frac{4e^2S}{h},
\label{eq:unitaryconductance}
\end{equation} 
i.e. each of the $N$ channels contributes one conductance quantum $2e^2/h$, and the coefficients
\begin{equation}\label{fermipar1}
\frac{c_T'}{\alpha_1^2 T_0^2} =  \pi^2 \frac{5+8 S}{9},\quad
\frac{c_V'}{\alpha_1^2 T_0^2} =  \frac{4+10 S}{6}.
\end{equation}
The Kondo scale $T_0$ used in Eq.~\eqref{eq:FLconductance} was defined via the static susceptibility in Eq.~\eqref{eq:defT0}. In the ratio of the FL coefficients the scale $T_0$, which is hard to determine experimentally, drops out and we find
\begin{equation}
\frac{c_V'}{c_T'} = \frac{3}{2 \pi^2} \, \frac{4+10 S}{5+8 S},
\label{eq:cVcT}
\end{equation}
which reproduces the known result\cite{Oguri05,Rincon-09,Mora-09,SelaMalecki09,Roura-Bas10,Kretinin-11,Munoz-13} $c_V'/c_T'=3/(2 \pi^2)$ for $S=1/2$.

The relation between the FL parameter $\alpha_1$ and the Kondo temperature $T_0$ can be made quantitative by computing the static susceptibility within the FL approach. We consider the zero temperature situation with a weak magnetic field $B$ splitting the chemical potential for the two spin species, $\mu_\sigma = \sigma B/2$. The relative densities are then calculated to be $\delta n_{i,\sigma} =\sigma B/2$ and the phase shifts, from Eq.~\eqref{pshift},
\begin{equation}
\delta_{i \sigma} (\varepsilon) = \frac{\pi}{2} + \alpha_1 \varepsilon + \phi_1 \sigma B \left( \frac{1}{2} + \frac{N-1}{3} \right)
\end{equation}
determine the spin populations on the dot through the Friedel sum rule. The dot magnetisation is thus given by $M= 1/(2 \pi) \sum_{i \sigma} \sigma \delta_{i \sigma} (\sigma B/2)$ and we obtain for the static susceptibility 
\begin{equation}
\chi =  \frac{4\alpha_1}{3 \pi} S (S+1).
\end{equation}
Comparing this expression with Eq.~\eqref{eq:defT0}, we find the relation $\alpha_1 = \pi/(4 T_0)$ and, from Eq.~\eqref{fermipar1}, the FL parameters 
\begin{equation}\label{eq:cT}
c_T' =  \frac{ \pi^4}{144}(5+8 S),
\end{equation}
and
\begin{equation}\label{eq:cV}
c_V' =  \frac{\pi^2}{96} (4+10 S),
\end{equation}
in agreement with Refs.~\onlinecite{Yoshimori76,MihalyZawadowski78} for the value of $c_T'$. The result Eq.~\eqref{eq:cVcT} and thus for $c_V'$ was, to the best of our knowledge, not obtained previously. We would like to add that recently\cite{Hanl-14} Hanl \emph{et al.} applied FL theory to derive the coefficients  characterising the magnetic-field dependence of the conductance and the curvature of the equilibrium Kondo resonance, respectively, for fully screened models, and compared them to numerical RG calculations. This work thus complements our derivation of the voltage dependence $c_V'$.

\section{Low-temperature behaviour of the static susceptibility}\label{sec:susceptibility}
The aim of this section is the derivation of the coefficient $a_T'$ in the FL expansion
\begin{equation}
\chi=\chi_0\left[1-a_T'\left(\frac{T}{T_0}\right)^2\right].
\label{eq:lowTsusc}
\end{equation}
We note that this coefficient cannot be calculated by the FL approach of Sec.~\ref{sec:FL} since the next-to-leading order perturbation\cite{LesageSaleur99,LesageSaleur99npb} around the fixed-point Hamiltonian would be required. However, we can use  the low-temperature behaviour of the dot magnetisation in an external magnetic field $B$ as our starting point, which has been derived using the Bethe Ansatz.\cite{TsvelickWiegmann84,TsvelickWiegmann85} For small fields the magnetisation is given by
\begin{equation}
\begin{split}
M=&\frac{1}{\sqrt{\pi}}\sum_{k=0}^\infty\frac{(-1)^k}{k!(2k+1)}\frac{\Gamma(k+\tfrac{1}{2})}{\Gamma\bigl(N(k+\tfrac{1}{2})\bigr)}\\
&\times\left(\frac{N(k+\tfrac{1}{2})}{e}\right)^{N(k+\tfrac{1}{2})}\left(\frac{B}{T_\text{H}}\right)^{2k+1},
\end{split}
\end{equation}
where, as usual, $N=2S$. The relation between the energy scale $T_\text{H}$ and the Kondo temperature $T_0$ defined via Eq.~\eqref{eq:defT0} is easily obtained using $\chi=\partial M/\partial B|_{B=0}$ to be
\begin{equation}
\frac{T_0}{T_\text{H}}=\frac{\Gamma(S)}{3}\left(\frac{e}{S}\right)^{S}S(S+1).
\end{equation}

To determine the second derivative of the susceptibility with respect to the temperature we start with the thermodynamic relation
\begin{equation}
\frac{\partial^2\chi}{\partial T^2}=\frac{1}{T}\frac{\partial^2 C}{\partial B^2},
\end{equation}
where the specific heat is linear at low temperatures $C=\gamma T$ with\cite{Tsvelick85} $\gamma=\pi^2 S/(3T_0)$. Thus the Wilson ratio is given by\cite{Hewson93,Yoshimori76,MihalyZawadowski78,NozieresBlandin80}
\begin{equation}
R=\frac{4\pi^2}{3}\frac{\chi}{\gamma}=\frac{2(N+2)}{3},
\label{eq:Wilsonratio} 
\end{equation}
where we have set $g\mu_\text{B}/k_\text{B}=1$. As can be shown in the FL approach the Wilson ratio is independent of the magnetic field (see App.~\ref{app:susceptibliity}), hence we obtain
\begin{equation}
\frac{\partial^2 \gamma}{\partial B^2}=\frac{2\pi^2}{N+2}\frac{\partial^2 \chi}{\partial B^2}.
\end{equation}
In total we thus have
\begin{equation}
a_T'=-\frac{T_0^2}{2\chi_0}\frac{\partial^2\chi}{\partial T^2}\bigg|_{T=0}=-\frac{T_0^2}{2\chi_0}\frac{2\pi^2}{N+2}\frac{\partial^3 M}{\partial B^3}\bigg|_{B=0}.
\end{equation}
Straightforward evaluation then gives 
\begin{equation}
a_T'=\frac{\pi^2 S^2(S+1)}{18}\frac{\Gamma(S)^3}{\Gamma(3S)}3^{3S},
\label{eq:aT}
\end{equation}
which for $S=1/2$ simplifies to\cite{Melnikov82,footnote4} $a_T'=\sqrt{3}\pi^3/8$. The result Eq.~\eqref{eq:aT} for $S=1$ is found to be consistent with recent numerical RG data.\cite{footnote6}  Unfortunately, since the Bethe Ansatz is not applicable in the presence of a finite bias voltage, it is not possible to derive the similarly defined coefficient $a_V'$ for the dependence of the susceptibility on small voltages.

After the investigation of the systems properties at small temperatures or voltages we now turn to the treatment of the crossover regime using the RTRG technique.

\section{Renormalisation-group treatment}\label{sec:method}
In this section we will present details of the calculation of the non-equilibrium transport properties of the fully screened Kondo model \eqref{eq:ham}. We will begin by reviewing the $E$-flow scheme\cite{PletyukhovSchoeller12} of the RTRG technique,\cite{Schoeller09} which was developed to study the differential conductance of the spin-1/2 model in the full crossover regime from weak to strong coupling and has been successfully applied to describe transport measurements\cite{Kretinin-12,Klochan-13} in quantum dots. We then extend this to the calculation of the dynamical spin-spin correlation functions as well as the static susceptibility. The resulting RG equations, presented in Sec.~\ref{sec:rgequation}, are solved numerically to obtain the results of Secs.~\ref{sec:correlations} and~\ref{sec:conductance}. 

We stress that the RTRG treatment presented here is restricted to fully screened models with $S=1/2$ or $S=1$ sketched in Fig.~\ref{fig:model}. While we always consider $N=2S$ we will keep the notations $N$ and $S$ simultaneously to clarify the origin of the different terms. Furthermore, we stress that the derivations below are based on a weak-coupling expansion in the renormalised exchange coupling between the spin on the dot and the electron spins in the leads. Thus weak-coupling results are intrinsically incorporated. An advantage of this expansion is that higher orders can be included in a systematic way, thus offering an internal consistency check when comparing observables in different orders of  truncation. We focus on the scaling limit of vanishing initial exchange interaction ($J_0\to 0$) and diverging bandwidth ($D\to\infty$) such that the Kondo scale $T_\text{K}$ defined below is kept constant, since in this limit universal behaviour is expected.

\subsection{General formalism}
For completeness we recall here the main steps in the $E$-flow scheme of the RTRG technique, for a more detailed presentation we refer to the original reference.\cite{PletyukhovSchoeller12} The central object of our study is the reduced density matrix of the dot denoted by $\rho$, which is obtained from the full density matrix of the system by tracing out the electronic degrees of freedom in the leads. After a Laplace transformation, 
\begin{equation}
  \rho(E)=\int_{t_0}^\infty \dd t\,e^{\ii E(t-t_0)}\,\rho(t)
\end{equation}
its evolution is governed by the effective Liouvillian $L(E)$ via\cite{footnote1}
\begin{equation}
  \rho(E)={\ii\over E-L(E)}\,\rho_0,
\label{laplace_rho}
\end{equation}
where $\rho_0$ denotes the initial density matrix of the dot at time $t_0$ and the leads are assumed to be initially in grandcanonical distributions incorporating the chemical potentials $\mu_\text{L/R}=\pm V/2$ or the temperature $T$. The stationary state is reached for $E=\ii 0^+$, which is equivalent to $t_0\to -\infty$. The stationary reduced density matrix is therefore given by
\begin{equation}
	\rho^\text{st}=\lim_{E\to \ii 0^+}\frac{E}{E-L(E)}\rho_0.
\end{equation}
In the following we will consider the stationary state only. Due to the absence of a magnetic field its non-vanishing elements are simply given by $\rho_{\uparrow\uparrow}=\rho_{\downarrow\downarrow}=1/2$ for the spin-1/2 model and $\rho_{11}=\rho_{00}=\rho_{-1-1}=1/3$ for the spin-1 model, respectively.

By expanding in the exchange interaction and performing the trace over the reservoir (i.e. lead) degrees of freedom one can derive\cite{Schoeller09} a series expansion for the effective Liouvillian, which consists of two-point interaction vertices $G_{12}(E,\bar\omega_1,\bar\omega_2)$ and propagators
\begin{equation}
	\Pi(E+\bar\omega)=\frac{1}{E+\bar\omega-L(E+\bar\omega)}.
	\label{eq:resolvent}
\end{equation}
The multi-index notation $1\equiv\eta i\alpha\sigma$ incorporates the channel and lead index $i$ and $\alpha$, the spin $\sigma$ and $\eta=\pm$ refers to the creation and annihilation operators of lead electrons. The frequency variable $\omega$ describes the energy of the electrons in the reservoir contractions. For convenience we define $\bar1\equiv-\eta i\alpha\sigma$, where the minus acts on $\eta$ only, as well as $\bar\omega=\eta\omega$. Both, vertex and propagator, are $E$-dependent due to the resummation of diagrams into effective quantities. Up to third order in the renormalised exchange coupling we can use the approximations
\begin{equation}
	G_{12}(E)\equiv G_{12}(E, \bar\omega_1=0, \bar\omega_{2}=0)
	\label{eq:vertex-approximation}
\end{equation}
and 
\begin{equation}
	\Pi(E+\bar\omega) \approx \sum_j\frac{Z_j(E)}{\bar\omega+\chi_j(E)}P_j(E).
	\label{eq:resolvent-approximation}
\end{equation}
Here $j$ runs over all eigenvalues $\lambda_j(E)$ of the effective Liouvillian and $P_j(E)$ are the corresponding projectors, $L(E)=\sum_j\lambda_j(E)P_j(E)$. Furthermore, we have introduced $Z_j(E)=1/[1-\frac{\partial}{\partial E}\lambda_j(E)]$ and $\chi_j(E)=Z_j(E)[E-\lambda_j(E)]$. Physically the eigenvalues $\lambda_j(E)$ of the Liouvillian correspond to the relaxation rates of the spin on the dot,\cite{SchoellerReininghaus09,HSA12} i.e. $\lambda_j(E)=\ii\,\Gamma_j(E)$. In the absence of an external magnetic field there is only one relaxation rate in the spin-1/2 model and two in the spin-1 model. Since in the latter only the triplet rate contributes in the following, we will consider only one rate $\Gamma$ for both models in the following.

In analogy to the effective dot Liouvillian one can introduce\cite{Schoeller09} a current kernel $\Sigma_\text{L}(E)$ from which the current \eqref{eq:current} follows as 
\begin{equation}
	I\equiv\average{\hat{I}_\text{L}}=-\ii\tr\Bigl[\Sigma_\text{L}(\ii 0^+)\rho^\text{st}\Bigr].
\end{equation}
Here the trace is taken over the Liouville space of the Hilbert space of the dot. In analogy to the interaction vertex $G_{12}$ the treatment of the current requires a current vertex which we denote by $I_{12}$.

The main goal of this paper is the computation of the local spin-spin correlation functions $S(\Omega)$ and $\chi(\Omega)$. In order to treat both on equal footing we introduce\cite{SS09} the auxiliary functions 
\begin{equation}
	C_{AB}^\pm(\Omega)=\int_{-\infty}^0\dd t\,e^{-\ii\Omega t}
	\average{\comm{A(0)_\text{H}}{B(t)_\text{H}}_\pm}_\text{st},
\end{equation}
where $A$ and $B$ are in principle two arbitrary operators with the corresponding superoperators in Liouville space given by
\begin{equation}
	\mathcal{A}=\frac{\ii}{2}\comm{A}{\cdot}_+,\quad
	\mathcal{B_\pm}=\ii\comm{B}{\cdot}_\pm.
\end{equation}
For the case at hand we have $A=B=S^z$ and 
\begin{equation}
	S(\Omega) = \re C^+_{S^zS^z}(\Omega), \quad \chi(\Omega) =\ii C^-_{S^zS^z}(\Omega).
	\label{eq:correlation-connection-C_AB}
\end{equation}
Again by expanding in the interaction part of the Liouvillian and resumming the resulting diagrams into irreducible kernels $\Sigma_{A}$ and $\Sigma_B^\pm$ we obtain
\begin{equation}
	C_{AB}^\pm(\Omega)=-\ii\tr\left[\Sigma_A(\Omega)\frac{1}{\Omega-L(\Omega)}
	\Sigma_B^\pm(\ii 0^+,\Omega)\rho^\text{st}\right].
	\label{eq:C_AB}
\end{equation}
Due to the simple structure of the spin operators on the dot the first kernel is given by its initial value, $\Sigma_A(\Omega)=\mathcal{A}$, while the second one will acquire a dependence on $E$ during the RG flow. In Sec.~\ref{sec:Eflowcorrelations} we will present a detailed derivation of the RG equations for\cite{footnote2} $\Sigma_B^\pm(E,\Omega)$.

We note that while the investigation of more general correlation functions like the finite-frequency current noise\cite{Korb-07,Moca-10,Moca-11,Basset-12,Chung-13} is in principle possible within the RTRG formalism,\cite{MPSA13} the analysis of the strong-coupling regime using the $E$-flow scheme will be considerably more complicated than the one of the spin-spin correlations worked out below.\cite{footnote3} 

\subsection{Parametrisation in Liouville space}\label{sec:parametrization}
Before deriving the RG equations we parametrise the various quantities introduced above using a suitable basis in Liouville space as discussed in App.~\ref{app:parametrization}. Specifically the Liouvillian is recast using a function $\Gamma(E)$, the interaction vertex is represented by the functions $J_{12}(E)$ and $K_{12}(E)$, the current kernel by $\Gamma_\text{L}(E)$, the current vertex by $I_{12}^\text{L}(E)$ and the correlation kernels $\Sigma_B^\pm(E,\Omega)$ by $\Gamma^\pm(E,\Omega)$.

With these quantities the observables of interest can be expressed as 
\begin{equation}
	I=\ii\pi\frac{2e^2}{h}\lim_{E\to\ii 0^+}\frac{1}{E}\Gamma_\text{L}(E),
\end{equation}
for the stationary current and 
\begin{equation}
	G=\frac{h}{2e^2}\lim_{E\to\ii 0^+}\frac{E}{\ii}\frac{\partial}{\partial V}I(E)
	=\pi\frac{\partial}{\partial V}\Gamma_\text{L}(\ii 0^+)
	\label{eq:conductance}
\end{equation}
for the differential conductance. Similarly the auxiliary correlation functions read
\begin{align}
	\label{eq:CPlus}
	C_{S^zS^z}^+(\Omega)&=\frac{2\ii}{3} \frac{S(S+1)}{\Omega+\ii\Gamma(\Omega)},\\
	\label{eq:CMinus}
	C_{S^zS^z}^-(\Omega)&=\frac12\frac{\Gamma^{-}(\ii 0^+,\Omega)}{\Omega+\ii\Gamma(\Omega)},
\end{align}
from which the spin-spin correlation function $S(\Omega)$ and dynamical susceptibility $\chi(\Omega)$ can be easily obtained via Eq.~\eqref{eq:correlation-connection-C_AB}. We note that $\Gamma^+(\ii 0^+,\Omega)$ does not appear in Eq.~\eqref{eq:CMinus}.

\subsection{\emph{E}-flow scheme}\label{sec:Eflow}
As is well known, the perturbative treatment of the Kondo model leads to logarithmic divergencies which makes a more careful RG approach necessary. In the $E$-flow scheme\cite{PletyukhovSchoeller12} the RG procedure is set up such that all integrals, which originate from performing the reservoir contractions using Wick's theorem, are UV-convergent in the limit $D\to\infty$. This is accomplished by taking the derivative of the full diagrammatic series with respect to the Laplace variable $E$, which serves as the natural flow parameter (hence the name). For the Liouvillian this requires taking the second derivative while for the vertex one derivative is sufficient to ensure UV-convergence. This yields self-consistent RG equations which are truncated systematically in orders of the interaction vertex up the third order, i.e. including $\mathcal{O}(G^3)$. At $T=V=0$ the Laplace variable can be written as $E=\ii\Lambda$. The RG-flow starts at high energy $E=\ii\Lambda_0$, with $\Lambda_0$ of the order of the bandwidth $D$, where the RG procedure agrees with perturbation theory. The initial values of all flowing quantities are fixed by the unitary conductance at low energies as we elaborate on in Sec.~\ref{sec:initial}. 

For example, at $T=V=\Omega=0$ the RG equation for the interaction vertex $G_{12}$ leads to (see App.~\ref{app:parametrization} for the parametrisation) 
\begin{equation}
	\label{eq:scaling-equation}
	Z\frac{\dd\tilde{J}}{\dd\Lambda}=-\frac{1}{\Lambda+\Gamma}2\tilde{J}^2(1-N\tilde{J}),
\end{equation}
where $\tilde{J}=ZJ$, $J\equiv J(E)$ is the effective coupling constant and $Z=1/(1+\frac{\dd\Gamma}{\dd\Lambda})$ is the Z-factor. The RG equation \eqref{eq:scaling-equation} possesses the scaling invariant 
\begin{equation}
	T_\text{K} = (\Lambda + \Gamma) \left(\frac{N \tilde{J}}{1 - N \tilde{J}}\right)^{N/2}\exp\left(-\frac{1}{2 \tilde{J}}\right) 
	\label{eq:kondo-temperature}
\end{equation}
which defines a dynamically generated energy scale---the Kondo temperature. This definition of the Kondo temperature is natural when studying the model using scaling equations like Eq.~\eqref{eq:scaling-equation}; of course, it is only defined up to a multiplicative prefactor. The standard poor-man's-scaling form of the Kondo temperature $T_\text{K}$ is obtained by neglecting the relaxation rate $\Gamma$ in Eq.~\eqref{eq:kondo-temperature}. However, we note that there exist other definitions frequently used in the literature which are more natural from an experimental point of view or when using other theoretical approaches, e.g. via the static susceptibility in Eq.~\eqref{eq:defT0}. In Sec.~\ref{sec:conductance} we will discuss these other definitions as well as the relations between them and collect the results in Tab.~\ref{tab:scales}.

We note the similarity of the RG equation \eqref{eq:scaling-equation} for $\tilde{J}$ with the scaling equation for the multichannel Kondo model.\cite{NozieresBlandin80,MitraRosch11,HS12} In particular, Eq.~\eqref{eq:scaling-equation} possesses a fixed point at $\tilde{J}=1/N=1/(2S)$. However, as we discuss at the end of Sec.~\ref{sec:initial} this fixed point is not reached since, starting in the weak-coupling regime $\tilde{J}\ll 1$, the relaxation rate $\Gamma$ cuts off the flow at the maximal value of $\tilde{J}$ corresponding to the unitary conductance (see Fig.~\ref{fig:flow}).

The RG equation for the Liouvillian translates into an equation governing the flow of the effective relaxation rate $\Gamma(E)$, 
\begin{equation}
\frac{\dd^2\Gamma}{\dd\Lambda^2}=-\frac{4N}{\Lambda+\Gamma}J^2.
\label{eq:rg-gamma}
\end{equation}
We note that $\Gamma$ stays finite during the whole procedure, see the inset of Fig.~\ref{fig:flow}. Therefore the Liouvillian remains analytic around the origin $E=0$ which results in FL behaviour discussed below. 

\subsection{\emph{E}-flow scheme for correlation functions}\label{sec:Eflowcorrelations}
After this brief overview of the $E$-flow scheme we now turn to the calculation of the correlation kernel. The starting point is its perturbation series, which has the diagrammatic representation\cite{SS09}
\begin{multline}
	\Sigma_B^\pm(E,\Omega) =
	\input{diagrams/PT/correlation-1.tikz}+
	\frac12\input{diagrams/PT/correlation-2.tikz}\\+
	\input{diagrams/PT/correlation-31.tikz}+
	\input{diagrams/PT/correlation-32.tikz}
	+\mathcal{O}(G^4).
	\label{diag:correlation-pt}
\end{multline}
All symbols and elements occurring here and in the following are summarised in Tab.~\ref{tab:rules}; furthermore see Refs.~\onlinecite{Schoeller09,PletyukhovSchoeller12} for a more detailed discussion of the notation. To achieve convergent integrals in Eq.~\eqref{diag:correlation-pt}, we take the derivative with respect to $E$. Afterwards, we replace the bare perturbative vertices by the effective vertices given by
\begin{equation}
	G_{12}(E,\omega_1,\omega_2)=
    \input{diagrams/PT/vertex-0.tikz}
    +\Bigg( \input{diagrams/PT/vertex-1.tikz} - (1\leftrightarrow2)\Bigg)+\mathcal{O}(G^3).
\end{equation}
This yields the effective diagrams for the correlation kernel
\begin{multline}
	\frac{\partial}{\partial E}\Sigma_B^\pm(E,\Omega) = 
		\frac12\input{diagrams/derivation/correlation-2-left.tikz}
		+\frac12\input{diagrams/derivation/correlation-2-right.tikz}\\
		+\frac12\input{diagrams/derivation/correlation-1-left.tikz}
		+\frac12\input{diagrams/derivation/correlation-1-right.tikz}
		+\mathcal{O}(G^4),
\end{multline}
where we have introduced the connected spin vertex $\mathcal{B}^\pm_{12}(E,\omega_1,\omega_2)$ satisfying
\begin{equation}
\mathcal{B}^\pm_{12}(E,\omega_1,\omega_2) \equiv 
		\input{diagrams/derivation/B-vertex-0.tikz} =
		\input{diagrams/derivation/B-vertex-1.tikz}
		-(1\leftrightarrow2)
		+\mathcal{O}(G^3).
		\label{diag:connected-spin-vertex}
\end{equation}
Furthermore, the bare spin vertex is not renormalised in second order,
\begin{equation}
	\mathcal{B}^\pm(E) =
		\input{diagrams/PT/correlation-1.tikz}
		+\mathcal{O}(G^2).
		\label{diag:isolated-spin-vertex}
\end{equation}
We note that the integrals in Eq.~\eqref{diag:connected-spin-vertex} are UV-convergent. Therefore taking derivatives with respect to $E$ is not necessary and the vertex $\mathcal{B}^\pm_{12}$ does not flow. The next step is to move the derivatives from the resolvent line to the contractions which is done via integration by parts
\begin{multline}
	\frac{\partial}{\partial E}\Sigma_B^\pm(E,\Omega) = 
		-\frac12\input{diagrams/derivation/correlation-2.tikz}
		-\frac12\input{diagrams/derivation/correlation-3-right.tikz}\\
		-\frac12\input{diagrams/derivation/correlation-3-left.tikz}
		+\mathcal{O}(G^4).
\end{multline}
Next we integrate out the frequency dependence of the effective vertex and thus obtain only vertices with zero frequency (depicted in the diagram by filled double dots). This integration introduces terms of the form $\Pi(E_{1\ldots n}+\bar{\omega}_{1\ldots n}+\bar\omega)-\Pi(E_{1\ldots n}+\bar{\omega}_{1\ldots n})$ which are denoted by bubbles on the corresponding contraction. Thus we find
\begin{widetext}
\begin{equation}
\begin{split}
	\label{diag:correlation-eflow}
	\frac{\partial}{\partial E}\Sigma_B^\pm(E,\Omega)=&
		-\frac12\input{diagrams/E-Flow/correlation-2.tikz}
		-\frac12\input{diagrams/E-Flow/correlation-311.tikz}
		-\frac12\input{diagrams/E-Flow/correlation-321.tikz}
		-\frac12\input{diagrams/E-Flow/correlation-312.tikz}
	\\
	&	-\frac12\input{diagrams/E-Flow/correlation-322.tikz}
		-\frac12\input{diagrams/E-Flow/correlation-313.tikz}
		-\frac12\input{diagrams/E-Flow/correlation-323.tikz}
		+\mathcal{O}(G^4).
\end{split}
\end{equation}
As shown in App.~\ref{app:integrals} the last four diagrams have the same form as the second and third one. Thus everything can be rewritten as
\begin{equation}
	\frac{\partial}{\partial E}\Sigma_B^\pm(E,\Omega) =
	-\frac12\input{diagrams/E-Flow/correlation-2.tikz}
	-\input{diagrams/E-Flow/correlation-311.tikz}
	-\input{diagrams/E-Flow/correlation-321.tikz}+\mathcal{O}(G^4).
	\label{eq:RGcorrelationsfinal}
\end{equation}

\begin{table}[tb]
\caption{\label{tab:rules} (Colour online) Summary of notations and objects used in the diagrammatic representation of the $E$-flow scheme. The chemical potentials in the leads appear in the resolvents together with the Laplace variable in the combinations $E_{1\ldots n}\equiv E+\bar{\mu}_1+\ldots+\bar{\mu}_n$. Indices $1$ and $2$ are chosen appropriately to left and right connections.}
		\begin{ruledtabular}
		\begin{tabular}{lll}
			Symbol & Name & Rule\\\hline
			$\input{diagrams/description/bare-vertex.tikz}$ 
				& bare two-point vertex $G_{12}^{(0)}$
				& indices connected by contraction are fixed to $1\leftrightarrow\bar1$\\
			$\input{diagrams/description/effective-vertex.tikz}$ 
				& effective two-point vertex $G_{12}(E_{12},\bar\omega_1, \bar\omega_{2})$
				& $E_{12}=E+\bar\mu_1+\bar\mu_{2}$ from resolvent to its left\\
			$\input{diagrams/description/effective-zero-vertex.tikz}$ 
				& two-point vertex $G_{12}(E_{12})$
				& $G_{12}(E_{12},\bar\omega_1=0, \bar\omega_{2}=0)$\\
			$\input{diagrams/description/B-vertex.tikz}$ 
				& bare spin vertex $\mathcal{B}_\pm$ 
				& add $\Omega$ to all $E$ left to $\mathcal{B}_\pm$\\
			$\input{diagrams/description/propagator.tikz}$ 
				& resolvent $\Pi(E_{12}+\bar\omega_{12})$ 
				& $E_{12}$ and $\bar\omega_{12}$ determined by vertical cut through contractions\\
			$\input{diagrams/description/derivative-propagator.tikz}$ 
				& derivative  $\frac{\partial}{\partial E}\Pi(E_{12}+\bar\omega_{12})$ 
				& derivative with respect to Laplace variable $E$\\
			$\input{diagrams/description/contraction-outgoing.tikz}$ 
				& contraction $\gamma_{12}^{pp'}(\omega)$
				& asymmetric Fermi function $f^a(\bar\omega)$\\
			$\input{diagrams/description/contraction-closed.tikz}$ 
				& contraction $\gamma_{12}^{pp'}(\omega)$
				& asymmetric Fermi function $f^a(\bar\omega)$, integrate over $\bar{\omega}$\\
			$\input{diagrams/description/derivative-contraction.tikz}$ 
				& derivative $\frac{\partial}{\partial\bar\omega}\gamma_{12}^{pp'}(\omega,\omega')$ 
				& derivative of Fermi function $f'(\bar\omega)$\\
			$\input{diagrams/description/difference-contraction.tikz}$ 
				& $\Pi(E_{1\ldots n}+\bar{\omega}_{1\ldots n}+\bar\omega)-\Pi(E_{1\ldots n}+\bar{\omega}_{1\ldots n})$
				&  $\bar\omega$ is the frequency of the contraction
		\end{tabular}
		\end{ruledtabular}
\end{table}
With the rules\cite{Schoeller09,PletyukhovSchoeller12} for translating the diagrammatic representation into ordinary expressions, which are summarised in Tab.~\ref{tab:rules}, we obtain 
	\begin{multline}
		\frac{\partial}{\partial E} \Sigma_B^\pm(E,\Omega)=
		-\frac12\iint \dd\bar\omega_{1}\,\dd\bar\omega_{2}\,f'(\bar\omega_1)f^a(\bar\omega_2)
		G_{12}(E+\Omega)\Pi(E_{12}+\Omega+\bar\omega_{12})\mathcal{B}_\pm\Pi(E_{12}+\bar\omega_{12})G_{\bar2\bar1}(E_{12})\\
		-\iiint \dd\bar\omega_{1}\,\dd\bar\omega_{2}\,\dd\bar\omega_{3}
		f'(\bar\omega_1)f^a(\bar\omega_2)f^a(\bar\omega_3)\\*
		\times\Bigl\{
		G_{12}(E+\Omega)\Bigl[\Pi(E_{12}+\Omega+\bar\omega_{12})-\Pi(E_{12}+\Omega+\bar\omega_2)\Bigr]G_{\bar23}(E_{12}+\Omega)\Pi(E_{13}+\Omega+\bar\omega_{13})\mathcal{B}_\pm\Pi(E_{13}+\bar\omega_{13})G_{\bar3\bar1}(E_{13})\\*
		+
		G_{12}(E+\Omega)\Pi(E_{12}+\Omega+\bar\omega_{12})\mathcal{B}_\pm\Pi(E_{12}+\bar\omega_{12})G_{\bar23}(E_{12})\Bigl[\Pi(E_{13}+\bar\omega_{13})-\Pi(E_{13}+\bar\omega_3)\Bigr]G_{\bar3\bar1}(E_{13}) \Bigr\},
		\label{eq:correlation-eflow}
\end{multline}
where we have dropped terms of $\mathcal{O}(G^4)$, the Fermi function is given by $f(\omega)=1/(1+e^{\omega/T})$ and its asymmetric part by $f^a(\omega)=f(\omega)-1/2$. As compared to the Liouvillian the diagrams for the correlation kernel contain one additional resolvent due to the presence of the spin vertex $\mathcal{B}_\pm$. This implies that the expressions are less UV divergent and that a single derivative is sufficient to render them convergent. 

\subsection{RG equation}\label{sec:rgequation}
As next step we employ the parametrisation given in App.~\ref{app:parametrization} to derive explicit RG equations. With the introduced notations we can summarise the RG equations for the correlation kernel $\Gamma^-(E,\Omega)$ and the variation of the current kernel $\delta\Gamma_\text{L}(E)$ for both $S=1/2$ and $S=1$ as
\begin{equation}
	\label{eq:correlation-rg-equation}
	\begin{split}
	&\frac{\partial}{\partial E} \Gamma^{-}(E,\Omega)=
	-\frac{4\pi}{3} S(S+1) N J_{12}(E+\Omega) K_{21}(E) F^{(a)}_{12}(E, E+\Omega)\\*
	&\quad+\frac{2\pi}{3} S(S+1) N J_{12}(E+\Omega) \left\{ 2 J_{23}(E_{12}) K_{31}(E) F^{(b)}_{12,13}(E,E+\Omega)
		+J_{23}(E_{12}+\Omega) K_{31}(E) F^{(b)}_{13,12}(E+\Omega,E)\right\}\\*
	&\quad+\frac{2\pi}{3} S(S+1) N J_{21}(E+\Omega)\left\{ 2 J_{32}(E_{21}) K_{13}(E) F^{(b)}_{21,31}(E,E+\Omega)
		+J_{32}(E_{21}+\Omega) K_{13}(E) F^{(b)}_{31,21}(E+\Omega,E)\right\}
	\end{split}
\end{equation}
and
\begin{equation}
	\label{eq:conductance-rg-equation}
	\begin{split}
	&\pi\frac{\partial}{\partial E}\delta\Gamma_\text{L}(E) =
		-\pi^2 S(S+1) N I^\text{L}_{12}(E)\Bigl\{K_{21}(E)
		\Bigl[\delta\mu_{12}+\ii Z_{12}\delta\Gamma(E_{12})\Bigr]F^ {(1)}_{12}(E)\\*
		&\quad-J_{23}(E_{12})K_{31}(E)
		\Bigl[F^{(2)}_{12,13}(E)\delta{\mu}_{12}+F^{(2)}_{13,12}(E)\delta{\mu}_{13}\Bigr]
		-J_{31}(E_{12})K_{23}(E)
		\Bigl[F^{(2)}_{12,32}(E)\delta{\mu}_{12}+F^{(2)}_{32,12}(E)\delta{\mu}_{32}\Bigr]\Bigr\},
	\end{split}
\end{equation}
where we write the number of channels $N=2S$ explicitly and thus the index $1\equiv\alpha$ contains the lead index only. Furthermore, we use the short-hand notations $Z_{12}=Z(E_{12})$, $\chi_{12}=\chi(E_{12})$, $\bar{Z}=Z(\bar{E})$ and $\bar{\chi}=\chi(\bar{E})$, where $\chi(E)=Z(E)[E+\ii\Gamma(E)]$, $E_{12}=E+\mu_{12}$ and $\mu_{12}=\mu_{\alpha_1}-\mu_{\alpha_2}$, while $\delta{\mu}_{12}$ denotes the infinitesimal variation of the chemical potentials in the leads. The occurring integrals are given by
\begin{eqnarray}
	\label{eq:Fa}
	F^{(a)}_{12}(E,\bar{E}) &=& 
		Z_{12}\bar{Z}_{12}\iint \dd\omega\dd\omega' 
		\frac{f'(\omega)f^a(\omega')}
		{(\omega+\omega'+\chi_{12})(\omega+\omega'+\bar{\chi}_{12})},\\
	\label{eq:Fb}
	F^{(b)}_{12,34}(E,\bar E) &=& 
		Z_{12} \bar{Z}_{12}\iint \dd\omega \dd\omega'
		\frac{\mathcal{F}_{34}(E,\omega) f'(\omega)f^a(\omega')}
		{(\omega+\omega'+\chi_{12})(\omega+\omega'+\bar{\chi}_{12})},\\
	\label{eq:F1}
	F^{(1)}_{12}(E) &=&
	Z_{12}\iint \dd\omega \dd\omega' \frac{f'(\omega)f'(\omega')}{
	\omega+\omega'+\chi_{12}}, \\
	\label{eq:F2}
	F^{(2)}_{12,34}(E) &=& 
	Z_{12}\iint \dd\omega \dd\omega'\frac{{\mathcal{F}}_{34}(E,\omega)f'(\omega)f'(\omega')}
	{\omega+\omega'+\chi_{12}},\\
	\label{eq:mathcalF}
	\mathcal{F}_{34}(E,\omega)&=&Z_{34}\int \dd\omega''\,f^a(\omega'')
	\left[\frac{1}{\omega+\omega''+\chi_{34}}-\frac{1}{\omega''+\chi_{34}}\right],
\end{eqnarray}
where we evaluate Eqs.~\eqref{eq:Fa} and \eqref{eq:Fb} in the App.~\ref{app:integrals}. We note that the RG equations~\eqref{eq:correlation-rg-equation} and~\eqref{eq:conductance-rg-equation} are valid for arbitrary temperature $T$, bias voltages $V$ and external frequencies $\Omega$.  
\end{widetext}

In order to obtain a closed set of RG equations, Eqs.~\eqref{eq:correlation-rg-equation} and~\eqref{eq:conductance-rg-equation} have to be supplemented by equations governing the flow of the remaining quantities $J_{12}(E)$, $K_{12}(E)$, $I^\text{L}_{12}(E)$, $\Gamma(E)$ and $\delta\Gamma(E)$. For the spin-1/2 model these were derived in detail in Refs.~\onlinecite{PletyukhovSchoeller12,Reininghaus-13}. The only difference in the spin-1 case is the appearance of additional factors $N$ if the trace over vertex indices is taken [similar to the explicit prefactor $N$ in Eqs.~\eqref{eq:correlation-rg-equation} and~\eqref{eq:conductance-rg-equation}]. 

The RG equations presented above were derived in a full two-loop or third-order treatment, i.e. on the right-hand side all terms containing up to three vertices $G$ were kept. This implies that upon integration the effective Liouvillian and kernels are obtained consistently including all logarithmic terms cubic in the effective coupling $\sim J^3\,\ln\Delta$, where $\Delta$ contains combinations of the energy scales. Cubic terms without logarithms are, however, not captured by the two-loop treatment. 

For later comparison we also derived the RG equations in second order. They are obtained from the third-order treatment by dropping all terms containing three vertices in the derivation. For example, the RG equation for the effective coupling and the corresponding Kondo temperature become
\begin{equation}
	\label{eq:scaling-equation2}
	\frac{\dd J}{\dd\Lambda}=-\frac{2J^2}{\Lambda+\Gamma},\quad
	T_\text{K}^{(2)} = (\Lambda + \Gamma) (NJ)^N \exp\left(-\frac{1}{2 J}\right),
\end{equation}
while the Z-factor is given by $Z=1/(1+2NJ)$. Unless explicitly stated otherwise all results present below were obtained in third order, in particular all data shown in the figures except for the inset of Fig.~\ref{fig:S1-GTGV}.

\subsection{Initial conditions}\label{sec:initial}
Finally we describe the procedure to solve the RG equations. For $T=V=\Omega=0$ we substitute $E=\ii\Lambda$ in Eqs.~\eqref{eq:correlation-rg-equation} and \eqref{eq:conductance-rg-equation} and obtain with Eq.~\eqref{eq:conductance} for the conductance 
\begin{eqnarray}
	\frac{\dd}{\dd \Lambda}\Gamma^{-} &=&
	 -\frac{\ii}{\Lambda+\Gamma}\frac{4\pi S(S+1)N}{3} Z^2 J K,\\
	\frac{\dd}{\dd \Lambda}G &=& - \frac{1}{\Lambda+\Gamma} 2\pi^2 S(S+1) N J_\text{I} K.
\end{eqnarray}
With $\tilde{J}=ZJ$, $\tilde{J}_\text{I}=ZJ_\text{I}=\tilde{J}(1-N\tilde{J})$, $Z=(1-N\tilde{J})^2$, $K=2\tilde{J}^2$, and using Eq.~\eqref{eq:scaling-equation} this yields
\begin{eqnarray}
	\Gamma^{-} &=& \frac{8\pi}{3} S(S+1) N \tilde{J}^2\left(1-\frac{2}{3}N\tilde{J}\right),\label{eq:GammaminusJ0}\\
	G &=& \pi^2 S(S+1) N \tilde{J}^2.\label{eq:Gpert}
\end{eqnarray}
Following Ref.~\onlinecite{PletyukhovSchoeller12} the initial conditions are fixed as follows: We calculate the current kernel, conductance etc. in perturbation theory in $J_0$ at the scale $\Lambda_0$ (see App.~\ref{app:perturbation-theory} for the initial conditions for $\Gamma^-$). Next we fix the numerical values of $J_0$ and $\Lambda_0$ such that at the end of the flow $\Lambda=0$ we recover the unitary conductance \eqref{eq:unitaryconductance}, 
\begin{equation}
	G(T=V=0)\Big|_{\Lambda=0}=\frac{2e^2N}{h}=\frac{4e^2S}{h}\equiv G_0.
	\label{eq:initialconditionG}
\end{equation}
We recall that we consider fully screened models with $S=1/2$ or $S=1$ only. With $J_0$ and $\Lambda_0$ fixed in this way the scaling invariant Kondo temperature \eqref{eq:kondo-temperature} or~\eqref{eq:scaling-equation2} as well as all the remaining initial conditions, e.g. for the rate $\Gamma$, are fixed by perturbation theory. The outlined procedure also fixes the renormalised exchange coupling $\tilde{J}$ at $E=0$ via [$G$ in Eq.~\eqref{eq:Gpert} is measured in units of the conductance quantum $2e^2/h$]
\begin{equation}
\tilde{J}(E=0)=\frac{1}{\pi\sqrt{S(S+1)}}.
\label{eq:JE0}
\end{equation}
We stress that we require only one condition, namely Eq.~\eqref{eq:initialconditionG}, to fix the initial values of the RG flow. In particular, the initial condition for $\Gamma^-$ is then fixed by Eq.~\eqref{eq:GammaminusJ0}, thus relating the conductance and the susceptibility.

We solve the RG equations starting from $E=0$ to $E=\ii\Lambda_0$ with $T=V=\Omega=0$ held fixed. At $E=\ii\Lambda_0$ we switch on $T$ or $V$ (the extension to simultaneously finite $T$ and $V$ is worked out in Ref.~\onlinecite{Reininghaus-13}) which is a negligible effect since $\Lambda_0\gg T,V$. We further incorporate the external frequency by evolving $J(E)$, $K(E)$ and $\Gamma(E)$ in Eq.~\eqref{eq:correlation-rg-equation} to finite $\Omega$ at $E=\ii\Lambda_0$ parallel to the real axis, i.e. having $E=\Omega+\ii\Lambda_0$ afterwards. Now with the energy scales $T$, $V$ and $\Omega$ at their physical values, we can solve the full RG equations, e.g. Eq.~\eqref{eq:correlation-rg-equation} for the correlation function, back to $E=0$ to obtain the observables in the stationary state. 

\begin{figure}[t]
	\centering
	\includegraphics[width=0.49\textwidth, trim=0 2cm 0 0]{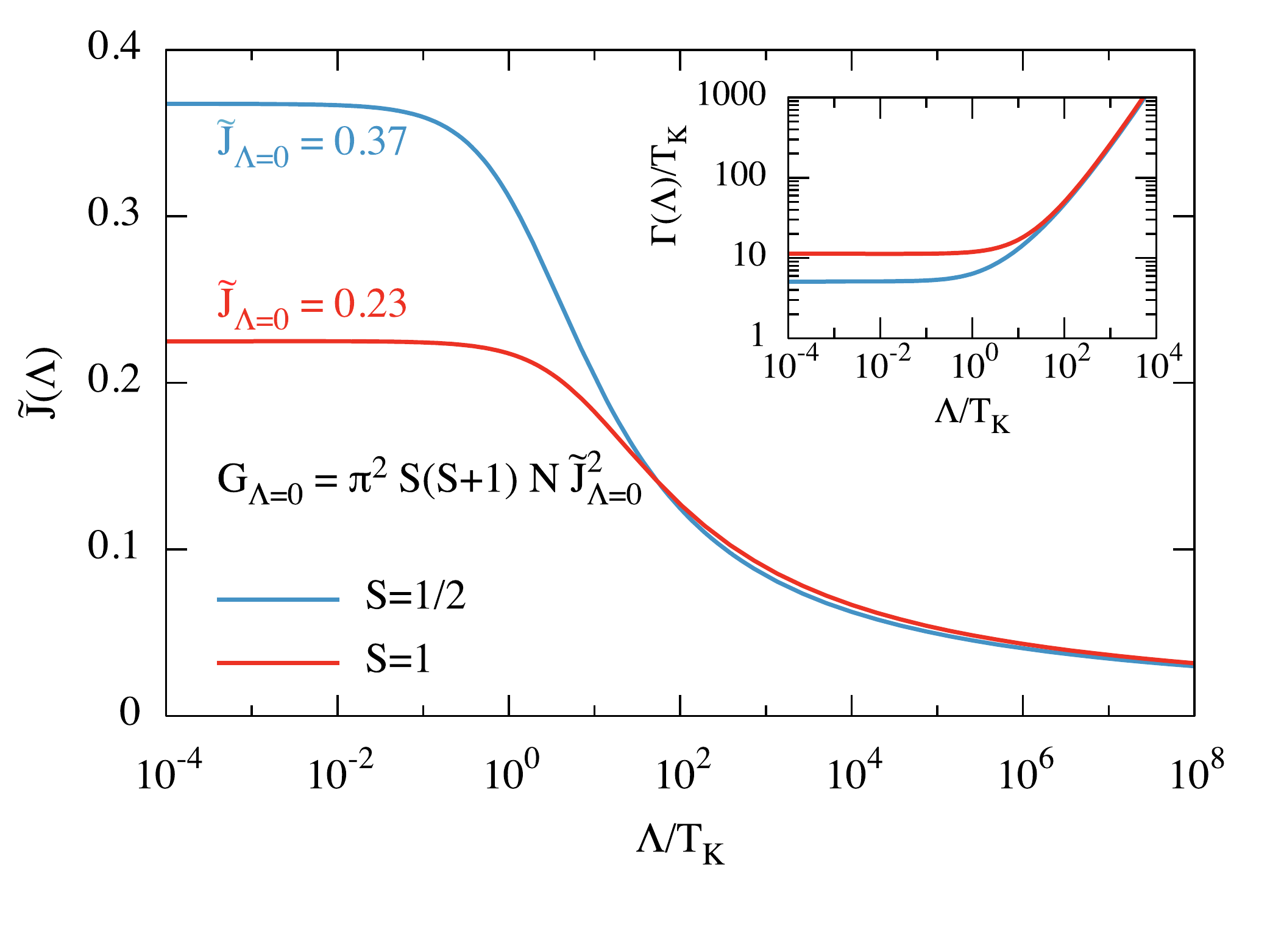}
	\caption{\label{fig:flow}(Colour online) RG flow of the renormalised exchange coupling $\tilde{J}$ governed by the scaling equation \eqref{eq:scaling-equation}. The maximum given by Eq.~\eqref {eq:JE0} is reached at $\Lambda=0$. In contrast to the poor-man's-scaling result the coupling does not diverge. Inset: RG flow of the effective relaxation rate $\Gamma$. We note that $\Gamma$ remains finite for $\Lambda\to 0$.}
\end{figure}
As an example, the RG flow of $\tilde{J}(E=\ii\Lambda)$ is shown in Fig.~\ref{fig:flow}. At sufficient high energies $\Lambda\gg T_\text{K}$ the system is in the perturbative regime $\tilde{J}\ll 1$ where a well-controlled, systematic and analytic solution is possible.\cite{SchoellerReininghaus09,SS09,PS11,HSA12} When lowering the energy scale $\Lambda$ the renormalised coupling $\tilde{J}$ increases. However, around $\Lambda\sim T_\text{K}$ this increase is cut-off by the finite relaxation rate $\Gamma$ (see inset of Fig.~\ref{fig:flow}). In contrast to the poor-man's-scaling situation the coupling does not diverge but reaches a maximum as $\Lambda\to 0$ which is fixed by the requirement of unitary conductance \eqref{eq:JE0}. In fact, for both models $\tilde{J}<1$ but, since $\tilde{J}\sim 0.3$ in the crossover regime $\Lambda\sim T_\text{K}$, it is a priory not clear whether the truncated RG equations yield reliable results. Thus it is essential to have benchmarks for the crossover and strong-coupling regime. Furthermore, one can compare different orders of truncation to gain insight into the reliability. We will come back to this when discussing our results in Sec.~\ref{sec:conclusion} below. As a side remark we note that the fixed point $\tilde{J}=1/(2S)$ of Eq.~\eqref{eq:scaling-equation} is not reached.

\section{Spin-spin correlation functions}\label{sec:correlations}
In this section we present our results on the spin-spin correlation functions of both the spin-1/2 and the spin-1 model. We first discuss the static spin susceptibility, which we use to define a second Kondo scale $T_0$, and then the dynamical correlations. In the next section we discuss the conductance with a particular focus on the FL coefficients.

\subsection{Static spin susceptibility}\label{sec:static}
The static spin susceptibility is, according to Eqs.~\eqref{eq:staticchi}, \eqref{eq:correlation-connection-C_AB} and~\eqref{eq:CMinus}, given by
\begin{equation}
		\chi=\lim_{\Omega\to0}\chi'(\Omega)=\frac{1}{2}
		\frac{\re\Gamma^{-}(\ii 0^+,0)}{\re\Gamma(\ii 0^+)}.
		\label{eq:static-susceptibility}
\end{equation}
The function $\Gamma^-(\ii 0^+,0)$ is obtained by solving the RG equation \eqref{eq:correlation-rg-equation} from the previous section numerically, while the rate $\Gamma(\ii 0^+)$ follows from the RG equation for the Liouvillian given in Ref.~\onlinecite{PletyukhovSchoeller12} [for $T=V=0$ it is given by Eq.~\eqref{eq:rg-gamma}]. We focus on the temperature dependence at zero bias voltage, $\chi(T)\equiv \chi(T,V=0)$, and the voltage dependence at zero temperature, $\chi(V)\equiv \chi(T=0,V)$. The static susceptibility can be used to define the Kondo scale $T_0$ via Eq.~\eqref{eq:defT0}, which is the definition usually adopted in numerical RG calculations.\cite{Hewson93,Bulla-08} Its relation to the Kondo temperature $T_\text{K}$ defined in Eqs.~\eqref{eq:kondo-temperature} and~\eqref{eq:scaling-equation2} is given in Tab.~\ref{tab:scales} in the next section.

\begin{figure}[t]
	\centering
	\includegraphics[width=0.49\textwidth, trim=0 2cm 0 0]{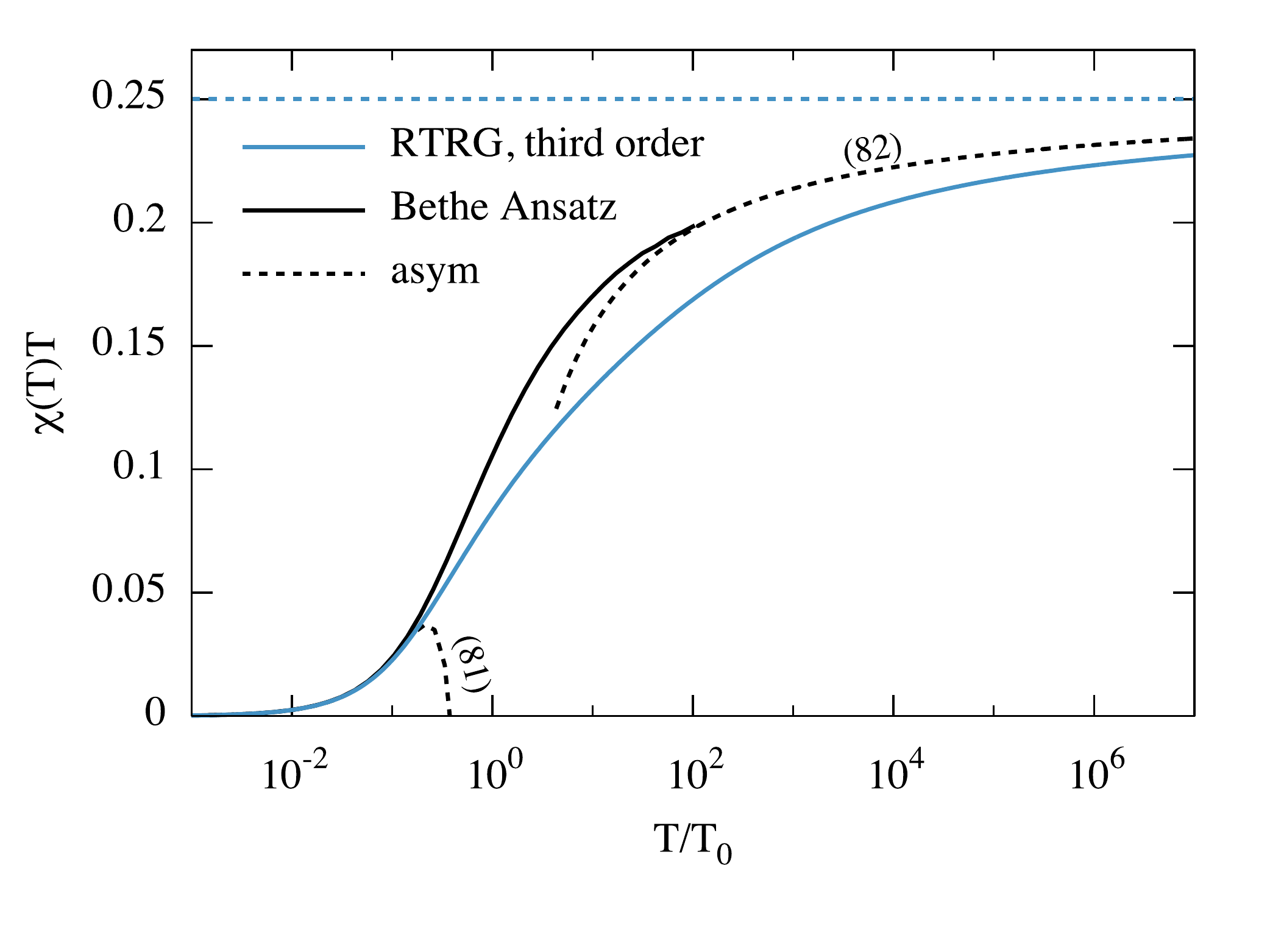}
	\caption{\label{fig:S12-susT}(Colour online) Static spin-susceptibility $\chi(T)T$ for the spin-1/2 model and $V=0$. We compare to the exact result from Bethe Ansatz [data are taken from Tab.~3 of Ref.~\onlinecite{TsvelickWiegmann83}] and asymptotic results Eqs.~\eqref{eq:chi-expansion-strong} and~\eqref{eq:chi-expansion-weak} respectively. There is no fit parameter in the RTRG result. The dashed line at $\chi(T)T=0.25$ shows the susceptibility of the asymptotically free impurity.}
\end{figure}
Our result for the temperature dependence at $V=0$ for the spin-1/2 model is shown in Fig.~\ref{fig:S12-susT}, where we have rescaled the temperature using the Kondo scale $T_0$. We compare our data to the exactly known result\cite{TsvelickWiegmann83} obtained by Bethe Ansatz as well as the asymptotic results at low and high temperatures given by\cite{Hewson93,Melnikov82,footnote4} 
\begin{equation} 
	\label{eq:chi-expansion-strong}
	\chi(T)\overset{T\ll T_0}{=}\chi(0)\left[1-\frac{\sqrt{3}\pi^3}{8}\left(\frac{T}{T_0}\right)^2
	+\mathcal{O}\left(\frac{T}{T_0}\right)^4\right],
\end{equation} 
and
\begin{multline}
	\label{eq:chi-expansion-weak}
	\chi(T)\overset{T\gg T_0}{=}\frac1{4T}\left[1-\frac{1}{\ln[T/(w T_0)]}
	-\frac{1}{2}\frac{\ln\bigl[\ln[T/(w T_0)]\bigr]}{\ln^2[T/(w T_0)]}\right.\\
	\left.+\mathcal{O}\left(\frac{1}{\ln^3[T/(w T_0)]}\right)\right],
\end{multline}
respectively. Here $w=0.41071\ldots$ denotes the Wilson number which is defined by the requirement that the term proportional to $1/\ln^2(T/wT_0)$ in Eq.~\eqref{eq:chi-expansion-weak} vanishes.\cite{Wilson75,AndreiLowenstein81,Andrei-83}. The combination $wT_0$ is also frequently used as Kondo scale in the literature. 

We first observe that our result shows reasonable agreement with the low-temperature behaviour \eqref{eq:chi-expansion-strong}; below we analyse this in more detail. In contrast, at high temperatures we observe clear deviations. While the asymptotic value $\chi(T)\to 1/(4T)$ is of course reproduced (as it is in the perturbative RTRG analysis\cite{SchoellerReininghaus09}), the logarithmic corrections $\propto 1/\ln[T/(wT_0)]\sim 1/\ln(T/T_\text{K})$ are not correctly captured. The reason for this is that the susceptibility is given by the ratio of the kernel $\Sigma_B^-$ and the Liouvillian $L$, see Eq.~\eqref{eq:static-susceptibility}. Since both start in $\mathcal{O}(J^2)$ the derivation of the contribution $\propto 1/\ln(T/T_\text{K})$ to the susceptibility would require a consistent calculation of $\Gamma^-(E,\Omega)$ and $\Gamma(E)$ including all terms in $\mathcal{O}(J^3)$. For this a full three-loop calculation including all terms with up to four vertices is necessary, which is, however, beyond the scope of this work.

\begin{figure}[t]
	\centering
	\includegraphics[width=0.49\textwidth, trim=0 2cm 0 0]{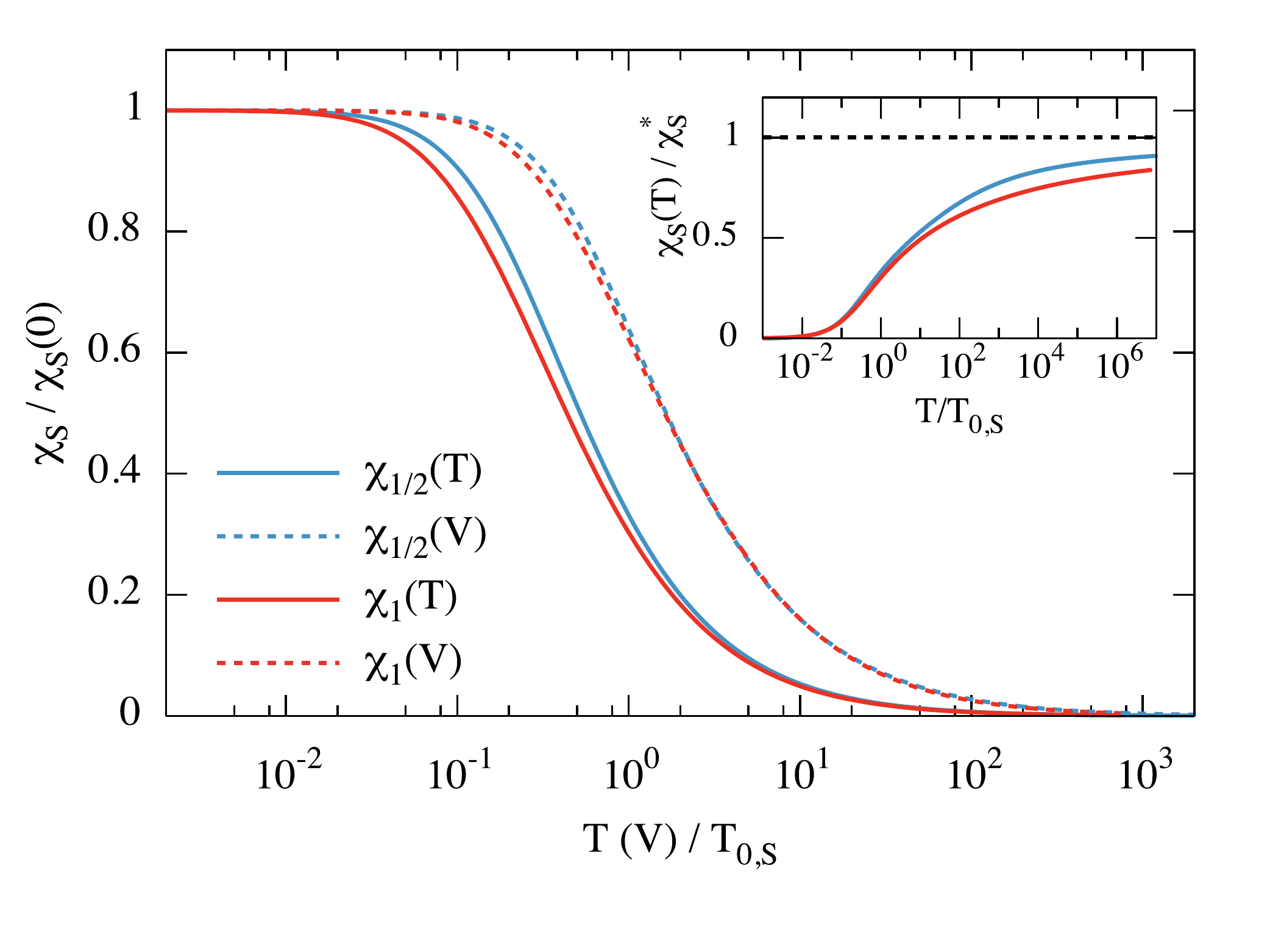}
	\caption{\label{fig:S12-S1-susTV}
	(Colour online) Static spin susceptibility $\chi_S$ of the $S=1/2$ (blue) and $S=1$ (red) model. The solid lines show the temperature dependence at $V=0$, $\chi_S(T)$, while the dashed lines show the voltage dependence at $T=0$, $\chi_S(V)$. All curves are normalised to $\chi_S(0)$, both temperature $T$ and voltage $V$ are rescaled with $T_{0,S}$ defined in Eq.~\eqref{eq:defT0}. Inset: Comparison of $\chi_S(T)/\chi_S^*$ for $S=1/2$ and $S=1$. The approach to the free spin susceptibility $\chi_S^*=S(S+1)/(3T)$ is slower in the spin-1 model.}
\end{figure}
In Fig.~\ref{fig:S12-S1-susTV} we plot the static susceptibility for the spin-1/2 and spin-1 model. We observe that, when plotted against the rescaled parameters $T/T_{S,0}$ and $V/T_{S,0}$ where $T_{S,0}$ is defined via Eq.~\eqref{eq:defT0}, $\chi_S(V)>\chi_S(T)$, i.e. thermal fluctuations lead to a stronger  suppression of the systems susceptibility to an external magnetic field then a finite bias voltage.  Asymptotically the susceptibility reaches the one of a free spin, $\chi_S^*=S(S+1)/(3T)$, in particular we find for the relative factor $\chi_{1/2}/\chi_1=3/8$ for $T\gg T_0$ or $V\gg T_0$. This factor also frequently appears in the RG equations discussed in Sec.~\ref{sec:method}. 

Finally, let us analyse the behaviour at low temperatures or small bias voltages in more detail. Specifically, we consider the coefficients $a_T'$ and $a_V'$ in the expansion [c.f. Eq.~\eqref{eq:lowTsusc}]
\begin{equation}
\chi=\chi_0\left[1-a_T'\left(\frac{T}{T_0}\right)^2-a_V'\left(\frac{eV}{T_0}\right)^2\right].
\end{equation}
We have extracted the coefficients from our RTRG calculation in second and third order; the results are shown in Tab.~\ref{tab:aT}. We observe that the value for $a_T'$ obtained from the RTRG treatment shows a significant deviation from the exact result and a rather strong dependence on the order of truncation of the RG equations. In contrast, for the ratio $a_V'/a_T'$ we do not observe such a drastic dependence on the truncation. Thus we would consider the result for $a_V'/a_T'$ to be more reliable (see the discussion for the FL coefficients of the conductance in Sec.~\ref{sec:conductance}).
\begin{table}[t]
	\begin{ruledtabular}
	\caption{Values of the coefficients $a_T'$ and $a_V'/a_T'$ extracted from the numerically obtained RTRG results in second and third order. The exact values for $a_T'$ are given by Eq.~\eqref{eq:aT}; the relative errors are stated in brackets.\label{tab:aT}}
	\begin{tabular}{llcc}
		model & method & $a_T'$ & $a_V'/a_T'$\\\hline
		$S=1/2$ & BA & 6.71 & -\\
		& RTRG 2nd & 4.89 (27\%) & 0.12\\
		& RTRG 3rd & 13.64 (103\%)& 0.10\\\hline
		$S=1$ & BA & 14.80 & -\\
		& RTRG 2nd & 7.38 (50\%) & 0.08\\
		& RTRG 3rd & 28.70 (94\%) & 0.07
	\end{tabular}	
	\end{ruledtabular}
\end{table}

\subsection{Dynamical correlation functions}\label{sec:dynamic}
From Eqs.~\eqref{eq:correlation-connection-C_AB}, \eqref{eq:CPlus} and~\eqref{eq:CMinus} we obtain the  dynamical correlation function 
\begin{equation}
	S(\Omega)
	=\frac{1}{2}\frac{\re\Gamma(\Omega)}{\Omega^2+\re\Gamma(\Omega)^2}
	\label{eq:S}
\end{equation}
as well as the imaginary part of the spin susceptibility 
\begin{equation}
	\chi''(\Omega)
	=\frac12\frac{\Omega\re\Gamma^{-}(\ii 0^+,\Omega)}{\Omega^2+\re\Gamma(\Omega)^2}.
	\label{eq:chi-prime-prime}
\end{equation}
We note that in deriving these expressions we have omitted terms of the form $\sim\Gamma(\Omega)\Gamma^-(\ii 0^+,\Omega)$, which contribute only to higher-order corrections and cannot be treated consistently. Furthermore, we have neglected the term $\im\Gamma(\Omega)$ in the denominator since it is much smaller than $\Omega$, $\abs{\im\Gamma(\Omega)}\ll\Omega$. We note that Eqs.~\eqref{eq:S} and~\eqref{eq:chi-prime-prime} are rather similar to the corresponding expressions\cite{SS09} in the weak-coupling regime, however, here we calculate $\Gamma(\Omega)$ and $\Gamma^-(\ii 0^+,\Omega)$ in the whole crossover regime.
 
\begin{figure}[t]
	\centering
	\includegraphics[width=0.49\textwidth, trim=0 2cm 0 0]{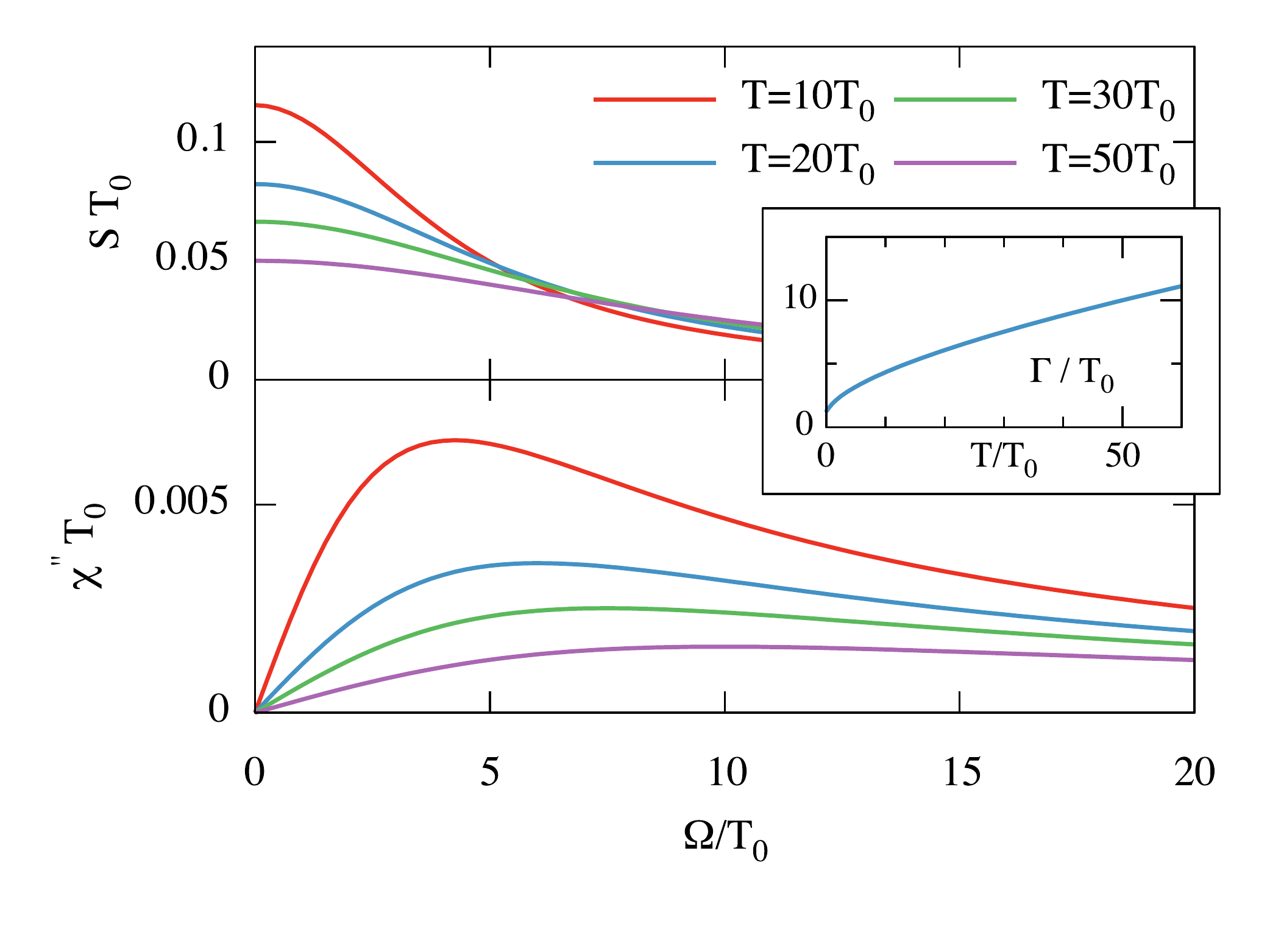}
	\caption{\label{fig:S12-dynamic-T}(Colour online) The spin-spin correlation function $S(\Omega)$ and the imaginary part of the spin susceptibility $\chi''(\Omega)$ for the spin-1/2 model and $V=0$. We observe the maximum of $\chi''(\Omega)$ at $\Omega\approx\Gamma$ and $S(\Omega=0)=1/(2\Gamma)$, where $\Gamma(T,V=0)$ is shown in the inset.}
\end{figure}
\begin{figure}[t]
	\centering
	\includegraphics[width=0.49\textwidth, trim=0 2cm 0 0]{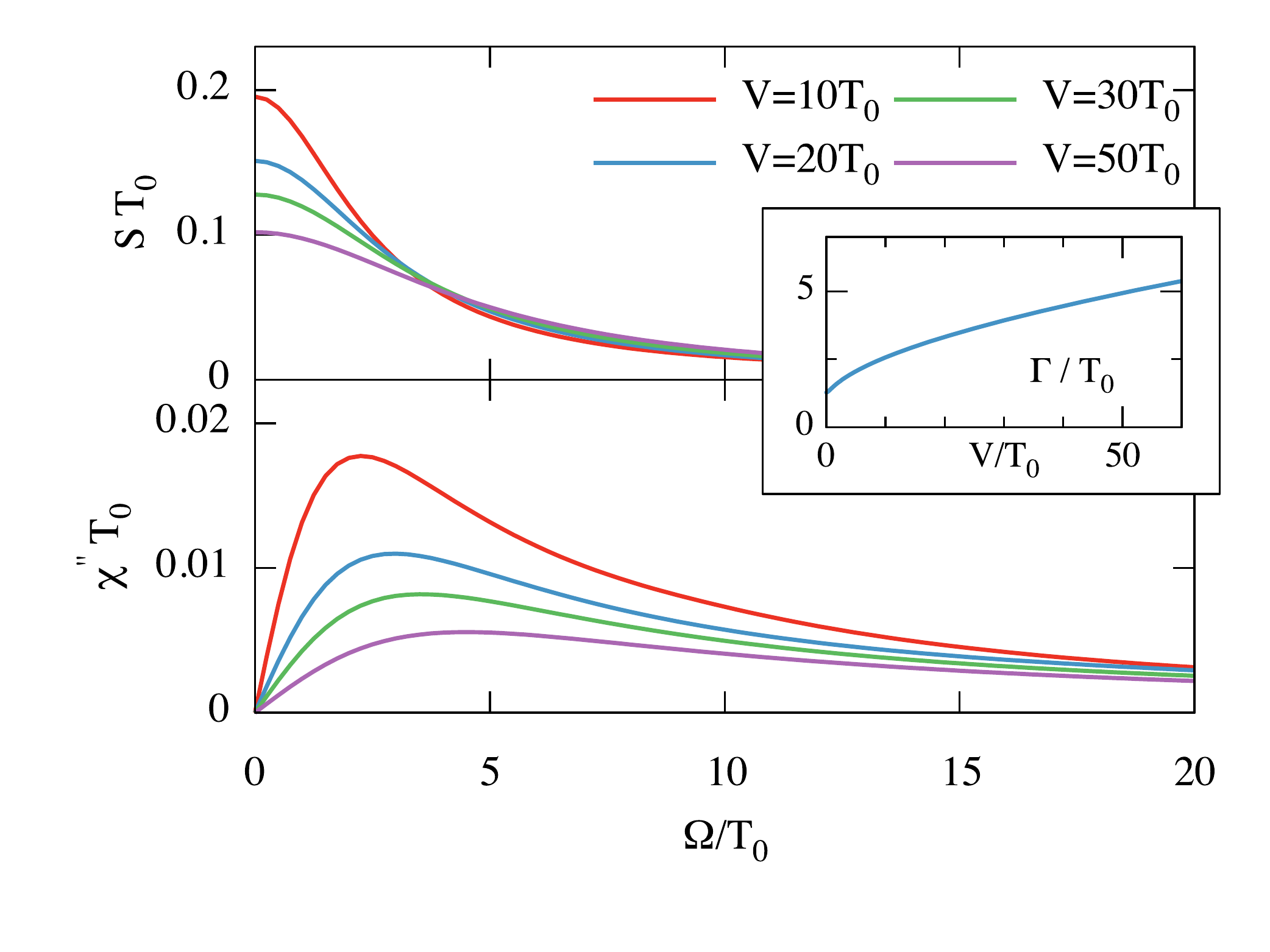}
	\caption{\label{fig:S12-dynamic-V}(Colour online) The spin-spin correlation function $S(\Omega)$ and the imaginary part of the spin susceptibility $\chi''(\Omega)$ for the spin-1/2 model and $T=0$. We observe that the maximum of $\chi''(\Omega)$ appears at lower frequencies as compared to the temperature-dependent susceptibility shown in Fig.~\ref{fig:S12-dynamic-T}, which was also observed in Ref.~\onlinecite{FritschKehrein09ap}. $\Gamma(T=0,V)$ is shown in the inset.}
\end{figure}
The dynamical correlation function and the imaginary part of the spin susceptibility for the spin-1/2 model are shown in Figs.~\ref{fig:S12-dynamic-T} and \ref{fig:S12-dynamic-V}. The behaviour agrees very well with the results obtained by Fritsch and Kehrein using the flow-equation method.\cite{FritschKehrein09ap,FritschKehrein10} We find in the low-frequency limit
\begin{equation}
S(\Omega\to 0)=\frac{1}{2\Gamma},
\label{eq:lowomega} 
\end{equation}
which holds for both $S=1/2$ and $S=1$. Here $\Gamma$ is the physical spin relaxation rate obtained from solving Eq.~\eqref{eq:rg-gamma} and taking $E\to \ii 0^+$. We show $\Gamma$ in the respective insets. For example, in the perturbative regime $T,T_0\ll V$ it is simply given by\cite{SchoellerReininghaus09,Paaske-04prb2,HSA12} $\Gamma=\pi J^2V$ with the renormalised exchange coupling $J=1/[2\ln(V/T_\text{K})]$. Physically the relaxation rate governs the real-time dynamics of the spin on the dot and the current through it.\cite{PSS10} The low-frequency behaviour \eqref{eq:lowomega} also agrees with results obtained by mapping the spin-spin correlation function of the spin-1/2 model to the one-particle Green function in an effective description in terms of Majorana fermions.\cite{Mao-03,ShnirmanMakhlin03} For large frequencies we recover the perturbative result\cite{SS09,FritschKehrein09ap} $S(\Omega)\propto 1/[\Omega\,\ln^2(\Omega/T_\text{K})]$.

Similarly the spin relaxation rate determines the maximum of the susceptibility, which is located at $\Omega\approx\Gamma$.  In agreement with Ref.~\onlinecite{FritschKehrein09ap} we find that this maximum is at lower values of $\Omega$ for the voltage-dependent susceptibility as compared to the temperature-dependent one since the rate is larger in the latter case. 

Finally we note that in equilibrium ($V=0$) the correlation function and the dynamical susceptibility are related to each other via the fluctuation-dissipation theorem\cite{CallenWelton51,Landau5} $\chi''(\Omega)/S(\Omega)=\tanh[\Omega/(2T)]$. Our results obtained by numerically solving the RG equations do not fully respect this relation, which seems to be due to inconsistently treated higher-order corrections in the derivations. For very large frequencies, however, one recovers the weak-coupling result\cite{SS09} including the fluctuation-dissipation relation.

\section{Differential conductance}\label{sec:conductance}
In this section we discuss our results for the conductance of the spin-1 model. We focus on the temperature dependence of the linear conductance, $G(T)\equiv G(T,V=0)$, and the voltage dependence of the differential conductance at zero temperature, $G(V)\equiv G(T=0,V)$. The corresponding analysis for the spin-1/2 case was performed previously in Ref.~\onlinecite{PletyukhovSchoeller12}. In principle it is also possible to extend the study to the full temperature and voltage dependence which is, however, considerably more involved (see Ref.~\onlinecite{Reininghaus-13} for the spin-1/2 case).

\begin{figure}[t]
	\centering
	\includegraphics[width=0.49\textwidth, trim=0 2cm 0 0]{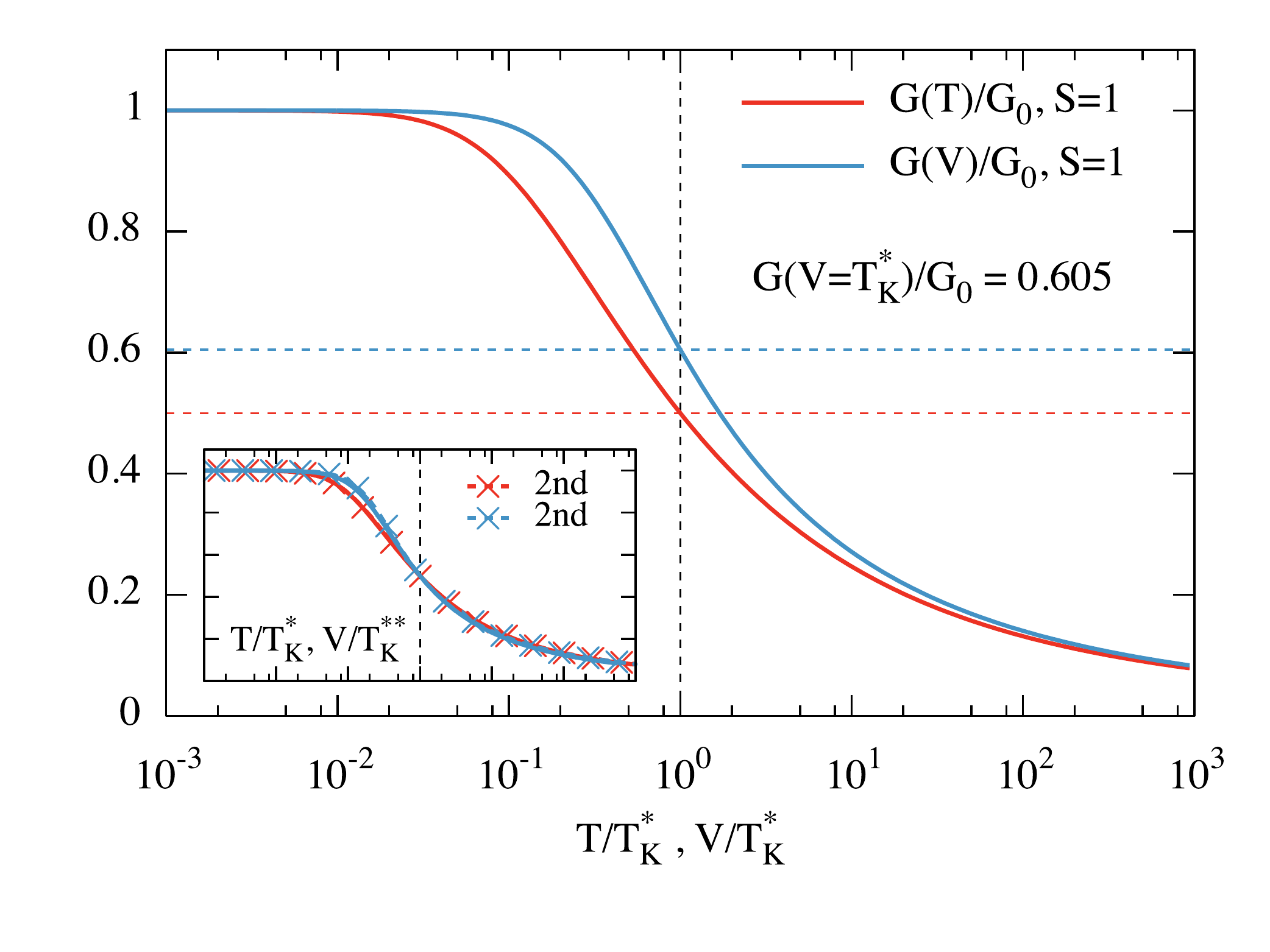}
	\caption{\label{fig:S1-GTGV}(Colour online) Linear and differential conductance $G(T)$ and $G(V)$ for $S=1$ scaled to the Kondo scale $T_\text{K}^*$ defined in Eq.~\eqref{eq:defTKstar}. The dashed lines are a guide to the eye to extract the relation $G(V=T_\text{K}^*)\approx 0.605\,G_0$. Inset: Second-order (crosses) and third-order (solid lines) RTRG results for $G(T)$ as a function of $T/T_\text{K}^*$ and $G(V)$ as a function of $V/T_\text{K}^{**}$. The results are almost independent of the truncation.}
\end{figure}
In Fig.~\ref{fig:S1-GTGV} we show the linear and differential conductance for the spin-1 model. The Kondo scale $T_\text{K}^*$ used to rescale the temperature and voltage, respectively, is defined by
\begin{equation}
G(T=T_\text{K}^*,V=0)=\frac{G_0}{2},
\label{eq:defTKstar}
\end{equation}
where $G_0=G(0,0)$ denotes the unitary conductance introduced in Eq.~\eqref{eq:initialconditionG}. Similarly we can define another scale by 
\begin{equation}
G(T=0,V=T_\text{K}^{**})=\frac{G_0}{2}.
\label{eq:defTKtwostar}
\end{equation}
In contrast to the scales $T_0$ and $T_\text{K}$ defined in Eqs.~\eqref{eq:defT0}, \eqref{eq:kondo-temperature} and~\eqref{eq:scaling-equation2}, respectively, which are convenient for theoretical purposes, the scales $T_\text{K}^*$ and $T_\text{K}^{**}$ are easier accessible in experiments. We stress that the notations used in the literature are not unique (e.g. in Ref.~\onlinecite{Kretinin-11} the scale $T_\text{K}^*$ is denoted by $T_\text{K}$ while Ref.~\onlinecite{Merker-13} uses $T_\text{K}^\text{expt}$). In the inset of Fig.~\ref{fig:S1-GTGV} we observe that both $G(T)$ and $G(V)$ are almost independent of the order of truncation when rescaled against the corresponding Kondo temperatures $T_\text{K}^*$ and $T_\text{K}^{**}$ respectively.

\begin{table}[t]
	\caption{Definitions of the Kondo scales $T_\text{K}$, $T_0$, $T_\text{K}^*$ and $T_\text{K}^{**}$ used in this article and the numerical relations between them as extracted from the RTRG analysis in second and third order respectively. We note that the numerical values depend on the order of truncation of the RG equations. We stress that the notations used in the literature are not unique.\label{tab:scales}}
	\begin{ruledtabular}
	\begin{tabular}{lcccl}
		scale & order & $S=1/2$ & $S=1$ & definition \\\hline
		$T_\text{K}$ & 2nd & - & - & scaling invariant~\eqref{eq:scaling-equation2} \\
		& 3rd & - & - & scaling invariant~\eqref{eq:kondo-temperature}\\\hline
		$T_{0}$ & 2nd & $9.17\,T_\text{K}$ & $56.07\,T_\text{K}$ & $S(S+1)/3\chi(T=0)$\\
		& 3rd & $3.99\,T_\text{K}$ & $12.64\,T_\text{K}$ & Eq.~\eqref{eq:defT0}\\\hline
		$T_\text{K}^*$ & 2nd &$10.58\,T_\text{K}$ & $98.98\,T_\text{K}$ & $G(T=T_\text{K}^*)=G_0/2$\\
		& 3rd &$2.07\,T_\text{K}$ & $10.29\,T_\text{K}$ & Eq.~\eqref{eq:defTKstar}\\\hline
		$T_\text{K}^{**}$ & 2nd & $10.89\,T_\text{K}$ & $115.44\,T_\text{K}$ & $G(V=T_\text{K}^{**})=G_0/2$\\
		& 3rd & $3.57\,T_\text{K}$ & $17.45\,T_\text{K}$ & Eq.~\eqref{eq:defTKtwostar}
	\end{tabular}	
	\end{ruledtabular}
\end{table}
In total we have thus four ways to define a Kondo scale which we collect in Tab.~\ref{tab:scales}. The scales differ by numerical prefactors which themselves depend on the order of truncation of the RG equations (see discussion at the end of Sec.~\ref{sec:rgequation}). We note that while the scale $T_\text{K}$ obviously depends on the truncation, also the ratios of the (in principle) observable scales $T_\text{K}^*/T_0$ and $T_\text{K}^{*}/T_\text{K}^{**}$ depend on it, i.e. the truncated RTRG equations are not able to yield reliable results for these quantities. For example, for the spin-1/2 model we find in second and third order $T_\text{K}^*/T_0=1.15$ and $T_\text{K}^*/T_0=0.52$, respectively, while recent numerical RG calculations~\cite{Merker-13,HanlWeichselbaum14} give $T_\text{K}^*/T_0\approx 1.04$ in the Kondo limit of the single-impurity Anderson model.

\begin{figure}[t]
	\centering
	\includegraphics[width=0.49\textwidth, trim=0 2cm 0 0]{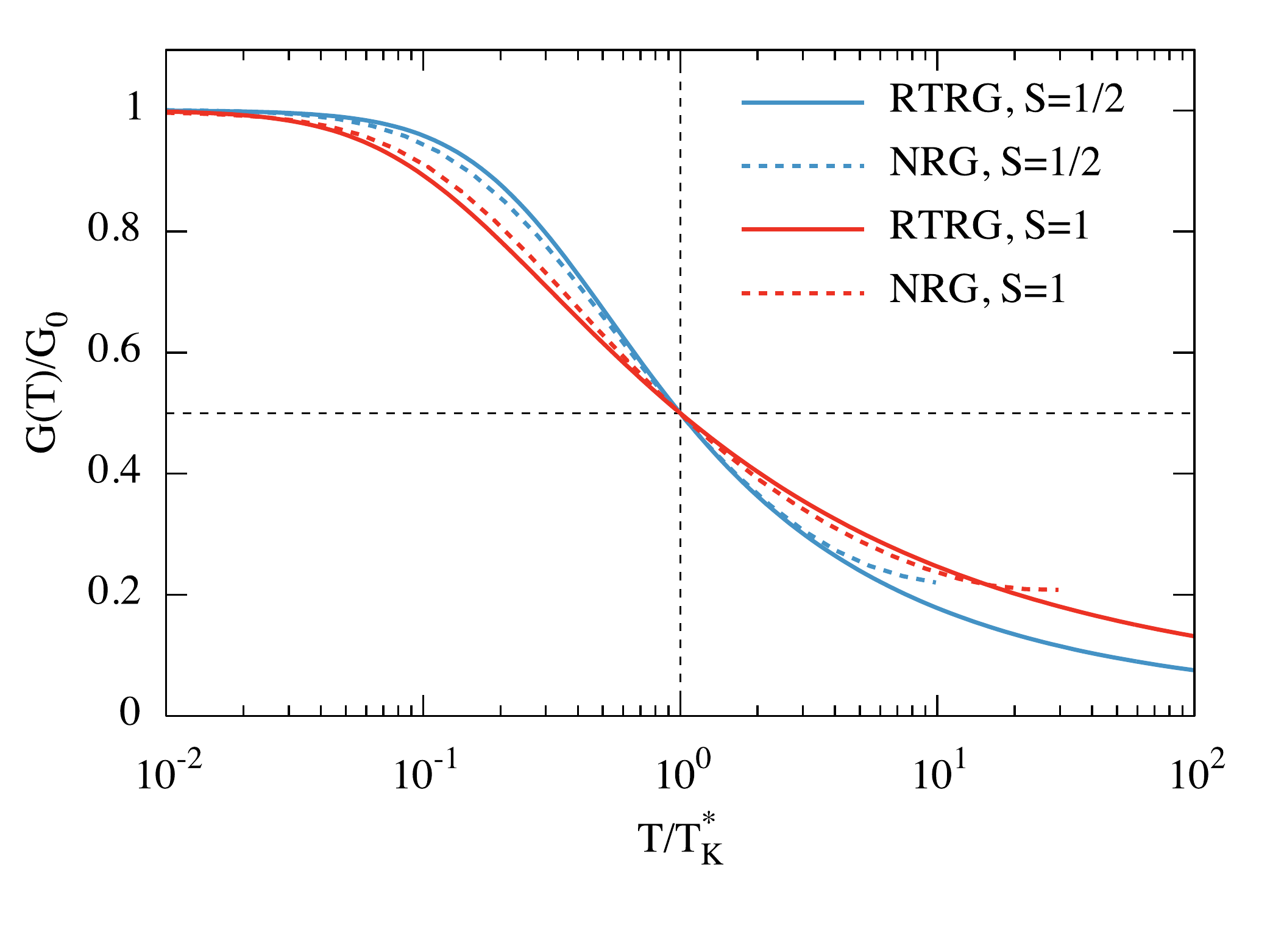}
	\caption{\label{fig:S12-S1-GT}(Colour online) Comparison of the linear conductance $G(T)$ for the fully screened Kondo model with $S=1/2$ and $S=1$, between the RTRG (solid lines) and numerical RG\cite{Costi-09} (dashed lines).}
\end{figure}
In Fig.~\ref{fig:S1-GTGV} we also observe that, when $T$ and $V$ are rescaled against the same scale, one finds $G(T)<G(V)$, which was also obtained in the spin-1/2 case. This finding is also supported by the fact that $T_\text{K}^{**}/T_\text{K}^*>1$ and $c_V'/c_T'<1$ (see below) independently of the used truncation of the RG equations. Furthermore, we can extract the differential conductance at $V=T_\text{K}^*$ and find in third order
\begin{equation}
G(V=T_\text{K}^*)\approx 0.605\,G_0
\label{eq:a0}
\end{equation} 
in the spin-1 model, while for $S=1/2$ the result $G(V=T_\text{K}^*)\approx 2/3\,G_0$ was observed.\cite{Kretinin-12,PletyukhovSchoeller12,SmirnovGrifoni13} We note, however, that the numerical value in \eqref{eq:a0} strongly depends on the considered order of truncation. 

In Fig.~\ref{fig:S12-S1-GT} we compare our results for the linear conductance with the corresponding ones obtained using numerical RG.\cite{Costi-09,Hanl-13} For both the spin-1/2 and spin-1 model we observe satisfactory agreement; the deviations at large temperatures originate from the fact that we directly treat the Kondo model \eqref{eq:ham} while Refs.~\onlinecite{Costi-09,Hanl-13} analysed the corresponding Anderson models, whose high-temperature physics deviates from the universal behaviour of the Kondo model.

\begin{figure}[t]
	\centering
	\includegraphics[width=0.49\textwidth, trim=0 2cm 0 0]{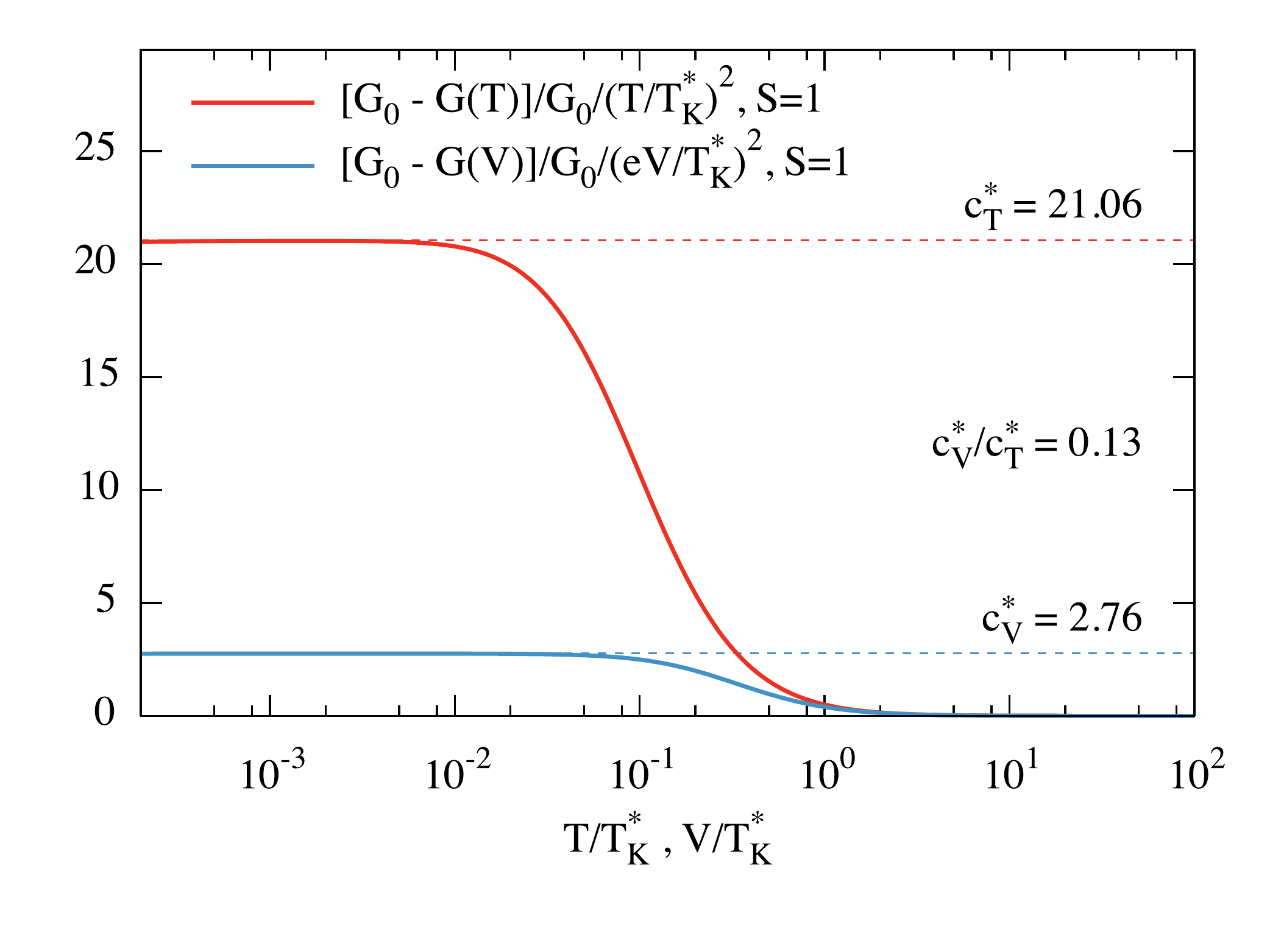}
	\caption{\label{fig:S1-cTcV}(Colour online) Linear and differential conductance $G(T)$ and $G(V)$ for the spin-1 model plotted in the form $\frac{G_0-G(T,V)}{G_0}/(\frac{T,eV}{T_\text{K}^*})^2$ such that the FL coefficients $c_T^*$ and $c_V^*$ can be readily determined by a quadratic fit (dashed line). Similarly for the spin-1/2 model we obtain $c_T^*=4.86$ and $c_V^*=0.89$ in agreement with Ref.~\onlinecite{PletyukhovSchoeller12}.}
\end{figure}
Beside the relations between the various Kondo scales we can also extract the FL coefficients from our RTRG calculation and in particular compare them with the results from FL theory derived in Sec.~\ref{sec:FL}. Using the scale $T_\text{K}^*$ as our energy unit we fit the RTRG results at low energies against (see Fig.~\ref{fig:S1-cTcV} for the spin-1 model)
\begin{equation}
G(T,V)=G_0\left[1-c_T^*\left(\frac{T}{T_\text{K}^*}\right)^2-c_V^*\left(\frac{eV}{T_\text{K}^*}\right)^2\right].
\end{equation}
We stress again that the notations used in the literature are not unique (e.g. in Ref.~\onlinecite{Kretinin-11} uses $c_{T,V}$ instead of $c_{T,V}^*$). The FL coefficients calculated in Sec.~\ref{sec:FL} are then obtained via
\begin{equation}
\frac{c_{T}'}{c_T^*}=\frac{c_{V}'}{c_V^*}=\left(\frac{T_0}{T_\text{K}^*}\right)^2,
\label{eq:cTcVrelations}
\end{equation}
we present our results in Tab.~\ref{tab:FL} together with the errors compared to the exact values are given by Eqs.~\eqref{eq:cT} and~\eqref{eq:cVcT} respectively. 

\begin{table}[t]
	\begin{ruledtabular}
	\caption{Values of the FL coefficients $c_T'$ and $c_V'/c_T'$ extracted from the numerically obtained RTRG results in second and third order. The exact values are given by Eqs.~\eqref{eq:cT} and~\eqref{eq:cVcT}; the relative errors are stated in brackets.\label{tab:FL}}
	\begin{tabular}{llcc}
		model & method & $c_T'$ & $c_V'/c_T'$\\\hline
		$S=1/2$ & FL theory & 6.088 & 0.152\\
		& RTRG 2nd & 5.02 ($18\%$) & 0.28 ($84\%$)\\
		& RTRG 3rd & 18.04 ($196\%$) & 0.18 ($18\%$)\\\hline
		$S=1$ & FL theory & 8.794 & 0.164\\
		& RTRG 2nd & 7.40 ($16\%$) & 0.18 ($10\%$)\\
		& RTRG 3rd & 31.77 ($261\%)$ & 0.13 ($21\%$)
	\end{tabular}	
	\end{ruledtabular}
\end{table}
We observe that the value for $c_T'$ obtained by numerically solving the second-order RG equations is in reasonable agreement with the FL results, i.e. the deviation is less than $20\%$. In contrast, when increasing the order of truncation the deviation increases considerably. Thus higher-order corrections do not improve the results for $c_T'$. In contrast, for the ratio $c_V'/c_T'$ we do not observe such a drastic dependence on the truncation. This is similar to the susceptibility discussed above, where the ratio $a_V'/a_T'$ also depends only weakly on the order of truncation.

The deviations can presumably be attributed to the fact that the FL coefficients are obtained in the RTRG treatment by expanding in $\tilde{J}(0)$ which takes the values $\tilde{J}(0)\approx 0.37$ and $\tilde{J}(0)\approx 0.23$ for $S=1/2$ and $S=1$ respectively [see Eq.~\eqref{eq:JE0}]. For the coefficient $c_T'$ [or the entering Kondo scales, see Eq.~\eqref{eq:cTcVrelations}] this expansion seems to be not reliable. Another aspect may lie in the fact that the RTRG method considers the RG flow of the effective quantities in the model \eqref{eq:ham}, i.e. the form, structure and symmetries of the Hamiltonian at low and high energies are the same. In contrast, as shown in Sec.~\ref{eq:FLham} (see also Refs.~\onlinecite{LesageSaleur99,LesageSaleur99npb}) the effective Hamiltonian describing the FL fixed point has a completely different structure. 

Furthermore, we note that the ratio $c_V'/c_T'$ can also be derived analytically within the RTRG formalism. For the spin-1/2 model this was done in Ref.~\onlinecite{PletyukhovSchoeller12} with the result $c_V'/c_T'=3/(2\pi^2)$ in prefect agreement with Eq.~\eqref{eq:cVcT}. Following the same steps in the spin-1 case we obtain the same result, i.e. the RTRG method does not capture the non-trivial spin dependence of the FL coefficients. 

Finally, we add that the weak-coupling solution $\tilde{J}\sim J\sim 1/\ln(\max\{T,V\}/T_\text{K})$ together with Eq.~\eqref{eq:Gpert} directly results in the perturbative result\cite{SchoellerReininghaus09,HSA12} for the differential conductance, $G\sim 1/\ln^2(T/T_\text{K})$ and $G\sim 1/\ln^2(V/T_\text{K})$ for $T\gg T_\text{K}$ and $V\gg T_\text{K}$ respectively.

\section{Conclusion and discussion}\label{sec:conclusion}
In this article we have studied the transport properties of fully screened Kondo quantum dots where the number of screening channels equals twice the spin on the dot. In the first part we employed FL theory to calculate the conductance at low temperatures and small bias voltages. In particular, we derived the FL coefficient for the voltage dependence of the conductance for models with arbitrary spin and found a non-trivial spin-dependence of the ratio $c_V'/c_T'$, see Eq.~\eqref{eq:cVcT}. We also determined the low-temperature behaviour of the static susceptibility from the Bethe Ansatz solution for the dot magnetisation with the main result given by Eq.~\eqref{eq:aT}.

In the second part we generalised the recently developed\cite{PletyukhovSchoeller12} $E$-flow scheme of the RTRG technique to study correlation functions and performed a two-loop analysis of the fully screened spin-1/2 and spin-1 Kondo models. In particular, this method allows to study the crossover from strong to weak coupling also in the non-equilibrium setup with a finite bias voltage. The starting point of the method in terms of an expansion in the renormalised exchange coupling offers an internal consistency check when comparing observables in different orders of truncation of this expansion. In the following we will briefly recall our main results and then turn to the comparison with other methods.

We calculated the static spin susceptibility for both models and found that thermal fluctuations lead to a more pronounced suppression of the susceptibility than a finite bias voltage, i.e. $\chi(T)<\chi(V)$ (see Fig.~\ref{fig:S12-S1-susTV}). We studied the behaviour at low temperatures or small voltages and extracted the corresponding coefficients $a_T'$ and $a_V'$, see Tab.~\ref{tab:aT}. For the dynamical spin-spin correlation functions we found good agreement with previous results obtained using the flow-equation method.\cite{FritschKehrein09ap,FritschKehrein10}

In addition we analysed the temperature and voltage dependence of the linear and differential conductance. We observed that $G(T)<G(V)$ provided $T$ and $V$ are rescaled against the same energy scale (see Fig.~\ref{fig:S1-GTGV}). We used the susceptibility and conductance to define various Kondo scales and discussed their relations; the results are summarised in Tab.~\ref{tab:scales}. We also extracted the FL coefficients $c_T'$ and $c_V'$ and compared them to the known results from FL theory (see Tab.~\ref{tab:FL}).

Now let us turn to a summary of comparisons with other methods: (i) The $E$-flow scheme of the RTRG method by construction correctly reproduces all perturbative results for high temperatures or large bias voltages. The failure to obtain the leading logarithmic corrections to the static susceptibility originates in the third-order truncation used here and can be cured by incorporating higher orders. The same is true for the fluctuation-dissipation theorem. (ii) The RTRG method reproduces the quadratic behaviour of the susceptibility and conductance for small temperatures or voltages. (iii) For the spin-1/2 model one can analytically derive\cite{PletyukhovSchoeller12} the correct ratio of the FL coefficients $c_V'/c_T'=3/(2\pi^2)$ in all orders of the truncation. (iv) As shown in Fig.~\ref{fig:S12-S1-GT} the temperature dependence of the linear conductance agrees well with numerical RG calculations.

However, there are some points which our RG treatment was not able to capture: (a) The ratios of the observable Kondo scales like $T_\text{K}^{*}/T_0$ cannot be determined reliably. In fact they strongly depend on the truncation of the RG procedure. (b) Similarly there is a strong dependence on the truncation for the of FL coefficients $a_T'$ and $c_T'$. (c) The non-trivial spin dependence of the ratio $c_V'/c_T'$ is not captured.

To put this into perspective we recall that the starting point of the RG procedure is a perturbative expansion in the renormalised exchange coupling. In the low-energy regime this is, however, not particularly small, i.e. $\tilde{J}(0)\approx 0.37$ and $\tilde{J}(0)\approx 0.23$ for $S=1/2$ and $S=1$ respectively. In light of this it is even somewhat surprising that the RTRG method correctly reproduces non-trivial aspects of the FL behaviour like point (iii) above. With the perturbative starting point in mind one may hope that going beyond the third-order truncation will improve points (a) and (b). This would, however, require much more involved calculations. 

In this context we stress that the truncation of the RG equations also offers an internal consistency check when comparing the results of different orders of truncation. Using this check we conclude that the absolute values of the FL coefficients $a_T'$ and $c_T'$ in the low-temperature regime are not reliable, as are the ratios of the Kondo scales like $T_\text{K}^{*}/T_\text{K}^{**}$. On the other hand there are several quantities that do not show a strong dependence on the truncation order: (i) The linear and differential conductance as shown in the inset of Fig.~\ref{fig:S1-GTGV}. (ii) The qualitative results $\chi(T)<\chi(V)$ and $G(T)<G(V)$ provided $T$ and $V$ are rescaled against the same energy scale. (iii) The ratios $a_V'/a_T'$ and $c_V'/c_T'$. We would thus expect these results to be reliable. For example, for the voltage dependence of the susceptibility at small voltages, $\chi(V)/\chi_0=1-a_V'(eV/T_0)^2$, we estimate the coefficients to be $a_V'\approx 0.7$ for $S=1/2$ and $a_V'\approx 1.1$ for $S=1$ respectively. 

An important aspect may also lie in the fact that the RTRG method considers the RG flow of the effective quantities in the model \eqref{eq:ham}, i.e. the form, structure and symmetries of the Hamiltonian at low and high energies are the same. In contrast, the effective Hamiltonian describing the FL fixed point has a completely different structure, which cannot be represented by the Kondo model \eqref{eq:ham}. Thus correctly connecting the high- and low-energy regimes within the RTRG framework seems quite intricate. One has to keep in mind, however, that currently there is no method available that masters this task in the presence of a bias voltage. From the equilibrium situation it is known that ultimately non-perturbative methods like numerical RG or Bethe Ansatz are required. A generalisation of these methods to the full non-equilibrium setup is still an open challenge, although there have been recent advances in the scattering state numerical RG\cite{Anders08prl,SchmittAnders10,SchmittAnders11,JovchevAnders13} and attempts to apply\cite{Konik-01,Konik-02,MehtaAndrei06,Nishino-09} the Bethe Ansatz method in the presence of a finite bias voltage.

To conclude, we have analysed the non-equlibrium transport properties of fully screened Kondo quantum dots. We employed FL theory to treat the low-energy regime as well as the $E$-flow scheme of the RTRG method to study the crossover from strong to weak coupling. Given the perturbative starting point of the latter approach the applicability in the strong-coupling regime is not guaranteed a priory and therefore should always be checked against alternative methods. On the other hand, in the weak-coupling regime the RTRG method allows a systematic analysis of a wide range of observables in Kondo quantum dots like the non-equilibrium transport properties,\cite{SchoellerReininghaus09,PS11,HSA12,HS12} the dynamical correlation functions\cite{SS09,MPSA13} and the relaxation dynamics.\cite{PSS10} Furthermore, it is possible to treat other problems like the transport properties, relaxation dynamics and adiabatic response of the interacting resonant level model\cite{KAPSBMS10,APSSB11,Kashuba-12} and the time evolution in the Ohmic spin-boson model.\cite{Kennes-13,KashubaSchoeller13,Kashuba-13,Kennes-13a}

\acknowledgements
We thank Sabine Andergassen, Edouard Boulat, Theo Costi, Serge Florens, Stefan G\"ottel, Mikhail Pletyukhov, Frank Reininghaus and Andreas Weichselbaum for useful comments and discussions. We are especially grateful to Herbert Schoeller for numerous and enlightening discussion on the topic of this work. Furthermore, we thank the authors of Ref.~\onlinecite{Costi-09} for providing the numerical RG data shown in Fig.~\ref{fig:S12-S1-GT}, and the authors of Refs.~\onlinecite{footnote6,HanlWeichselbaum14,Merker-12} for the numerical RG data to extract the low-temperature behaviour of the static susceptibility. DS thanks the Institute for Theory of Statistical Physics, RWTH Aachen University, where substantial parts of this work have been performed. This work is part of the D-ITP consortium, a program of the Netherlands Organisation for Scientific Research (NWO) that is funded by the Dutch Ministry of Education, Culture and Science (OCW). CBMH acknowledges support by the JSPS Postdoctoral Fellowship Program (Short-term). CBMH and DS were supported by the German Research Foundation (DFG) through the Emmy-Noether Program under SCHU 2333/2-1.

\appendix
\section{Alternative expression for the current}\label{appen-current}
We discuss an expression for the current, alternative to Eq.~\eqref{eq:current}, that is more adapted to the FL approach. In addition, we prefer to work in a symmetric/antisymmetric basis, instead of the left/right basis, to take advantage of the symmetric setup.

The operators $a_{i k \sigma} = (c_{i \text{L} k \sigma} - c_{i \text{R} k \sigma})/\sqrt{2}$ are odd combinations of the original left and right electrons for the channel $i$ and spin $\sigma$, and they are decoupled from the dot variables from the outset. The operators $b_{i k \sigma}$ derive from the even combination $\tilde{b}_{i k \sigma} = (c_{i \text{L} k \sigma} + c_{i \text{R} k \sigma})/\sqrt{2}$, with an additional phase shift $\delta_0 = \pi/2$, namely (omitting $i$ and $\sigma$ indices
for clarity)
\begin{equation}
\tilde{b} (x) \equiv \sum_k e^{\ii k x} \tilde{b}_k = 
\begin{cases} e^{2\ii \delta_0} b(x) \qquad {x>0} \\
b(x) \qquad {x<0}
\end{cases}
\end{equation}
where the Fourier expansion $b(x) \equiv \sum_k e^{\ii k x} b_k$ defines the operators $b_{i k \sigma}$. The screening of the dot spin implies that the field $\tilde{b} (x)$ becomes discontinuous at $x=0$ at low energies while $b(x)$ is a continuous field even at $x=0$. Note that the $x$-axis is obtained by unfolding the outgoing part of wave functions into the half-space of positive $x$, while negative $x$ correspond to incoming states.

We introduce a general expression for the current operator 
\begin{equation}\label{current-alt}
\hat{I} (y) = \frac{1}{2\ii} \sum_{i,\sigma} \left( \psi_{i \sigma}^\dagger (y) \partial_y  \psi_{i \sigma} (y) 
- \bigl[\partial_y \psi_{i\sigma}^\dagger (y)\bigr] \psi_{i\sigma} (y) \right).
\end{equation}
$y$ is the position on a one-dimensional line where $y<0$ is the left lead, $y>0$ is the right lead and the dot is placed at $y=0$. Hence, for $y<0$, $\hat{I}_\text{L} = \hat{I} (y)$ and $\hat{I}_\text{R} = \hat{I} (-y)$. The field operator $\psi_{i \sigma} (y)$ can be related to the operators $c_{i \text{L/R} k \sigma}$, for example for $y<0$,
\begin{equation}
\psi_{i \sigma} (y) = \sum_k \left( e^{\ii (k_\text{F}+k) y} - e^{-\ii (k_\text{F}+k) y} \right) c_{i \text{L} k \sigma},
\end{equation}
ensuring that electrons are fully backscattered in the lead when the coupling to the dot is absent. The $y$- and $x$-axis are physically different, as the $y$ one-dimensional line runs across the dot while the $x$-line is reflected at the dot. The expression Eq.~\eqref{current} for the current operator is obtained by expanding the field operator $\psi$ in Eq.~\eqref{current-alt} onto the relevant fields $a$ and $b$, and by taking the symmetrised combination of the left current at $y<0$ with the right current at $-y$.

In contrast to the fields $\tilde{b}_{i k \sigma}$, the fields $b_{i k \sigma}$ are free at the Kondo fixed point. Their populations are fixed by the left and right chemical potentials, namely
\begin{subequations}\label{steady}
\begin{align}
\langle b_k^\dagger b_{k'} \rangle & = \frac{\delta_{k,k'}}{2} \Bigl[
 f_{\rm L} (\varepsilon_k) 
+   f_{\rm R} (\varepsilon_k) \Bigr], \\[1mm]
 \langle a_k^\dagger a_{k'} \rangle &  = \frac{\delta_{k,k'}}{2} \Bigl[
 f_{\rm L} (\varepsilon_k) 
+  f_{\rm R} (\varepsilon_k) \Bigr], \\[1mm]
\langle a_k^\dagger b_{k'} \rangle & = \langle  b_k^\dagger a_{k'} \rangle =
\frac{\delta_{k,k'}}{2}  \Bigl[ f_{\rm L} (\varepsilon_k) 
- f_{\rm R} (\varepsilon_k) \Bigr], \\[1mm]
f_{\rm L/R} (\varepsilon) & = f(\varepsilon - \mu_{\rm L/R}),
\end{align}
\end{subequations}
where  $f (\varepsilon)$ is the Fermi function.

\section{Wilson ratio at finite magnetic field}\label{app:susceptibliity}
In this appendix we show that the Wilson ratio \eqref{eq:Wilsonratio} is independent of the applied magnetic field $H$. We start with the spin-1/2 case. The arguments that follow are based on a simple observation: the ground state of the Kondo model is a FL for all values of the magnetic field. In particular, a FL description requires $T,V \ll T_\text{K}$ but in no way $B \ll T_\text{K}$. A straightforward consequence is that the Korringa-Shiba formula, which indicates that energy dissipation is caused by elementary FL particle-hole excitations, holds at arbitrary magnetic field. This prediction has been confirmed by numerical RG calculations.~\cite{Garst-05}

Using a FL description, we can write the phase shift for bulk electrons as ($\sigma = \uparrow/\downarrow$, $\bar{\sigma} = \downarrow/\uparrow$)
\begin{equation}\label{psh}
\delta_{\sigma} (\varepsilon,n_{\bar{\sigma}})  = \delta^0_\sigma 
+ \alpha_{1,\sigma} (\varepsilon - \varepsilon_\sigma)
- \phi_{1,\sigma} \sum_{\varepsilon'} 
 \bigl[ n_{\bar{\sigma}} (\varepsilon') - n_{0,\bar{\sigma}} (\varepsilon') \bigr]
\end{equation}
where $n_{0,\bar{\sigma}} (\varepsilon) = \Theta (\varepsilon_{\bar{\sigma}}-\varepsilon)$ are the zero-temperature Fermi distributions for spin $\bar{\sigma}$. $\varepsilon_\sigma$ denotes the Fermi energy for the spin species $\sigma$, i.e. at finite magnetic field $\varepsilon_\sigma = \varepsilon_\text{F} + \sigma B/2$. The zero-energy phase shifts $\delta^0_\sigma$ are related to the spin-dependent occupations on the dot through the Friedel sum rule, $\delta^0_\sigma = \pi \langle d_\sigma^\dagger d_\sigma \rangle$. In Eq.~\eqref{psh}, the phase shift of an incoming bulk electron depends on its energy  $\varepsilon$ and on the energy distribution $n_{\bar{\sigma}} (\varepsilon)$ of the bulk electrons with opposite spin. The expansion Eq.~\eqref{psh} defines \emph{a priori} four parameters, $\alpha_{1,\sigma}$ and $\phi_{1,\sigma}$. These parameters, and the reference phase shifts $\delta^0_\sigma$, change with the Fermi energies $\varepsilon_\sigma$ or, more precisely - and this is the key to the FL invariance - they depend only on the energy difference $\varepsilon_\uparrow - \varepsilon_\downarrow$ which in our case is the magnetic field $B$.

Therefore, the ground state of the model is unchanged upon adding electrons in a narrow slice of energies between $\varepsilon_\sigma$ and $\varepsilon_\sigma + \delta \varepsilon$ (with infinitesimal $\delta \varepsilon$) for both spin species. This invariance reads in Eq.~\eqref{psh},
\begin{equation}\label{psh2}
\begin{split}
\delta_{\sigma} (\varepsilon_\sigma+ \delta \varepsilon,n^1_{\bar{\sigma}})  = &\delta^0_\sigma =  \delta^0_\sigma + \alpha_{1,\sigma} \delta \varepsilon\\
&- \phi_{1,\sigma} \sum_{\varepsilon'} 
 \bigl[ n^1_{\bar{\sigma}} (\varepsilon') - n_{0,\bar{\sigma}} (\varepsilon') \bigr],
 \end{split}
\end{equation}
where  $n^1_{\bar{\sigma}} (\varepsilon') = \Theta (\varepsilon_{\bar{\sigma}}+\delta \varepsilon-\varepsilon)$. Performing the energy summation in Eq.~\eqref{psh2}, one finds the identities $\alpha_{1,\sigma} =  \phi_{1,\sigma}$ for $\sigma = \uparrow/\downarrow$. A third identity is obtained by noting that changing the magnetic field $B$ redistributes electrons from spin down to spin up but does not change\cite{footnote5} the sum of phase shifts taken at the Fermi energies $\varepsilon_\sigma$, hence $\alpha_{1,\uparrow} +  \phi_{1,\uparrow} = \alpha_{1,\downarrow} +  \phi_{1,\downarrow}$. From these three relations, we conclude that the four coefficients are in fact equal,
\begin{equation}\label{iden}
\alpha_{1,\uparrow} =  \phi_{1,\uparrow} = \alpha_{1,\downarrow} =  \phi_{1,\downarrow}
\end{equation}
and the low energy model can be parametrised by a single coupling constant $\alpha_1 (B)$ which depends on the magnetic field.

Equipped with our low energy FL model, we compute the impurity specific heat
\begin{equation}
C_\text{i} (B) = \frac{\pi T}{3} \left[ \alpha_{1,\uparrow} (B) + \alpha_{1,\downarrow} (B) \right] = \frac{2 \pi T}{3} \alpha_1 (B),
\end{equation}
and the spin susceptibility 
\begin{equation}
\chi_\text{i} (B) = \frac{\partial M}{\partial B} (B) = \frac{\alpha_{1,\uparrow} + \phi_{1,\uparrow} + \alpha_{1,\downarrow} + \phi_{1,\downarrow}}{4 \pi} = \frac{\alpha_1 (B)}{\pi}.
\end{equation}
Hence, using Eq.~\eqref{iden} we have 
\begin{equation}
\frac{1}{T} \, C_i (B) = \frac{2 \pi^2}{3}  \chi_i (B),
\end{equation}
and the Wilson ratio is found to be independent of the magnetic field. The same arguments can be reproduced for arbitrary $S$ with the same conclusion that the Wilson ratio is the same for all magnetic field. One finds in particular
\begin{equation}
C_\text{i} (B) = \frac{2 N \pi T}{3} \alpha_1 (B) \qquad 
\chi_\text{i} (B) = \frac{4 \alpha_1 (B)}{3 \pi} S(S+1).
\end{equation}
With $2 S = N$, one obtains the generalised expression
\begin{equation}
\frac{1}{T} \, C_\text{i} (B) = \frac{2 \pi^2 N}{4 S (S+1)}  \chi_\text{i} (B) = \frac{2 \pi^2}{N+2}  \chi_\text{i} (B)
\end{equation}
and hence the Wilson ratio~\eqref{eq:Wilsonratio}.

\section{Parametrisation in Liouville space}\label{app:parametrization}
\subsection{Spin-1/2}\label{app:parametrization_spin-12}
In the spin-1/2 model a convenient basis in Liouville space was given in Refs.~\onlinecite{SchoellerReininghaus09,SS09}. It consists of two scalar superoperators $L^{a}$ and $L^{b}$ as well as three vector superoperators $\vec{L}^{1}$, $\vec{L}^{2}$ and $\vec{L}^{3}$. In the absence of a magnetic field conservation laws then yield a convenient parametrisation for the Liouvillian, current kernel and vertices as 
\begin{align}
	L(E) &= -\ii\,\Gamma(E) L^a,
	\label{eq:s12-liouvillian} \\
	G_{12}(E)&= - J_{12}(E)\vec{L}^2\cdot\vec\sigma + \ii\frac{\pi}{2} K_{12}(E) \vec{L}^3\cdot\vec\sigma,
	\label{eq:s12-vertex}\\
	\Sigma_\text{L}E)&=\ii\,\Gamma_\text{L}(E) L^b,
	\label{eq:s12-current-kernel}\\
	I^\text{L}_{12}(E)&=-\frac{1}{4}I^\text{L}_{12}(E) \vec{L}^1\cdot\vec\sigma,
	\label{eq:s12-current-vertex}
\end{align}
where the vertices are given for $\eta_1=-\eta_2=+$. The vertices for $\eta_1=-\eta_2=-$ are obtained via the relations $J_{12}(E)=-J_{21}(-E^*)^*$, $K_{12}(E)=-K_{21}(-E^*)^*$ and $I^\text{L}_{12}(E)=-I^\text{L}_{21}(-E^*)^*$. Furthermore, we assume symmetric couplings to the leads, i.e. for $V=0$ we can drop the lead indices from the vertices and parametrize $J_{12}(E)=J(E)$ and so on. For the correlation kernels we use\cite{SS09}
\begin{align}
	\Sigma_{B}^+(E,\Omega)&=\mathcal{B}_{+} + \Gamma^{+}(E,\Omega) L^{3z},
	\label{eq:s12-sigma-plus}\\
	\Sigma_{B}^-(E,\Omega)&=\mathcal{B}_{-} + \Gamma^{-}(E,\Omega) L^{3z},
	\label{eq:s12-sigma-minus}
\end{align}
with $\mathcal{B}_+=\ii(L^{1z}+L^{3z})$ and $\mathcal{B}_-=-2\ii L^{2z}$. We note that a term proportional to $L^{2z}$ occurs in both kernels, but due to the trace to be taken in Eq.~\eqref{eq:C_AB} together with $\rho_{\uparrow\uparrow}=\rho_{\downarrow\downarrow}=1/2$ does not contribute to the correlation functions. Furthermore, we neglect $\Gamma^{+}(E,\Omega)$ throughout this work since it is only generated in higher orders.

\subsection{Spin-1}\label{app:parametrization_spin-1}
For spin-1 a significantly larger algebra of superoperators is required.\cite{HSA12} It consists of three scalar superoperators $L^{s}$, $L^{t}$ and $L^{q}$ and six vector superoperators $\vec{K}^{j}$, $j=1,\ldots,6$, which are defined by
\begin{gather}
	\label{eq:lscalar_singlet}
	L^{s(inglet)}
		=-\frac{1}{3}\openone +\frac{1}{3}(\vec L^+ \cdot \vec L^-)^2, \\
	\label{eq:lscalar_triplet}
	L^{t(riplet)}
		=\openone- \frac{1}{2}\vec L^+ \cdot \vec L^- -\frac{1}{2}(\vec L^+ \cdot \vec L^-)^2,\\
	\label{eq:lscalar_quinet}
	L^{q(uintet)}
		=\frac{1}{3}\openone+ \frac{1}{2}\vec L^+ \cdot\vec L^-+\frac{1}{6}(\vec L^+ \cdot \vec L^-)^2.
\end{gather}
Here $\vec{L}^+A=\vec{S}A$ and $\vec{L}^-A=-A\vec{S}$ for an arbitrary operator $A$ and $\vec{S}$ is the spin operator on the dot. The six vector superoperators are given by
\begin{align}
	\vec{K}^1
		&=\vec{L}^+ +\vec{L}^- +i \vec{L}^{+-} \times\vec{L}^{+-},\\
	\vec{K}^2
		&=\vec{L}^+ +\vec{L}^- -i \vec{L}^{+-} \times\vec{L}^{+-},\\
	\vec{K}^3
		&=\vec{L}^+ -\vec{L}^- +2\vec{L}^{+-}-(\vec{L}^{+-} \times\vec{L}^{+-})\times\vec{L}^{+-},\\
	\vec{K}^4
		&=\vec{L}^+ -\vec{L}^- -2\vec{L}^{+-}+\vec{L}^{+-}\times(\vec{L}^{+-} \times\vec{L}^{+-}),\\
	\vec{K}^5
		&=\vec{L}^+ -\vec{L}^- +\vec{L}^{+-}+\frac32(\vec{L}^{+-} \times\vec{L}^{+-})\times\vec{L}^{+-}\notag\\&
		\quad-\frac12\vec{L}^{+-}\times(\vec{L}^{+-} \times\vec{L}^{+-}),\\
	\vec{K}^6
		&=\vec{L}^+ -\vec{L}^- -\vec{L}^{+-}+\frac12(\vec{L}^{+-} \times\vec{L}^{+-})\times\vec{L}^{+-}\notag\\&
		\quad-\frac32\vec{L}^{+-}\times(\vec{L}^{+-} \times\vec{L}^{+-}),
\end{align}
where $\vec{L}^{+-}=\ii\vec{L}^+\times\vec{L}^-$. The scalars and vectors fulfill the symmetries
\begin{equation}
	\tr{L^j}=0\text{ for }j=t,q,\quad\tr{K^j}=0\text{ for }j\neq 4,
\end{equation}
where the trace is taken over the dot space.

Applying the algebra we find that the RG equations for the differential conductance and the correlation functions decouple into triplet and quintet sector. Up to the third order in the interaction vertex both depend solely on the triplet sector spanned by $L^t, \vec{K}^2$ and $\vec{K}^3$, while mixtures between the two sectors are generated only in higher orders, which we do not consider in this work. Therefore it is sufficient to consider 
\begin{align}
	L(E) &= -\ii\,\Gamma(E) L^t,
	\label{eq:s1-liouvillian}\\
	G_{12}(E) &=  \frac14 J_{12}(E) \vec{K}^2\cdot\vec\sigma + \ii\frac{\pi}{6}K_{12}(E) \vec{K}^3\cdot\vec\sigma,
	\label{eq:s1-vertex}\\
	\Sigma_\text{L}(E)&=\ii\,\Gamma_\text{L}(E) L^s,
	\label{eq:s1-current-kernel} \\
	I^L_{12}(E)&=-\frac{1}{12}I^\text{L}_{12}(E) \vec{K}^4\cdot\vec\sigma,
	\label{eq:s1-current-vertex}\\
	\Sigma_{B}^+(E,\Omega)&=\mathcal{B}_{+} + \Gamma^{+}(E,\Omega) K^{3z},
	\label{eq:s1-sigma-plus}\\
	\Sigma_{B}^-(E,\Omega)&=\mathcal{B}_{-} + \frac18 \Gamma^{-}(E,\Omega) K^{3z},
	\label{eq:s1-sigma-minus}
\end{align}
where ${\mathcal{B}_+=\frac{\ii}{3}(K^{3z}+K^{4z})+\frac{\ii}{6}(K^{5z}+K^{6z})}$ and ${\mathcal{B}_-=\frac{\ii}{2}(K^{1z}+K^{2z})}$. Similar to $L^{2z}$ above here terms proportional to $K^{2z}$ do not contribute to the correlation functions. We note that in order to keep the notation between the spin-1/2 and spin-1 case as consistent as possible we have used the same notations as in Eqs.~\eqref{eq:s12-liouvillian}--\eqref{eq:s12-sigma-minus}. 

\section{Correlation integral}\label{app:integrals}
In this appendix we further analyse the integrals \eqref{eq:Fa}--\eqref{eq:mathcalF}. The Fermi function is given by $f(\omega)=f^a(\omega)+1/2$ where the asymmetric part reads 
\begin{equation}
	f^a(\omega) =  -T\sum_{n\in\mathbb{Z}}\frac{1}{\omega-\ii\omega_n},
\end{equation}
with ${\omega_n=2\pi T(n+1/2)}$ denoting the Matusbara frequencies. Furthermore, we will use the polygamma function
\begin{equation}
\psi(z)=-\sum_{n=0}^\infty\frac{1}{n+z}.
\end{equation}

\emph{Rewriting integrals:}
We first show that
\begin{multline}
	\input{diagrams/E-Flow/correlation-311.tikz}=\\ \input{diagrams/E-Flow/correlation-312.tikz}+\input{diagrams/E-Flow/correlation-313.tikz}.
	\label{eq:appdiagram}
\end{multline}
With $f^a(\omega)f'(\omega')\to f'(\omega)f^a(\omega')$ we obtain for the first diagram on the right-hand side by integrating by parts twice
\begin{align}
	&\iint \dd\omega\dd\omega'
		\frac{\mathcal{F}_{34}(E,\omega)f^a(\omega)f'(\omega')}
		{(\omega+\omega'+\chi_{12})(\omega+\omega'+\bar{\chi}_{12})}\notag\\
		= & \iint \dd\omega\dd\omega'\left[f'(\omega)\mathcal{F}_{34}(E,\omega) + f^a(\omega)\frac{\dd}{\dd\omega}\mathcal{F}_{34}(E,\omega)\right]\notag\\*
	& \qquad\times 
		\frac{f^a(\omega')}
		{(\omega+\omega'+\chi_{12})(\omega+\omega'+\bar{\chi}_{12})}.
		\label{eq:partial-integration-Fc}
\end{align}
Now the second diagram exactly cancels the last term in Eq.~\eqref{eq:partial-integration-Fc},
\begin{multline*}
	\int \dd\omega''
		\frac{Z_{34}f'(\omega'') }{\omega+\omega''+\chi_{34}}\\
	= \int \dd\omega''
		\frac{Z_{34}f^a(\omega'')}{(\omega+\omega''+\chi_{34})^2}
		=  -\frac{\dd}{\dd\omega}\mathcal{F}_{34}(E,\omega)
\end{multline*}
such that Eq.~\eqref{eq:appdiagram} and thus Eq.~\eqref{eq:RGcorrelationsfinal} follow.

\emph{Integrals in static case:}
For $\bar{E}\to E$ we recover the integrals obtained for the Liouvillian
\begin{gather}
	F^{(a)}_{12}(E,E) = Z_{12}F^{(1)}_{12},\\
	F^{(b)}_{12,34}(E,E) = Z_{12}F^{(2)}_{12,34}.
\end{gather}

\emph{Integrals at zero temperature:}
We obtain in the zero-temperature limit
\begin{gather}
	F^{(a)}_{12}(E,\bar{E}) = 
		\frac{Z_{12} \bar{Z}_{12}}{\chi_{12}-\bar{\chi}_{12}}\ln\frac{\chi_{12}}{\bar{\chi}_{12}},\\
	F^{(b)}_{12,34}(E,\bar{E}) \propto \mathcal{F}_{34}(E,0) = 0.
\end{gather}

\emph{Integrals at finite temperature:}
Introducing the notations $\gamma=\chi_{12}/(2\pi\ii T)$ and $\bar{\gamma}=\chi_{34}/(2\pi\ii T)$ we obtain after a straightforward calculation
\begin{widetext}
\begin{eqnarray}
	F^{(a)}_{12}(E,\bar{E}) &=& \frac{Z_{12} \bar{Z}_{12}}{\chi_{12}-\bar{\chi}_{12}}
	\Bigl[\psi(\gamma_{12}+1)-\psi(\bar{\gamma}_{12}+1)+
	\gamma\psi'(\gamma_{12}+1)-\bar{\gamma}_{12}\psi'(\bar{\gamma}_{12}+1)\Bigr],\\
	F^{(b)}_{12,34}(E,\bar E) &=& -\frac{Z_{12} Z_{34} \bar{Z}_{12}}{2\pi \ii T} \frac{1}{\bar{\gamma}_{12} - \gamma_{12}}\nonumber\\
		& &\qquad\times\sum_{k=0}^{\infty}
		\frac{\dd}{\dd k}
			\Bigl[\psi(k+\bar{\gamma}_{12}+1)-\psi(k+\gamma_{12}+1)\Bigr]
	    	\Bigl[\psi(k+\gamma_{34}+1)-\psi(\gamma_{34}+\tfrac12)\Bigr]
		\label{eq:Fb-result}.
\end{eqnarray}
\end{widetext}
We note that the sum in Eq.~\eqref{eq:Fb-result} is convergent. We evaluate the first $k_0$ terms explicitly and approximate the remaining sum by an integral
\begin{multline}
	\sum_{k=k_0}^{\infty}
		\frac{\dd}{\dd k}
			\Bigl[\psi(k+\bar{\gamma}_{12}+1)-\psi(k+\gamma_{12}+1)\Bigr]\\
	    	\times
	    	\Bigl[\psi(k+\gamma_{34}+1)-\psi(\gamma_{34}+\tfrac12)\Bigr]\\
		\approx-\Bigl[\psi(k_0+\bar{\gamma}_{12}+\tfrac12)-\psi(k_0+\gamma_{12}+\tfrac12)\Bigr]\\
	    	\times\Bigl[\psi(k_0+\gamma_{34}+\tfrac12)-\psi(\gamma_{34}+\tfrac12)\Bigr]
\end{multline} 
The value of $k_0$ is determined such that the introduced error $\epsilon$ is sufficiently small. At $V=0$ it can be roughly estimated to be
\begin{equation*}
	\epsilon \sim \frac{1}{a^2},\quad a=\min\left\{\abs{k_0+\gamma + \frac12},\abs{k_0+\bar\gamma + \frac12}\right\}.
\end{equation*}

\section{Perturbation theory for the correlation kernel \label{app:perturbation-theory}}
In order to fix the initial conditions for the RG flow at high energies we evaluate the perturbative series for the correlation kernel at $T=V=\Omega=0$. To leading order the relevant diagram is given by
\begin{equation}
	\Sigma_B^\pm(E,\Omega) =
		\input{diagrams/PT/correlation-1.tikz}+
		\frac12\input{diagrams/PT/correlation-2.tikz}
		+\mathcal{O}(G^3).	
\end{equation}
The bare interaction vertex for the Kondo model is given by
\begin{equation}
	\hat{G}_{11'}^{pp'(0)}=\delta_{pp'}\frac12J^{(0)}_{\alpha\alpha'}\vec{L}^p\cdot\vec{\sigma}_{\sigma\sigma'}
\end{equation}
where $p,p'$ are the Keldysh indices. The reservoir contraction can be split into symmetric $\gamma^s$ and antisymmetric parts $\gamma^a$ with respect to the Fermi function where $\gamma=p\gamma^s+\gamma^a$ with $\gamma^s(\omega)=\tfrac12 N(\omega)$, $\gamma^a(\omega)=-\tfrac12 N(\omega)\sign(\omega)$ and the density of states $N(\omega)$. Inserting the contractions yields
\begin{multline}
	\Sigma_B^\pm(E,0) - \mathcal{B}_\pm = \frac12 G_{11'}^{(0)}\mathcal{B}_\pm G_{\bar1'\bar1}^{pp(0)}\frac{\gamma_1^p\gamma_{1'}^p}{(E_{11'}+\bar\omega_{11'}+\ii 0)^2} \\
			=\hat{G}_{11'}^{(0)}\mathcal{B}_\pm\hat{G}_{1'1}^{(0)}\iint\dd\omega\dd\omega'\,\frac{\gamma^s(\omega)\gamma^s(\omega')+\gamma^a(\omega)\gamma^a(\omega')}{(\hat E_{11'}+\omega+\omega'+\ii 0)^2}\\
			+\hat{G}_{11'}^{(0)}\mathcal{B}_\pm\tilde{G}_{1'1}^{(0)}\iint \dd\omega\dd\omega'\,\frac{2\gamma^s(\omega)\gamma^a(\omega')}{(\hat E_{11'}+\omega+\omega'+\ii 0)^2},
\end{multline}
where $\hat{G}_{11'}^{(0)}=\sum_p\hat{G}_{11'}^{pp(0)}$ and $\tilde{G}_{11'}^{(0)}=\sum_p p\hat{G}_{11'}^{pp(0)}$. Only the second part will contribute to $\Gamma^-(E,0)$. Thus for large bandwidth $D$ the integral involving one symmetric and one antisymmetric contraction is given by ($z=E+\ii 0$)
\begin{equation}
	\int \dd\omega\dd\omega'\,\frac{\rho(\omega)\rho(\omega')\sign(\omega)}{(\omega+\omega'+z)^2}\cong-\ii\pi+\mathcal{O}\left(\frac1D\right).\label{eq:symmeticasymmetic}
\end{equation}
For $S=1/2$ the bare vertices are explicitly given by 
\begin{gather*}
	\hat{G}_{11'}^{(0)}=-J_{\alpha\alpha'}^{(0)}\hat{L}^2_{\sigma\sigma'},\\
	\tilde{G}_{11'}^{(0)}=\frac12J_{\alpha\alpha'}^{(0)}\left(\hat{L}^1_{\sigma\sigma'}+\hat{L}^3_{\sigma\sigma'}\right),
\end{gather*}
thus we obtain in leading order in the exchange coupling
\begin{equation}
	\Gamma^{-(0)}(E,0)=\frac{\pi}{2}J^{(0)}_{\alpha\alpha'}J^{(0)}_{\alpha'\alpha}.
\end{equation}
Similarly for $S=1$ we obtain using 
\begin{gather*}
	\hat{G}_{11'}^{(0)}
	=\frac14 J_{\alpha\alpha'}^{(0)}\left(\hat{K}^1_{\sigma\sigma'}+\hat{K}^2_{\sigma\sigma'}\right),\\
	\tilde{G}_{11'}^{(0)}
	=\frac16 J_{\alpha\alpha'}^{(0)}\left(\hat{K}^3_{\sigma\sigma'}+\hat{K}^4_{\sigma\sigma'}\right)+\frac1{12} J_{\alpha\alpha'}^{(0)}\left(\hat{K}^5_{\sigma\sigma'}+\hat{K}^6_{\sigma\sigma'}\right)
\end{gather*}
the result ($N=2$)
\begin{equation}
	\Gamma^{-(0)}(E,0)=\frac{4\pi}{3}N J^{(0)}_{\alpha\alpha'}J^{(0)}_{\alpha'\alpha}.
\end{equation}


\end{document}

%% file: diagrams/tikz-settings.tikz
\usepackage{tikz}
\usepackage{color}
\usetikzlibrary{shapes}

\definecolor{darkgreen}{RGB}{0,152,10}
\definecolor{darkblue}{RGB}{35,0,145}
\definecolor{darkred}{RGB}{158,0,0}

\def\LOWERLINE{1.0}
\def\UPPERLINE{1.5}
\def\DISTANCETOVERTEX{2.5}
\def\DISTANCETOB{1.25}
\def\DERIVATIVEBONUS{1.0}
\def\SECONDBUBBLE{0.5}
\def\OUTGOING{0.5}

\def\FONTSIZE{\scriptsize}

\pgfdeclarelayer{edgelayer}
\pgfdeclarelayer{nodelayer}
\pgfsetlayers{edgelayer,nodelayer,main}

\tikzstyle{none}=[inner sep=0pt]

\tikzstyle{bare-vertex}=[circle, densely dotted, fill=white, draw=black, scale=0.6, thick]
\tikzstyle{B-vertex}=[circle, fill=white, draw=black, scale=0.6, thick]
\tikzstyle{effective-vertex}=[circle, fill=white, draw=black, scale=0.6, thick]
\tikzstyle{effective-zero-vertex}=[circle, fill=black, draw=black, scale=0.6, thick]

\tikzstyle{derivative-contraction}=[cross out, draw=darkblue, scale=0.8, thick]
\tikzstyle{derivative-propagator}=[strike out, draw=darkred, scale=0.8, thick]
\tikzstyle{difference-contraction}=[circle, fill=white, draw=black, scale=0.6, thick]

\tikzstyle{propagator}=[draw=black, thick]
\tikzstyle{closed-contraction}=[draw=darkgreen, thick]
\tikzstyle{outgoing-contraction}=[draw=darkblue, thick]

\tikzset{%
    baseline=4 + 2*\LOWERLINE,
    every picture/.style={scale=0.4}
}

%% file: diagrams/PT/correlation-1.tikz
\begin{tikzpicture}
	\begin{pgfonlayer}{nodelayer}
		\node [style=B-vertex] at (0, 1.0) {\Large${B_\pm}$};
	\end{pgfonlayer}
	\begin{pgfonlayer}{edgelayer}
	\end{pgfonlayer}
\end{tikzpicture}

%% file: diagrams/PT/correlation-2.tikz
\begin{tikzpicture}
	\begin{pgfonlayer}{nodelayer}
		\node [style=bare-vertex] (0) at (0, 0) {};
		\node [style=bare-vertex] (1) at (\SECONDBUBBLE, 0) {};
		\node [style=bare-vertex] (2) at (\SECONDBUBBLE + \DISTANCETOVERTEX + \SECONDBUBBLE, 0) {};
		\node [style=bare-vertex] (3) at (\SECONDBUBBLE + \DISTANCETOVERTEX, 0) {};
		\node [style=none] (4) at (0, \UPPERLINE) {};
		\node [style=none] (5) at (\SECONDBUBBLE + \DISTANCETOVERTEX, \LOWERLINE) {};
		\node [style=none] (6) at (\SECONDBUBBLE, \LOWERLINE) {};
		\node [style=none] (7) at (\SECONDBUBBLE + \DISTANCETOVERTEX + \SECONDBUBBLE, \UPPERLINE) {};
		\node [style=B-vertex] at (\SECONDBUBBLE + \DISTANCETOB, 0) {$B_\pm$};
	\end{pgfonlayer}
	\begin{pgfonlayer}{edgelayer}
		\draw [style=closed-contraction] (0) to (4.center) to (7.center) to (2);
		\draw [style=closed-contraction] (1) to (6.center) to (5.center) to (3);
		\draw [style=propagator] (0) to (2);
	\end{pgfonlayer}
\end{tikzpicture}

%% file: diagrams/PT/correlation-31.tikz
\begin{tikzpicture}
	\begin{pgfonlayer}{nodelayer}
		\node [style=bare-vertex] (0) at (0, 0) {};
		\node [style=bare-vertex] (1) at (\SECONDBUBBLE, 0) {};
		\node [style=bare-vertex] (2) at (\SECONDBUBBLE + 2*\DISTANCETOB + \SECONDBUBBLE, 0) {};
		\node [style=bare-vertex] (3) at (\SECONDBUBBLE + 2*\DISTANCETOB, 0) {};
		\node [style=none] (4) at (0, \UPPERLINE) {};
		\node [style=none] (5) at (\SECONDBUBBLE + 2*\DISTANCETOB, \LOWERLINE) {};
		\node [style=none] (6) at (\SECONDBUBBLE, \LOWERLINE) {};
		\node [style=none] (7) at (\SECONDBUBBLE + 2*\DISTANCETOB + \SECONDBUBBLE, \LOWERLINE) {};
		\node [style=B-vertex] at (\SECONDBUBBLE + \DISTANCETOB, 0) {$B_\pm$};
		\node [style=bare-vertex] (8) at (\SECONDBUBBLE + \DISTANCETOVERTEX + \SECONDBUBBLE + 2*\DISTANCETOB, 0) {};
		\node [style=bare-vertex] (9) at (\SECONDBUBBLE + \DISTANCETOVERTEX + \SECONDBUBBLE + 2*\DISTANCETOB + \SECONDBUBBLE, 0) {};
		\node [style=none] (10) at (\SECONDBUBBLE + \DISTANCETOVERTEX + \SECONDBUBBLE + 2*\DISTANCETOB, \LOWERLINE) {};
		\node [style=none] (11) at (\SECONDBUBBLE + \DISTANCETOVERTEX + \SECONDBUBBLE + 2*\DISTANCETOB + \SECONDBUBBLE, \UPPERLINE) {};
	\end{pgfonlayer}
	\begin{pgfonlayer}{edgelayer}
		\draw [style=closed-contraction] (0) to (4.center) to (11.center) to (9);
		\draw [style=closed-contraction] (1) to (6.center) to (5.center) to (3);
		\draw [style=closed-contraction] (2) to (7.center) to (10.center) to (8);
		\draw [style=propagator] (0) to (9);
	\end{pgfonlayer}
\end{tikzpicture}

%% file: diagrams/PT/correlation-32.tikz
\begin{tikzpicture}
	\begin{pgfonlayer}{nodelayer}
		\node [style=bare-vertex] (0) at (0, 0) {};
		\node [style=bare-vertex] (1) at (\SECONDBUBBLE, 0) {};
		\node [style=bare-vertex] (2) at (\SECONDBUBBLE + \DISTANCETOVERTEX + \SECONDBUBBLE, 0) {};
		\node [style=bare-vertex] (3) at (\SECONDBUBBLE + \DISTANCETOVERTEX, 0) {};
		\node [style=none] (4) at (0, \UPPERLINE) {};
		\node [style=none] (5) at (\SECONDBUBBLE + \DISTANCETOVERTEX, \LOWERLINE) {};
		\node [style=none] (6) at (\SECONDBUBBLE, \LOWERLINE) {};
		\node [style=none] (7) at (\SECONDBUBBLE + \DISTANCETOVERTEX + \SECONDBUBBLE, \LOWERLINE) {};
		\node [style=B-vertex] at (\SECONDBUBBLE + \DISTANCETOVERTEX + \SECONDBUBBLE + \DISTANCETOB, 0) {$B_\pm$};
		\node [style=bare-vertex] (8) at (\SECONDBUBBLE + \DISTANCETOVERTEX + \SECONDBUBBLE + 2*\DISTANCETOB, 0) {};
		\node [style=bare-vertex] (9) at (\SECONDBUBBLE + \DISTANCETOVERTEX + \SECONDBUBBLE + 2*\DISTANCETOB + \SECONDBUBBLE, 0) {};
		\node [style=none] (10) at (\SECONDBUBBLE + \DISTANCETOVERTEX + \SECONDBUBBLE + 2*\DISTANCETOB, \LOWERLINE) {};
		\node [style=none] (11) at (\SECONDBUBBLE + \DISTANCETOVERTEX + \SECONDBUBBLE + 2*\DISTANCETOB + \SECONDBUBBLE, \UPPERLINE) {};
	\end{pgfonlayer}
	\begin{pgfonlayer}{edgelayer}
		\draw [style=closed-contraction] (0) to (4.center) to (11.center) to (9);
		\draw [style=closed-contraction] (1) to (6.center) to (5.center) to (3);
		\draw [style=closed-contraction] (2) to (7.center) to (10.center) to (8);
		\draw [style=propagator] (0) to (9);
	\end{pgfonlayer}
\end{tikzpicture}

%% file: diagrams/PT/vertex-0.tikz
\begin{tikzpicture}
	\begin{pgfonlayer}{nodelayer}
		\node [style=bare-vertex] (0) at (0, 0) {};
		\node [style=bare-vertex] (1) at (\SECONDBUBBLE, 0) {};
		\node [below] at (0, 0) {\FONTSIZE 1};
		\node [below] at (\SECONDBUBBLE, 0) {\FONTSIZE 2};
		\node [style=none] (2) at (0, \UPPERLINE) {};
		\node [style=none] (3) at (\SECONDBUBBLE, \LOWERLINE) {};
		\node [style=none] (4) at (\SECONDBUBBLE + \OUTGOING, \UPPERLINE) {};
		\node [style=none] (5) at (\SECONDBUBBLE + \OUTGOING, \LOWERLINE) {};
	\end{pgfonlayer}
	\begin{pgfonlayer}{edgelayer}
		\draw [style=outgoing-contraction] (0) to (2.center) to (4.center);
		\draw [style=outgoing-contraction] (1) to (3.center) to (5.center);
	\end{pgfonlayer}
\end{tikzpicture}

%% file: diagrams/PT/vertex-1.tikz
\begin{tikzpicture}
	\begin{pgfonlayer}{nodelayer}
		\node [style=bare-vertex] (0) at (0, 0) {};
		\node [style=bare-vertex] (1) at (\SECONDBUBBLE, 0) {};
		\node [below] at (0, 0) {\FONTSIZE 1};
		\node [style=bare-vertex] (2) at (\SECONDBUBBLE + \DISTANCETOVERTEX + \SECONDBUBBLE, 0) {};
		\node [style=bare-vertex] (3) at (\SECONDBUBBLE + \DISTANCETOVERTEX, 0) {};
		\node [below] at (\SECONDBUBBLE + \DISTANCETOVERTEX + \SECONDBUBBLE, 0) {\FONTSIZE 2};
		\node [style=none] (4) at (0, \UPPERLINE) {};
		\node [style=none] (5) at (\SECONDBUBBLE + \DISTANCETOVERTEX, \LOWERLINE) {};
		\node [style=none] (6) at (\SECONDBUBBLE, \LOWERLINE) {};
		\node [style=none] (7) at (2*\SECONDBUBBLE + \DISTANCETOVERTEX + \OUTGOING, \UPPERLINE) {};
		\node [style=none] (8) at (2*\SECONDBUBBLE + \DISTANCETOVERTEX + \OUTGOING, \LOWERLINE) {};
		\node [style=none] (9) at (\SECONDBUBBLE + \DISTANCETOVERTEX + \SECONDBUBBLE, \LOWERLINE) {};
	\end{pgfonlayer}
	\begin{pgfonlayer}{edgelayer}
		\draw [style=outgoing-contraction] (0) to (4.center) to (7.center);
		\draw [style=closed-contraction] (1) to (6.center) to (5.center) to (3);
		\draw [style=outgoing-contraction] (2) to (9.center) to (8.center);
		\draw [style=propagator] (0) to (2);
	\end{pgfonlayer}
\end{tikzpicture}

%% file: diagrams/derivation/correlation-2-left.tikz
\begin{tikzpicture}
	\begin{pgfonlayer}{nodelayer}
		\node [style=effective-vertex] (0) at (0, 0) {};
		\node [style=effective-vertex] (1) at (\SECONDBUBBLE, 0) {};
		\node [style=effective-vertex] (2) at (\SECONDBUBBLE + 2*\DISTANCETOB + \DERIVATIVEBONUS + \SECONDBUBBLE, 0) {};
		\node [style=effective-vertex] (3) at (\SECONDBUBBLE + 2*\DISTANCETOB + \DERIVATIVEBONUS, 0) {};
		\node [style=none] (4) at (0, \UPPERLINE) {};
		\node [style=none] (5) at (\SECONDBUBBLE + 2*\DISTANCETOB + \DERIVATIVEBONUS, \LOWERLINE) {};
		\node [style=none] (6) at (\SECONDBUBBLE, \LOWERLINE) {};
		\node [style=none] (7) at (\SECONDBUBBLE + 2*\DISTANCETOB + \DERIVATIVEBONUS + \SECONDBUBBLE, \UPPERLINE) {};
		\node [style=B-vertex] at (\SECONDBUBBLE + \DISTANCETOB + \DERIVATIVEBONUS, 0) {$B_\pm$};
		\node [style=derivative-propagator] at (\SECONDBUBBLE + \DISTANCETOB / 2.0 + \DERIVATIVEBONUS / 4.0 , 0) {};
	\end{pgfonlayer}
	\begin{pgfonlayer}{edgelayer}
		\draw [style=closed-contraction] (0) to (4.center) to (7.center) to (2);
		\draw [style=closed-contraction] (1) to (6.center) to (5.center) to (3);
		\draw [style=propagator] (0) to (2);
	\end{pgfonlayer}
\end{tikzpicture}

%% file: diagrams/derivation/correlation-2-right.tikz
\begin{tikzpicture}
	\begin{pgfonlayer}{nodelayer}
		\node [style=effective-vertex] (0) at (0, 0) {};
		\node [style=effective-vertex] (1) at (\SECONDBUBBLE, 0) {};
		\node [style=effective-vertex] (2) at (\SECONDBUBBLE + 2*\DISTANCETOB + \DERIVATIVEBONUS + \SECONDBUBBLE, 0) {};
		\node [style=effective-vertex] (3) at (\SECONDBUBBLE + 2*\DISTANCETOB + \DERIVATIVEBONUS, 0) {};
		\node [style=none] (4) at (0, \UPPERLINE) {};
		\node [style=none] (5) at (\SECONDBUBBLE + 2*\DISTANCETOB + \DERIVATIVEBONUS, \LOWERLINE) {};
		\node [style=none] (6) at (\SECONDBUBBLE, \LOWERLINE) {};
		\node [style=none] (7) at (\SECONDBUBBLE + 2*\DISTANCETOB + \DERIVATIVEBONUS + \SECONDBUBBLE, \UPPERLINE) {};
		\node [style=B-vertex] at (\SECONDBUBBLE + \DISTANCETOB, 0) {$B_\pm$};
		\node [style=derivative-propagator] at (\SECONDBUBBLE + \DISTANCETOB + \DERIVATIVEBONUS * 3.0 / 4.0 + \DISTANCETOB / 2.0, 0) {};
	\end{pgfonlayer}
	\begin{pgfonlayer}{edgelayer}
		\draw [style=closed-contraction] (0) to (4.center) to (7.center) to (2);
		\draw [style=closed-contraction] (1) to (6.center) to (5.center) to (3);
		\draw [style=propagator] (0) to (2);
	\end{pgfonlayer}
\end{tikzpicture}

%% file: diagrams/derivation/correlation-1-left.tikz
\begin{tikzpicture}
	\begin{pgfonlayer}{nodelayer}
		\node [style=none] (0) at (0, 0) {};
		\node [style=none] (1) at (\SECONDBUBBLE, 0) {};
		\node [style=effective-vertex] (2) at (\SECONDBUBBLE + \DISTANCETOVERTEX + \SECONDBUBBLE, 0) {};
		\node [style=effective-vertex] (3) at (\SECONDBUBBLE + \DISTANCETOVERTEX, 0) {};
		\node [style=none] (4) at (0, \UPPERLINE) {};
		\node [style=none] (5) at (\SECONDBUBBLE + \DISTANCETOVERTEX, \LOWERLINE) {};
		\node [style=none] (6) at (\SECONDBUBBLE, \LOWERLINE) {};
		\node [style=none] (7) at (\SECONDBUBBLE + \DISTANCETOVERTEX + \SECONDBUBBLE, \UPPERLINE) {};
		\node [style=B-vertex] at (\SECONDBUBBLE / 2.0, 0) {$B_\pm$};
		\node [style=derivative-propagator] at (\SECONDBUBBLE + \DISTANCETOVERTEX / 2.0, 0) {};
	\end{pgfonlayer}
	\begin{pgfonlayer}{edgelayer}
		\draw [style=closed-contraction] (0) to (4.center) to (7.center) to (2);
		\draw [style=closed-contraction] (1) to (6.center) to (5.center) to (3);
		\draw [style=propagator] (0) to (2);
	\end{pgfonlayer}
\end{tikzpicture}

%% file: diagrams/derivation/correlation-1-right.tikz
\begin{tikzpicture}
	\begin{pgfonlayer}{nodelayer}
		\node [style=effective-vertex] (0) at (0, 0) {};
		\node [style=effective-vertex] (1) at (\SECONDBUBBLE, 0) {};
		\node [style=none] (2) at (\SECONDBUBBLE + \DISTANCETOVERTEX + \SECONDBUBBLE, 0) {};
		\node [style=none] (3) at (\SECONDBUBBLE + \DISTANCETOVERTEX, 0) {};
		\node [style=none] (4) at (0, \UPPERLINE) {};
		\node [style=none] (5) at (\SECONDBUBBLE + \DISTANCETOVERTEX, \LOWERLINE) {};
		\node [style=none] (6) at (\SECONDBUBBLE, \LOWERLINE) {};
		\node [style=none] (7) at (\SECONDBUBBLE + \DISTANCETOVERTEX + \SECONDBUBBLE, \UPPERLINE) {};
		\node [style=B-vertex] at (\SECONDBUBBLE + \DISTANCETOVERTEX + \SECONDBUBBLE / 2.0, 0) {$B_\pm$};
		\node [style=derivative-propagator] at (\SECONDBUBBLE + \DISTANCETOVERTEX / 2.0, 0) {};
	\end{pgfonlayer}
	\begin{pgfonlayer}{edgelayer}
		\draw [style=closed-contraction] (0) to (4.center) to (7.center) to (2);
		\draw [style=closed-contraction] (1) to (6.center) to (5.center) to (3);
		\draw [style=propagator] (0) to (2);
	\end{pgfonlayer}
\end{tikzpicture}

%% file: diagrams/derivation/B-vertex-0.tikz
\begin{tikzpicture}
	\begin{pgfonlayer}{nodelayer}
		\node [style=none] (0) at (0, 0) {};
		\node [style=none] (1) at (\SECONDBUBBLE, 0) {};
		\node [style=none] (2) at (0, \UPPERLINE) {};
		\node [style=none] (3) at (\SECONDBUBBLE, \LOWERLINE) {};
		\node [style=none] (4) at (\SECONDBUBBLE + \OUTGOING, \UPPERLINE) {};
		\node [style=none] (5) at (\SECONDBUBBLE + \OUTGOING, \LOWERLINE) {};
		\node [style=B-vertex] at (0.25, 0) {$B_\pm$};
	\end{pgfonlayer}
	\begin{pgfonlayer}{edgelayer}
		\draw[style=outgoing-contraction] (0) to (2.center) to (4.center);
		\draw[style=outgoing-contraction] (1) to (3.center) to (5.center);
	\end{pgfonlayer}
\end{tikzpicture}

%% file: diagrams/derivation/B-vertex-1.tikz
\begin{tikzpicture}
	\begin{pgfonlayer}{nodelayer}
		\node [style=effective-vertex] (0) at (0, 0) {};
		\node [style=effective-vertex] (1) at (\SECONDBUBBLE, 0) {};
		
		\node [style=effective-vertex] (2) at (\SECONDBUBBLE + 2*\DISTANCETOB, 0) {};
		\node [style=effective-vertex] (3) at (\SECONDBUBBLE + 2*\DISTANCETOB + \SECONDBUBBLE, 0) {};
		\node [style=B-vertex] at (\SECONDBUBBLE + \DISTANCETOB, 0) {$B_\pm$};

		\node [below] at (0, 0) {\FONTSIZE 1};
		\node [below] at (\SECONDBUBBLE + 2*\DISTANCETOB + \SECONDBUBBLE, 0) {\FONTSIZE 2};

		\node [style=none] (4) at (0, \UPPERLINE) {};
		\node [style=none] (5) at (\SECONDBUBBLE, \LOWERLINE) {};
		\node [style=none] (6) at (\SECONDBUBBLE + 2*\DISTANCETOB, \LOWERLINE) {};
		\node [style=none] (7) at (\SECONDBUBBLE + 2*\DISTANCETOB + \SECONDBUBBLE, \LOWERLINE) {};
		\node [style=none] (8) at (2*\SECONDBUBBLE + 2*\DISTANCETOB + \OUTGOING, \LOWERLINE) {};
		\node [style=none] (9) at (2*\SECONDBUBBLE + 2*\DISTANCETOB + \OUTGOING, \UPPERLINE) {};

	\end{pgfonlayer}
	\begin{pgfonlayer}{edgelayer}
		\draw[style=propagator] (0) to (3);
		\draw[style=outgoing-contraction] (0) to (4.center) to (9.center);
		\draw[style=closed-contraction] (1) to (5.center) to (6.center) to (2);
		\draw[style=outgoing-contraction] (3) to (7.center) to (8.center);
	\end{pgfonlayer}
\end{tikzpicture}

%% file: diagrams/derivation/correlation-2.tikz
\begin{tikzpicture}
	\begin{pgfonlayer}{nodelayer}
		\node [style=effective-vertex] (0) at (0, 0) {};
		\node [style=effective-vertex] (1) at (\SECONDBUBBLE, 0) {};
		\node [style=effective-vertex] (2) at (\SECONDBUBBLE + \DISTANCETOVERTEX + \SECONDBUBBLE, 0) {};
		\node [style=effective-vertex] (3) at (\SECONDBUBBLE + \DISTANCETOVERTEX, 0) {};
		\node [style=none] (4) at (0, \UPPERLINE) {};
		\node [style=none] (5) at (\SECONDBUBBLE + \DISTANCETOVERTEX, \LOWERLINE) {};
		\node [style=none] (6) at (\SECONDBUBBLE, \LOWERLINE) {};
		\node [style=none] (7) at (\SECONDBUBBLE + \DISTANCETOVERTEX + \SECONDBUBBLE, \UPPERLINE) {};
		\node [style=B-vertex] at (\SECONDBUBBLE + \DISTANCETOB, 0) {$B_\pm$};
		\node [style=derivative-contraction, scale=1.0] at (\SECONDBUBBLE + \DISTANCETOVERTEX / 2.0, \UPPERLINE) {};
	\end{pgfonlayer}
	\begin{pgfonlayer}{edgelayer}
		\draw [style=closed-contraction] (0) to (4.center) to (7.center) to (2);
		\draw [style=closed-contraction] (1) to (6.center) to (5.center) to (3);
		\draw [style=propagator] (0) to (2);
	\end{pgfonlayer}
\end{tikzpicture}

%% file: diagrams/derivation/correlation-3-right.tikz
\begin{tikzpicture}
	\begin{pgfonlayer}{nodelayer}
		\node [style=effective-vertex] (0) at (0, 0) {};
		\node [style=effective-vertex] (1) at (\SECONDBUBBLE, 0) {};
		\node [style=effective-vertex] (2) at (\SECONDBUBBLE + 2*\DISTANCETOB + \SECONDBUBBLE, 0) {};
		\node [style=effective-vertex] (3) at (\SECONDBUBBLE + 2*\DISTANCETOB, 0) {};
		\node [style=none] (4) at (0, \UPPERLINE) {};
		\node [style=none] (5) at (\SECONDBUBBLE + 2*\DISTANCETOB, \LOWERLINE) {};
		\node [style=none] (6) at (\SECONDBUBBLE, \LOWERLINE) {};
		\node [style=none] (7) at (\SECONDBUBBLE + 2*\DISTANCETOB + \SECONDBUBBLE, \LOWERLINE) {};
		\node [style=B-vertex] at (\SECONDBUBBLE + \DISTANCETOB, 0) {$B_\pm$};
		\node [style=effective-vertex] (8) at (\SECONDBUBBLE + \DISTANCETOVERTEX + \SECONDBUBBLE + 2*\DISTANCETOB, 0) {};
		\node [style=effective-vertex] (9) at (\SECONDBUBBLE + \DISTANCETOVERTEX + \SECONDBUBBLE + 2*\DISTANCETOB + \SECONDBUBBLE, 0) {};
		\node [style=none] (10) at (\SECONDBUBBLE + \DISTANCETOVERTEX + \SECONDBUBBLE + 2*\DISTANCETOB, \LOWERLINE) {};
		\node [style=none] (11) at (\SECONDBUBBLE + \DISTANCETOVERTEX + \SECONDBUBBLE + 2*\DISTANCETOB + \SECONDBUBBLE, \UPPERLINE) {};
		\node [style=derivative-contraction] at (2*\SECONDBUBBLE + 2*\DISTANCETOB + \DISTANCETOVERTEX / 2.0, \LOWERLINE) {};
	\end{pgfonlayer}
	\begin{pgfonlayer}{edgelayer}
		\draw [style=closed-contraction] (0) to (4.center) to (11.center) to (9);
		\draw [style=closed-contraction] (1) to (6.center) to (5.center) to (3);
		\draw [style=closed-contraction] (2) to (7.center) to (10.center) to (8);
		\draw [style=propagator] (0) to (9);
	\end{pgfonlayer}
\end{tikzpicture}

%% file: diagrams/derivation/correlation-3-left.tikz
\begin{tikzpicture}
	\begin{pgfonlayer}{nodelayer}
		\node [style=effective-vertex] (0) at (0, 0) {};
		\node [style=effective-vertex] (1) at (\SECONDBUBBLE, 0) {};
		\node [style=effective-vertex] (2) at (\SECONDBUBBLE + \DISTANCETOVERTEX + \SECONDBUBBLE, 0) {};
		\node [style=effective-vertex] (3) at (\SECONDBUBBLE + \DISTANCETOVERTEX, 0) {};
		\node [style=none] (4) at (0, \UPPERLINE) {};
		\node [style=none] (5) at (\SECONDBUBBLE + \DISTANCETOVERTEX, \LOWERLINE) {};
		\node [style=none] (6) at (\SECONDBUBBLE, \LOWERLINE) {};
		\node [style=none] (7) at (\SECONDBUBBLE + \DISTANCETOVERTEX + \SECONDBUBBLE, \LOWERLINE) {};
		\node [style=B-vertex] at (\SECONDBUBBLE + \DISTANCETOVERTEX + \SECONDBUBBLE + \DISTANCETOB, 0) {$B_\pm$};
		\node [style=effective-vertex] (8) at (\SECONDBUBBLE + \DISTANCETOVERTEX + \SECONDBUBBLE + 2*\DISTANCETOB, 0) {};
		\node [style=effective-vertex] (9) at (\SECONDBUBBLE + \DISTANCETOVERTEX + \SECONDBUBBLE + 2*\DISTANCETOB + \SECONDBUBBLE, 0) {};
		\node [style=none] (10) at (\SECONDBUBBLE + \DISTANCETOVERTEX + \SECONDBUBBLE + 2*\DISTANCETOB, \LOWERLINE) {};
		\node [style=none] (11) at (\SECONDBUBBLE + \DISTANCETOVERTEX + \SECONDBUBBLE + 2*\DISTANCETOB + \SECONDBUBBLE, \UPPERLINE) {};
		\node [style=derivative-contraction] at (\SECONDBUBBLE + \DISTANCETOVERTEX / 2.0, \LOWERLINE) {};
	\end{pgfonlayer}
	\begin{pgfonlayer}{edgelayer}
		\draw [style=closed-contraction] (0) to (4.center) to (11.center) to (9);
		\draw [style=closed-contraction] (1) to (6.center) to (5.center) to (3);
		\draw [style=closed-contraction] (2) to (7.center) to (10.center) to (8);
		\draw [style=propagator] (0) to (9);
	\end{pgfonlayer}
\end{tikzpicture}

%% file: diagrams/E-Flow/correlation-2.tikz
\begin{tikzpicture}
	\begin{pgfonlayer}{nodelayer}
		\node [style=effective-zero-vertex] (0) at (0, 0) {};
		\node [style=effective-zero-vertex] (1) at (\SECONDBUBBLE, 0) {};
		\node [style=effective-zero-vertex] (2) at (\SECONDBUBBLE + 2*\DISTANCETOB + \SECONDBUBBLE, 0) {};
		\node [style=effective-zero-vertex] (3) at (\SECONDBUBBLE + 2*\DISTANCETOB, 0) {};
		\node [style=none] (4) at (0, \UPPERLINE) {};
		\node [style=none] (5) at (\SECONDBUBBLE + 2*\DISTANCETOB, \LOWERLINE) {};
		\node [style=none] (6) at (\SECONDBUBBLE, \LOWERLINE) {};
		\node [style=none] (7) at (\SECONDBUBBLE + 2*\DISTANCETOB + \SECONDBUBBLE, \UPPERLINE) {};
		\node [style=B-vertex] at (\SECONDBUBBLE + \DISTANCETOB, 0) {$B_\pm$};
		\node [style=derivative-contraction] at (\SECONDBUBBLE + \DISTANCETOB, \UPPERLINE) {};
	\end{pgfonlayer}
	\begin{pgfonlayer}{edgelayer}
		\draw [style=closed-contraction] (0) to (4.center) to (7.center) to (2);
		\draw [style=closed-contraction] (1) to (6.center) to (5.center) to (3);
		\draw [style=propagator] (0) to (2);
	\end{pgfonlayer}
\end{tikzpicture}

%% file: diagrams/E-Flow/correlation-311.tikz
\begin{tikzpicture}
	\begin{pgfonlayer}{nodelayer}
		\node [style=effective-zero-vertex] (0) at (0, 0) {};
		\node [style=effective-zero-vertex] (1) at (\SECONDBUBBLE, 0) {};
		\node [style=effective-zero-vertex] (2) at (\SECONDBUBBLE + 2*\DISTANCETOB + \SECONDBUBBLE, 0) {};
		\node [style=effective-zero-vertex] (3) at (\SECONDBUBBLE + 2*\DISTANCETOB, 0) {};
		\node [style=none] (4) at (0, \UPPERLINE) {};
		\node [style=none] (5) at (\SECONDBUBBLE + 2*\DISTANCETOB, \LOWERLINE) {};
		\node [style=none] (6) at (\SECONDBUBBLE, \LOWERLINE) {};
		\node [style=none] (7) at (\SECONDBUBBLE + 2*\DISTANCETOB + \SECONDBUBBLE, \LOWERLINE) {};
		\node [style=B-vertex] at (\SECONDBUBBLE + \DISTANCETOB, 0) {$B_\pm$};
		\node [style=effective-zero-vertex] (8) at (\SECONDBUBBLE + \DISTANCETOVERTEX + \SECONDBUBBLE + 2*\DISTANCETOB, 0) {};
		\node [style=effective-zero-vertex] (9) at (\SECONDBUBBLE + \DISTANCETOVERTEX + \SECONDBUBBLE + 2*\DISTANCETOB + \SECONDBUBBLE, 0) {};
		\node [style=none] (10) at (\SECONDBUBBLE + \DISTANCETOVERTEX + \SECONDBUBBLE + 2*\DISTANCETOB, \LOWERLINE) {};
		\node [style=none] (11) at (\SECONDBUBBLE + \DISTANCETOVERTEX + \SECONDBUBBLE + 2*\DISTANCETOB + \SECONDBUBBLE, \UPPERLINE) {};
		\node [style=derivative-contraction] at (\SECONDBUBBLE + \SECONDBUBBLE / 2.0 + \DISTANCETOB + \DISTANCETOVERTEX / 2.0, \UPPERLINE) {};
		\node [style=difference-contraction] at (2*\SECONDBUBBLE + 2*\DISTANCETOB + \DISTANCETOVERTEX / 2.0, \UPPERLINE) {};
	\end{pgfonlayer}
	\begin{pgfonlayer}{edgelayer}
		\draw [style=closed-contraction] (0) to (4.center) to (11.center) to (9);
		\draw [style=closed-contraction] (1) to (6.center) to (5.center) to (3);
		\draw [style=closed-contraction] (2) to (7.center) to (10.center) to (8);
		\draw [style=propagator] (0) to (9);
	\end{pgfonlayer}
\end{tikzpicture}

%% file: diagrams/E-Flow/correlation-321.tikz
\begin{tikzpicture}
	\begin{pgfonlayer}{nodelayer}
		\node [style=effective-zero-vertex] (0) at (0, 0) {};
		\node [style=effective-zero-vertex] (1) at (\SECONDBUBBLE, 0) {};
		\node [style=effective-zero-vertex] (2) at (\SECONDBUBBLE + \DISTANCETOVERTEX + \SECONDBUBBLE, 0) {};
		\node [style=effective-zero-vertex] (3) at (\SECONDBUBBLE + \DISTANCETOVERTEX, 0) {};
		\node [style=none] (4) at (0, \UPPERLINE) {};
		\node [style=none] (5) at (\SECONDBUBBLE + \DISTANCETOVERTEX, \LOWERLINE) {};
		\node [style=none] (6) at (\SECONDBUBBLE, \LOWERLINE) {};
		\node [style=none] (7) at (\SECONDBUBBLE + \DISTANCETOVERTEX + \SECONDBUBBLE, \LOWERLINE) {};
		\node [style=B-vertex] at (\SECONDBUBBLE + \DISTANCETOVERTEX + \SECONDBUBBLE + \DISTANCETOB, 0) {$B_\pm$};
		\node [style=effective-zero-vertex] (8) at (\SECONDBUBBLE + \DISTANCETOVERTEX + \SECONDBUBBLE + 2*\DISTANCETOB, 0) {};
		\node [style=effective-zero-vertex] (9) at (\SECONDBUBBLE + \DISTANCETOVERTEX + \SECONDBUBBLE + 2*\DISTANCETOB + \SECONDBUBBLE, 0) {};
		\node [style=none] (10) at (\SECONDBUBBLE + \DISTANCETOVERTEX + \SECONDBUBBLE + 2*\DISTANCETOB, \LOWERLINE) {};
		\node [style=none] (11) at (\SECONDBUBBLE + \DISTANCETOVERTEX + \SECONDBUBBLE + 2*\DISTANCETOB + \SECONDBUBBLE, \UPPERLINE) {};
		\node [style=derivative-contraction] at (\SECONDBUBBLE + \SECONDBUBBLE / 2.0 + \DISTANCETOB + \DISTANCETOVERTEX / 2.0, \UPPERLINE) {};
		\node [style=difference-contraction] at (\SECONDBUBBLE + \DISTANCETOVERTEX / 2.0, \UPPERLINE) {};
	\end{pgfonlayer}
	\begin{pgfonlayer}{edgelayer}
		\draw [style=closed-contraction] (0) to (4.center) to (11.center) to (9);
		\draw [style=closed-contraction] (1) to (6.center) to (5.center) to (3);
		\draw [style=closed-contraction] (2) to (7.center) to (10.center) to (8);
		\draw [style=propagator] (0) to (9);
	\end{pgfonlayer}
\end{tikzpicture}

%% file: diagrams/E-Flow/correlation-312.tikz
\begin{tikzpicture}
	\begin{pgfonlayer}{nodelayer}
		\node [style=effective-zero-vertex] (0) at (0, 0) {};
		\node [style=effective-zero-vertex] (1) at (\SECONDBUBBLE, 0) {};
		\node [style=effective-zero-vertex] (2) at (\SECONDBUBBLE + 2*\DISTANCETOB + \SECONDBUBBLE, 0) {};
		\node [style=effective-zero-vertex] (3) at (\SECONDBUBBLE + 2*\DISTANCETOB, 0) {};
		\node [style=none] (4) at (0, \UPPERLINE) {};
		\node [style=none] (5) at (\SECONDBUBBLE + 2*\DISTANCETOB, \LOWERLINE) {};
		\node [style=none] (6) at (\SECONDBUBBLE, \LOWERLINE) {};
		\node [style=none] (7) at (\SECONDBUBBLE + 2*\DISTANCETOB + \SECONDBUBBLE, \LOWERLINE) {};
		\node [style=B-vertex] at (\SECONDBUBBLE + \DISTANCETOB, 0) {$B_\pm$};
		\node [style=effective-zero-vertex] (8) at (\SECONDBUBBLE + \DISTANCETOVERTEX + \SECONDBUBBLE + 2*\DISTANCETOB, 0) {};
		\node [style=effective-zero-vertex] (9) at (\SECONDBUBBLE + \DISTANCETOVERTEX + \SECONDBUBBLE + 2*\DISTANCETOB + \SECONDBUBBLE, 0) {};
		\node [style=none] (10) at (\SECONDBUBBLE + \DISTANCETOVERTEX + \SECONDBUBBLE + 2*\DISTANCETOB, \LOWERLINE) {};
		\node [style=none] (11) at (\SECONDBUBBLE + \DISTANCETOVERTEX + \SECONDBUBBLE + 2*\DISTANCETOB + \SECONDBUBBLE, \UPPERLINE) {};
		\node [style=derivative-contraction] at (\SECONDBUBBLE + \DISTANCETOB, \LOWERLINE) {};
		\node [style=difference-contraction] at (2*\SECONDBUBBLE + 2*\DISTANCETOB + \DISTANCETOVERTEX / 2.0, \UPPERLINE) {};
	\end{pgfonlayer}
	\begin{pgfonlayer}{edgelayer}
		\draw [style=closed-contraction] (0) to (4.center) to (11.center) to (9);
		\draw [style=closed-contraction] (1) to (6.center) to (5.center) to (3);
		\draw [style=closed-contraction] (2) to (7.center) to (10.center) to (8);
		\draw [style=propagator] (0) to (9);
	\end{pgfonlayer}
\end{tikzpicture}

%% file: diagrams/E-Flow/correlation-322.tikz
\begin{tikzpicture}
	\begin{pgfonlayer}{nodelayer}
		\node [style=effective-zero-vertex] (0) at (0, 0) {};
		\node [style=effective-zero-vertex] (1) at (\SECONDBUBBLE, 0) {};
		\node [style=effective-zero-vertex] (2) at (\SECONDBUBBLE + \DISTANCETOVERTEX + \SECONDBUBBLE, 0) {};
		\node [style=effective-zero-vertex] (3) at (\SECONDBUBBLE + \DISTANCETOVERTEX, 0) {};
		\node [style=none] (4) at (0, \UPPERLINE) {};
		\node [style=none] (5) at (\SECONDBUBBLE + \DISTANCETOVERTEX, \LOWERLINE) {};
		\node [style=none] (6) at (\SECONDBUBBLE, \LOWERLINE) {};
		\node [style=none] (7) at (\SECONDBUBBLE + \DISTANCETOVERTEX + \SECONDBUBBLE, \LOWERLINE) {};
		\node [style=B-vertex] at (\SECONDBUBBLE + \DISTANCETOVERTEX + \SECONDBUBBLE + \DISTANCETOB, 0) {$B_\pm$};
		\node [style=effective-zero-vertex] (8) at (\SECONDBUBBLE + \DISTANCETOVERTEX + \SECONDBUBBLE + 2*\DISTANCETOB, 0) {};
		\node [style=effective-zero-vertex] (9) at (\SECONDBUBBLE + \DISTANCETOVERTEX + \SECONDBUBBLE + 2*\DISTANCETOB + \SECONDBUBBLE, 0) {};
		\node [style=none] (10) at (\SECONDBUBBLE + \DISTANCETOVERTEX + \SECONDBUBBLE + 2*\DISTANCETOB, \LOWERLINE) {};
		\node [style=none] (11) at (\SECONDBUBBLE + \DISTANCETOVERTEX + \SECONDBUBBLE + 2*\DISTANCETOB + \SECONDBUBBLE, \UPPERLINE) {};
		\node [style=derivative-contraction] at (2*\SECONDBUBBLE + \DISTANCETOVERTEX + \DISTANCETOB, \LOWERLINE) {};
		\node [style=difference-contraction] at (\SECONDBUBBLE + \DISTANCETOVERTEX / 2.0, \UPPERLINE) {};
	\end{pgfonlayer}
	\begin{pgfonlayer}{edgelayer}
		\draw [style=closed-contraction] (0) to (4.center) to (11.center) to (9);
		\draw [style=closed-contraction] (1) to (6.center) to (5.center) to (3);
		\draw [style=closed-contraction] (2) to (7.center) to (10.center) to (8);
		\draw [style=propagator] (0) to (9);
	\end{pgfonlayer}
\end{tikzpicture}

%% file: diagrams/E-Flow/correlation-313.tikz
\begin{tikzpicture}
	\begin{pgfonlayer}{nodelayer}
		\node [style=effective-zero-vertex] (0) at (0, 0) {};
		\node [style=effective-zero-vertex] (1) at (\SECONDBUBBLE, 0) {};
		\node [style=effective-zero-vertex] (2) at (\SECONDBUBBLE + 2*\DISTANCETOB + \SECONDBUBBLE, 0) {};
		\node [style=effective-zero-vertex] (3) at (\SECONDBUBBLE + 2*\DISTANCETOB, 0) {};
		\node [style=none] (4) at (0, \UPPERLINE) {};
		\node [style=none] (5) at (\SECONDBUBBLE + 2*\DISTANCETOB, \LOWERLINE) {};
		\node [style=none] (6) at (\SECONDBUBBLE, \LOWERLINE) {};
		\node [style=none] (7) at (\SECONDBUBBLE + 2*\DISTANCETOB + \SECONDBUBBLE, \LOWERLINE) {};
		\node [style=B-vertex] at (\SECONDBUBBLE + \DISTANCETOB, 0) {$B_\pm$};
		\node [style=effective-zero-vertex] (8) at (\SECONDBUBBLE + \DISTANCETOVERTEX + \SECONDBUBBLE + 2*\DISTANCETOB, 0) {};
		\node [style=effective-zero-vertex] (9) at (\SECONDBUBBLE + \DISTANCETOVERTEX + \SECONDBUBBLE + 2*\DISTANCETOB + \SECONDBUBBLE, 0) {};
		\node [style=none] (10) at (\SECONDBUBBLE + \DISTANCETOVERTEX + \SECONDBUBBLE + 2*\DISTANCETOB, \LOWERLINE) {};
		\node [style=none] (11) at (\SECONDBUBBLE + \DISTANCETOVERTEX + \SECONDBUBBLE + 2*\DISTANCETOB + \SECONDBUBBLE, \UPPERLINE) {};
		\node [style=derivative-contraction] at (2*\SECONDBUBBLE + 2*\DISTANCETOB + \DISTANCETOVERTEX / 2.0, \LOWERLINE) {};
	\end{pgfonlayer}
	\begin{pgfonlayer}{edgelayer}
		\draw [style=closed-contraction] (0) to (4.center) to (11.center) to (9);
		\draw [style=closed-contraction] (1) to (6.center) to (5.center) to (3);
		\draw [style=closed-contraction] (2) to (7.center) to (10.center) to (8);
		\draw [style=propagator] (0) to (9);
	\end{pgfonlayer}
\end{tikzpicture}

%% file: diagrams/E-Flow/correlation-323.tikz
\begin{tikzpicture}
	\begin{pgfonlayer}{nodelayer}
		\node [style=effective-zero-vertex] (0) at (0, 0) {};
		\node [style=effective-zero-vertex] (1) at (\SECONDBUBBLE, 0) {};
		\node [style=effective-zero-vertex] (2) at (\SECONDBUBBLE + \DISTANCETOVERTEX + \SECONDBUBBLE, 0) {};
		\node [style=effective-zero-vertex] (3) at (\SECONDBUBBLE + \DISTANCETOVERTEX, 0) {};
		\node [style=none] (4) at (0, \UPPERLINE) {};
		\node [style=none] (5) at (\SECONDBUBBLE + \DISTANCETOVERTEX, \LOWERLINE) {};
		\node [style=none] (6) at (\SECONDBUBBLE, \LOWERLINE) {};
		\node [style=none] (7) at (\SECONDBUBBLE + \DISTANCETOVERTEX + \SECONDBUBBLE, \LOWERLINE) {};
		\node [style=B-vertex] at (\SECONDBUBBLE + \DISTANCETOVERTEX + \SECONDBUBBLE + \DISTANCETOB, 0) {$B_\pm$};
		\node [style=effective-zero-vertex] (8) at (\SECONDBUBBLE + \DISTANCETOVERTEX + \SECONDBUBBLE + 2*\DISTANCETOB, 0) {};
		\node [style=effective-zero-vertex] (9) at (\SECONDBUBBLE + \DISTANCETOVERTEX + \SECONDBUBBLE + 2*\DISTANCETOB + \SECONDBUBBLE, 0) {};
		\node [style=none] (10) at (\SECONDBUBBLE + \DISTANCETOVERTEX + \SECONDBUBBLE + 2*\DISTANCETOB, \LOWERLINE) {};
		\node [style=none] (11) at (\SECONDBUBBLE + \DISTANCETOVERTEX + \SECONDBUBBLE + 2*\DISTANCETOB + \SECONDBUBBLE, \UPPERLINE) {};
		\node [style=derivative-contraction] at (\SECONDBUBBLE + \DISTANCETOVERTEX / 2.0, \LOWERLINE) {};
	\end{pgfonlayer}
	\begin{pgfonlayer}{edgelayer}
		\draw [style=closed-contraction] (0) to (4.center) to (11.center) to (9);
		\draw [style=closed-contraction] (1) to (6.center) to (5.center) to (3);
		\draw [style=closed-contraction] (2) to (7.center) to (10.center) to (8);
		\draw [style=propagator] (0) to (9);
	\end{pgfonlayer}
\end{tikzpicture}

%% file: diagrams/description/bare-vertex.tikz
\begin{tikzpicture}[baseline=-3]
	\begin{pgfonlayer}{nodelayer}
		\node [style=bare-vertex] (0) at (0, 0) {};
		\node [style=bare-vertex] (1) at (\SECONDBUBBLE, 0) {};
		\node [style=none] (2) at (-1, 0) {};
		\node [style=none] (3) at (\SECONDBUBBLE + 1.0, 0) {};
		\node [style=none] (4) at (0, 0.5) {};
		\node [style=none] (5) at (\SECONDBUBBLE, 0.5) {};
	\end{pgfonlayer}
	\begin{pgfonlayer}{edgelayer}
		\draw [style=propagator] (2) to (3);
		\draw [style=propagator] (0) to (4);
		\draw [style=propagator] (1) to (5);
	\end{pgfonlayer}
\end{tikzpicture}

%% file: diagrams/description/effective-vertex.tikz
\begin{tikzpicture}[baseline=-3]
	\begin{pgfonlayer}{nodelayer}
		\node [style=effective-vertex] (0) at (0, 0) {};
		\node [style=effective-vertex] (1) at (\SECONDBUBBLE, 0) {};
		\node [style=none] (2) at (-1, 0) {};
		\node [style=none] (3) at (\SECONDBUBBLE + 1.0, 0) {};
		\node [style=none] (4) at (0, 0.5) {};
		\node [style=none] (5) at (\SECONDBUBBLE, 0.5) {};
	\end{pgfonlayer}
	\begin{pgfonlayer}{edgelayer}
		\draw [style=propagator] (2) to (3);
		\draw [style=propagator] (0) to (4);
		\draw [style=propagator] (1) to (5);
	\end{pgfonlayer}
\end{tikzpicture}

%% file: diagrams/description/effective-zero-vertex.tikz
\begin{tikzpicture}[baseline=-3]
	\begin{pgfonlayer}{nodelayer}
		\node [style=effective-zero-vertex] (0) at (0, 0) {};
		\node [style=effective-zero-vertex] (1) at (\SECONDBUBBLE, 0) {};
		\node [style=none] (2) at (-1, 0) {};
		\node [style=none] (3) at (\SECONDBUBBLE + 1.0, 0) {};
		\node [style=none] (4) at (0, 0.5) {};
		\node [style=none] (5) at (\SECONDBUBBLE, 0.5) {};
	\end{pgfonlayer}
	\begin{pgfonlayer}{edgelayer}
		\draw [style=propagator] (2) to (3);
		\draw [style=propagator] (0) to (4);
		\draw [style=propagator] (1) to (5);
	\end{pgfonlayer}
\end{tikzpicture}

%% file: diagrams/description/B-vertex.tikz
\begin{tikzpicture}[baseline=-3]
	\begin{pgfonlayer}{nodelayer}
		\node [style=B-vertex] at (0.25, 0) {$B_\pm$};
		\node [style=none] (2) at (-1,0) {};
		\node [style=none] (3) at (1.5,0) {};
	\end{pgfonlayer}
	\begin{pgfonlayer}{edgelayer}
		\draw [style=propagator] (2) to (3);
	\end{pgfonlayer}
\end{tikzpicture}

%% file: diagrams/description/propagator.tikz
\begin{tikzpicture}[baseline=-3]
	\begin{pgfonlayer}{nodelayer}
	\node [style=none] (0) at (-1,0) {};
	\node [style=none] (1) at (1.5,0) {};
	\end{pgfonlayer}
	\begin{pgfonlayer}{edgelayer}
		\draw [style=propagator] (0) to (1);
	\end{pgfonlayer}
\end{tikzpicture}

%% file: diagrams/description/derivative-propagator.tikz
\begin{tikzpicture}[baseline=-3]
	\begin{pgfonlayer}{nodelayer}
		\node [style=none] (0) at (-1,0) {};
		\node [style=none] (1) at (1.5,0) {};
		\node [style=derivative-propagator] at (0.25,0) {};
	\end{pgfonlayer}
	\begin{pgfonlayer}{edgelayer}
		\draw [style=propagator] (0) to (1);
	\end{pgfonlayer}
\end{tikzpicture}

%% file: diagrams/description/contraction-outgoing.tikz
\begin{tikzpicture}[baseline=-3]
	\begin{pgfonlayer}{nodelayer}
		\node [style=none] (0) at (-1,0.0) {};
		\node [style=none] (1) at (1.5,0.0) {};
	\end{pgfonlayer}
	\begin{pgfonlayer}{edgelayer}
		\draw [style=outgoing-contraction] (0) to (1);
	\end{pgfonlayer}
\end{tikzpicture}

%% file: diagrams/description/contraction-closed.tikz
\begin{tikzpicture}[baseline=-3]
	\begin{pgfonlayer}{nodelayer}
		\node [style=none] (0) at (-1,0) {};
		\node [style=none] (1) at (1.5,0) {};
	\end{pgfonlayer}
	\begin{pgfonlayer}{edgelayer}
		\draw [style=closed-contraction] (0) to (1);
	\end{pgfonlayer}
\end{tikzpicture}

%% file: diagrams/description/derivative-contraction.tikz
\begin{tikzpicture}[baseline=-3]
	\begin{pgfonlayer}{nodelayer}
		\node [style=none] (0) at (-1, 0) {};
		\node [style=none] (1) at (1.5, 0) {};
		\node [style=derivative-contraction] at (0.25, 0) {};
	\end{pgfonlayer}
	\begin{pgfonlayer}{edgelayer}
		\draw [style=closed-contraction] (0) to (1);
	\end{pgfonlayer}
\end{tikzpicture}

%% file: diagrams/description/difference-contraction.tikz
\begin{tikzpicture}[baseline=-3]
	\begin{pgfonlayer}{nodelayer}
		\node [style=none] (0) at (-1, 0) {};
		\node [style=none] (1) at (1.5, 0) {};
		\node [style=difference-contraction] at (0.25, 0) {};
	\end{pgfonlayer}
	\begin{pgfonlayer}{edgelayer}
		\draw [style=closed-contraction] (0) to (1);
	\end{pgfonlayer}
\end{tikzpicture}

%% file: SK-paper2.bbl
\begin{thebibliography}{100}
\bibitem{footnote4}
We note that our result differs by a factor $1/2$ from Eq.~(6.79) in Ref.~\onlinecite{Hewson93}. Our result is consistent with the low-temperature behaviour of the susceptibility obtained from numerical RG   calculations.\cite{HanlWeichselbaum14,Merker-12}
\bibitem{footnote1}
We note that in previous works\cite{Schoeller09,SchoellerReininghaus09,SS09,HSA12} the reduced density matrix and effective Liouvillian were denoted by $\rho_\text{S}(E)$ and $L_\text{S}^\text{eff}(E)$ respectively.
\bibitem{footnote2}
We explicitly add the flow parameter $E$ as the first variable and drop the
  second variable $\xi=\ii 0^+$ of $\Sigma_B^\pm$ introduced in
  Ref.~\onlinecite{SS09} since this limit is always implicitly taken.
\bibitem{footnote3}
We note that for general correlation functions a third kernel, $\Sigma_{AB}$,
  appears.\cite{SS09} For the spin-spin correlations on the dot this is,
  however, not the case.
\bibitem{footnote5}
The reason is the Friedel sum rule with a frozen occupancy of $1$ on the dot.
\end{thebibliography}
